\documentclass[a4paper,usenatbib]{mnras}

\usepackage[T1]{fontenc}
\usepackage{ae,aecompl}

\usepackage{amssymb}	% Extra maths symbols
\usepackage{natbib}
\usepackage{graphicx}
\usepackage{rotating,times,graphicx,latexsym}
\usepackage{color}
\usepackage{longtable}
\usepackage{amsmath}
\usepackage{lscape}
\usepackage{lipsum} % To generate text
\usepackage{array}
\usepackage{soul} % To add highlighting of text
\usepackage{xargs} % Use more than one optional parameter in a new commands
\usepackage[pdftex,dvipsnames]{xcolor}  % Coloured text etc.
\usepackage{amsmath}
\usepackage{siunitx}
\usepackage{subcaption}
\usepackage{longtable}

\newcommand{\mic}{\,$\mu$m~}

\newcommand{\dg}{$^{\circ}$}

\newcommand{\Msolar}{\,M$_{\odot}$~}

\newcommand{\hii}{H{\sc ii}~}

\newcommand{\msun}{\ensuremath{\,\mathrm{M}_{\odot}}}
\newcommand{\lsun}{\ensuremath{\mathrm{L}_{\odot}}}
\newcommand{\rsun}{\ensuremath{\mathrm{R}_{\odot}}}
 
\newcommand{\paperIt}{\citetalias{2018MNRAS.477.5486C}}
\newcommand{\paperIp}{\citepalias{2018MNRAS.477.5486C}}
\title[Shape Analysis of \hii Regions -- II]{Shape Analysis of \hii Regions -- II. Synthetic Observations}

\author[J. Campbell-White et al.]{
Justyn Campbell-White,$^{1,2}$\thanks{E-mail: \href{mailto:JCampbellWhite001@dundee.ac.uk}{JCampbellWhite001@dundee.ac.uk}}
Ahmad A. Ali,$^{3}$
Dirk Froebrich,$^{2}$
Alfred Kume$^{4}$
\\
$^{1}$SUPA, School of Science and Engineering, University of Dundee, Nethergate, Dundee DD1 4HN, U.K.\\
$^{2}$Centre for Astrophysics and Planetary Science, The University of Kent, Canterbury, CT2 7NH, U.K.\\
$^{3}$Department of Physics and Astronomy, University of Exeter, Stocker Road, Exeter EX4 4QL, U.K.\\
$^{4}$School of Mathematics, Statistics and Actuarial Sciences, The University of Kent, Canterbury, CT2 7FS, U.K.\\
}

\date{Accepted XXX. Received YYY; in original form ZZZ}

\pubyear{2020}

\begin{document}
\label{firstpage}
\pagerange{\pageref{firstpage}--\pageref{lastpage}}
\maketitle

\begin{abstract}
The statistical shape analysis method developed for probing the link between physical parameters and morphologies of Galactic \hii regions is applied here to a set of synthetic observations (SOs) of a numerically modelled \hii region. The systematic extraction of \hii region shape, presented in the first paper of this series, allows for a quantifiable confirmation of the accuracy of the numerical simulation, with respect to the real observational counterparts of the resulting SOs. A further aim of this investigation is to determine whether such SOs can be used for direct interpretation of the observational data, in a future supervised classification scheme based upon \hii region shape. The numerical \hii region data was the result of photoionisation and radiation pressure feedback of a 34 \Msolar star, in a 1000 \Msolar cloud. The SOs analysed herein comprised four evolutionary snapshots (0.1, 0.2, 0.4 and 0.6\,Myr), and multiple viewing projection angles. The shape analysis results provided conclusive evidence of the efficacy of the numerical simulations. When comparing the shapes of the synthetic regions to their observational counterparts, the SOs were grouped in amongst the Galactic \hii regions by the hierarchical clustering procedure. There was also an association between the evolutionary distribution of regions and the respective groups. This suggested that the shape analysis method could be further developed for morphological classification of \hii regions by using a synthetic data training set, with differing initial conditions of well-defined parameters.

\end{abstract}


%%%%%%%%%%%%%%%%%%%%%%%%%%%%%%%%%%%%%%%%%%%%%%%%%%

%%%%%%%%%%%%%%%%% BODY OF PAPER %%%%%%%%%%%%%%%%%%
%\listoftodos %uncomment if any \todos are added in body

%%%%%%%%%%%%%%%%%%%%%%%%%%%%%%%%%%%%%%%%%%%%%%%%%
%%%%%%%%%%%%%%%%%%%%%%%%%%%%%%%%%%%%%%%%%%%%%%%%%
\section{Introduction}
\defcitealias{2018MNRAS.477.5486C}{Paper I}
\defcitealias{2018MNRAS.477.5422A}{AHD18}

Since the advance in high performance computing in the latter part of the 20th century, astrophysicists have utilised these tools to perform numerical simulations of all aspects of the Universe. From modelling cloud collapse, star formation (SF) and feedback \citep[e.g.][]{2011A&A...536A..79R,2007ApJ...667..275B,2005A&A...434..167S}, to galaxy formation and evolution \citep[e.g.][]{2019MNRAS.487.2753W,2011ApJS..196...22B}, to entire cosmological models that reflect the largest scale structure astronomers have ever observed \citep[e.g.][]{2005MNRAS.364.1105S}. Radiation plays an important role in astrophysics. The transport of radiation through the interstellar medium (ISM) is therefore one of the most fundamental processes to be considered when modelling stellar objects and galactic structures. Analysing the radiation from an object not only tells us about the nature of the radiation source, but also the medium through which it has travelled to reach us. Interstellar dust therefore also plays an important role in the study of radiation, since it scatters and re-radiates UV through to IR photons \citep{2001ApJ...548..296W}. 

 Quantifiable results can be obtained directly from numerical simulations and compared with observational results, such as the stellar initial mass function \citep{1997MNRAS.288..145P}. However, there exists many important reasons why the production of synthetic observations (SOs) from the numerical models are necessary \citep[see the extensive review by][]{2018NewAR..82....1H}. In the last decade, a number of radiative transfer (RT) models have been used to generate synthetic observations (SOs) of the numerical simulation they relate to \citep{steinacker2013three}. The RT codes work by sampling the simulation at every grid point, for the given dimensionality of the simulations. Given some density and temperature, the emissivity can be computed, which is then integrated to obtain the flux. Flux images can be generated from any viewing angle the user specifies. Such SOs are referred to as `ideal synthetic observations'  \citep{2017ApJ...849....3K}, which must then be further processed in order to account for observational effects when detecting and processing astronomical radiation. Such resultant `realistic' SOs are then directly comparable to their real observational counterparts. This allows us to test observational diagnostics that are well defined in the simulations, and hence produce bespoke models for direct interpretation of the observational data. In this work, we use the statistical shape analysis method of \hii regions, developed in \citet{2018MNRAS.477.5486C} to directly compare realistic SOs of an \hii region produced by the numerical simulations in \citet{2018MNRAS.477.5422A} to radio continuum observational data.

\hii regions are the result of photoionisation from massive stars (> 8\,\msun). Due to their significant role in providing feedback to the giant molecular clouds (GMCs), in which these stars are born, they have been extensively modelled in order to probe the varying physical processes and mechanisms associated with such feedback; such as their role in altering gas dynamics and star formation. \hii regions have been shown to be fundamental in calculations of the observed Galactic star formation efficiency (SFE) \citep{2015ASSL..412...43K}. Photoionisation feedback from \hii regions can have both negative and positive effects in terms of the local SFE of GMCs. \citet{2007MNRAS.375.1291D,2007MNRAS.377..535D} showed via numerical models of GMCs irradiated by ionising stars, both internally and externally, that some stars formed earlier, compared to the control runs without photoionisation feedback. Furthermore, evidence for triggered star formation was noted, such that the overall SFE of the cloud was increased. Conversely, simulations by \citet{2013MNRAS.435..917W} found that although triggering was effective on small timescales, larger timescales resulted in a reduced SFE due to the dispersal of the gas, a further feedback mechanism of \hii region evolution. More recently, the ionising radiation models of \citet{2017MNRAS.471.4844G} displayed a low SFE that was consistent with the Galactic observations \citep[of the order a few percent,][]{2003ARA&A..41...57L}. 

In \citet[][hereafter, \paperIt]{2018MNRAS.477.5486C} we successfully applied our shape analysis methodology to a selection of 1.4\,Ghz radio continuum images of \hii regions from the MAGPIS survey \citep{2006AJ....131.2525H}. The mathematical description of the shape of each \hii region was systematically extracted from the contoured radio continuum images. By determining the local curvature values along each \hii region boundary, curvature distributions were obtained and compared pairwise using the Anderson-Darling non-parametric test statistic. The resulting test statistic distance matrix was then the subject of hierarchical clustering, allowing for the identification of groups of \hii regions that share a common morphology. From investigation of potential associations between assigned group and physical parameters, the results showed evidence for \hii regions of a given shape to have similar dynamical ages. We also found indication of ionising cluster mass to be associated to the shape groupings. We suggested that the application of this shape analysis method to SOs of \hii regions would not only give a direct quantifiable test of the efficacy of the SOs, but also allow us to further refine the methodology and reduce errors with a well defined sample set. SOs from differing initial conditions could then be used as a training set in a machine learning supervised classification scheme of \hii regions, via this shape analysis method.

This work is organised as follows: The numerical simulation of the \hii region, and corresponding SOs that are analysed in this investigation, from \citet{2018MNRAS.477.5422A} are detailed in section \ref{sec:ali_data}. In section \ref{sec:shape_ext}, we cover the shape extraction from the SOs and compare the shapes of these synthetic \hii regions to those of the MAGPIS observational sample from \citetalias{2018MNRAS.477.5486C}. In section \ref{sec:so_discussion} we investigate how different parameters such as noise and projection angle influence the identified shape of the \hii regions and their resultant groupings. We also discern whether such SOs can be used as a training set in a supervised morphological classification scheme of \hii regions and discuss further potential applications of the methodology. Our summaries and conclusions are given in section \ref{sec:conc}.

%%%%%%%%%%%%%%%%%%%%%%%%%%%%%%%%%%%%%%%%%%%%%%%%%
%%%%%%%%%%%%%%%%%%%%%%%%%%%%%%%%%%%%%%%%%%%%%%%%%
\section{The Synthetic Observation Data Sample}
\label{sec:ali_data}

\begin{table}
	\centering
	\caption[Initial Parameters of Numerical Simulation]{Table 2 from \citetalias{2018MNRAS.477.5422A}: Initial parameters of the massive star in the numerical simulation.}
	\label{tab:starparameters}
	\begin{tabular}{lc} 
		\hline
		Parameter & Value \\
		\hline
		Mass & 33.7\,\msun \\
		Luminosity & \SI{1.49e5}\,\lsun  \\
		Radius & \SI{7.59}\,\rsun  \\
		Effective temperature & \SI{41189}{K} \\
		Ionizing flux ($h\nu \geq \SI{13.6}{\eV}$) & \SI{7.36e48}{\per\s} \\
		
		\hline
	\end{tabular}
\end{table}

\begin{figure}
	\centering
	\includegraphics[width=\columnwidth]{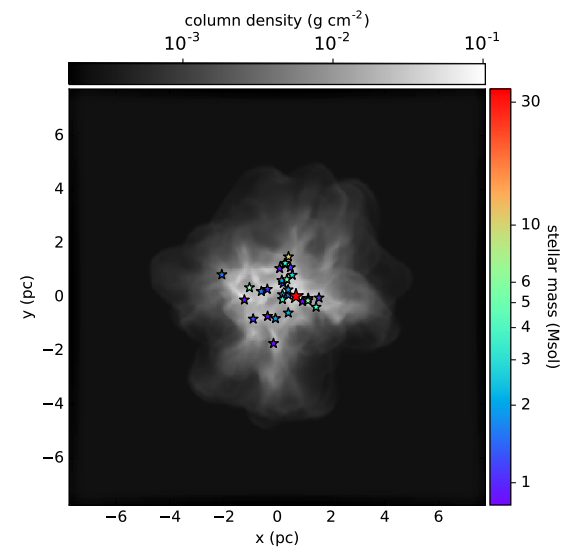}
	\caption[Column Density and Star Positions of Numerical Simulation]{Figure\,1 from \citetalias{2018MNRAS.477.5422A}: Positions of stars at the onset of feedback, with stellar mass in colour scale, overlaid on column density in greyscale (both are logarithmic). The most massive star is \SI{33.7}\,\msun\ in red. The second highest is \SI{11.3}\,\msun. The third is \SI{5.7}\,\msun. The least massive is \SI{0.82}\,\msun.}
	\label{fig:col_dens_stars}
\end{figure}

The synthetic observations of \hii regions used in this work are from the numerical simulations of \citet[][hereafter, \citetalias{2018MNRAS.477.5422A}]{2018MNRAS.477.5422A}, specifically the model including both photoionisation and radiation pressure. This simulation was performed using the Monte Carlo radiative transfer (MCRT) and hydrodynamics (HD) code \textsc{torus} \citep{2019A&C....27...63H}. The comprehensive level of detail in the radiative transfer means the resulting \hii region has accurate temperatures, ion fractions, and size.

\begin{figure*}
	\centering 
	%row 1  
	\includegraphics[width=0.24\textwidth]{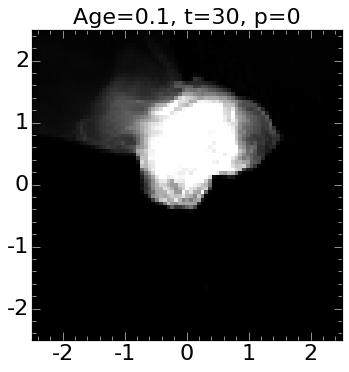}
	\includegraphics[width=0.24\textwidth]{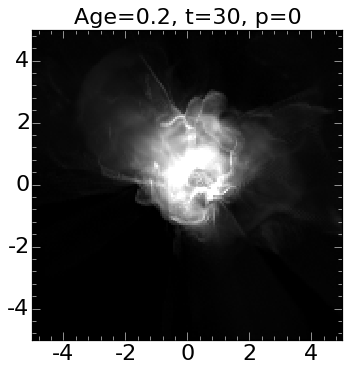} 
	\includegraphics[width=0.24\textwidth]{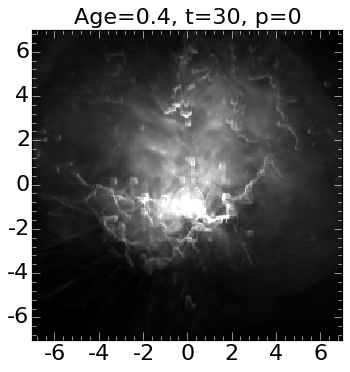} 
	\includegraphics[width=0.24\textwidth]{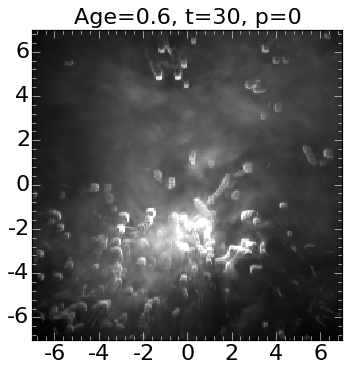} 
	
	%row 2
	\includegraphics[width=0.24\textwidth]{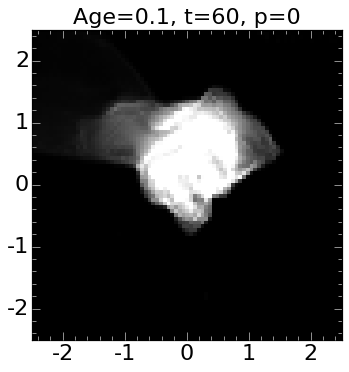}	
	\includegraphics[width=0.24\textwidth]{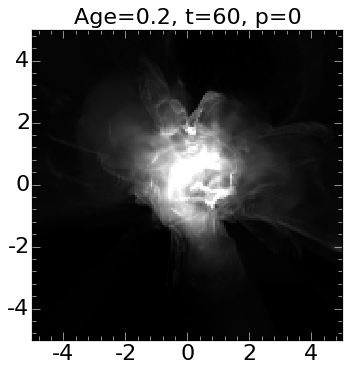}
	\includegraphics[width=0.24\textwidth]{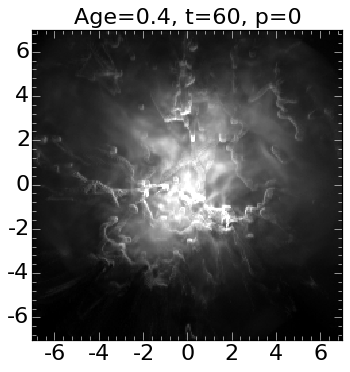}
	\includegraphics[width=0.24\textwidth]{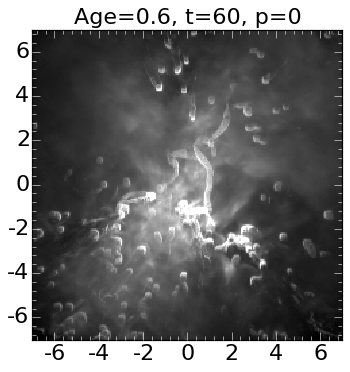}
	
	%row 3
	\includegraphics[width=0.24\textwidth]{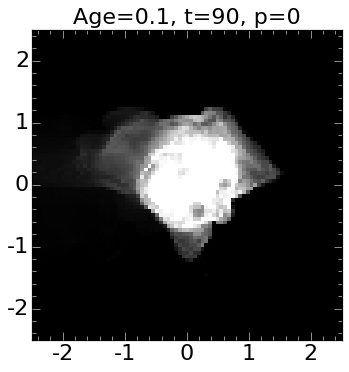}          
	\includegraphics[width=0.24\textwidth]{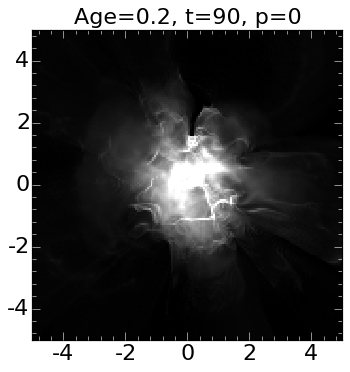}          
	\includegraphics[width=0.24\textwidth]{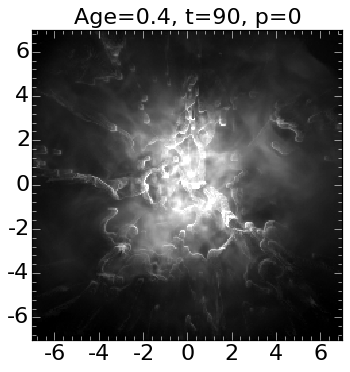}          
	\includegraphics[width=0.24\textwidth]{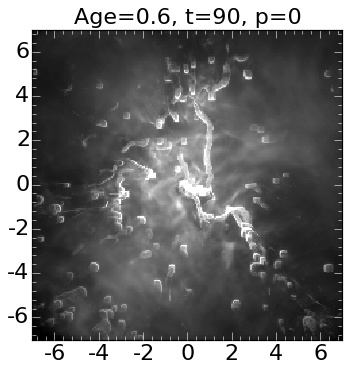}          
	
	\caption[Overview of the Synthetic Observations from the Numerical Simulation]{Overview of the 20\,cm, 1.4\,GHz synthetic observations from the numerical simulation in \citet{2018MNRAS.477.5422A}. Three snapshots of $\phi = p = 0$, $\theta = t = 30, 60$ and $90 deg$ projection angle are shown for the four evolutionary time-steps of 0.1, 0.2, 0.4 and 0.6\,Myr. Coordinates are shown in pc scale.}
	\label{fig:raw_so}
\end{figure*}

For each hydrodynamical time step, an MCRT calculation was carried out in order to compute photoionisation balance (giving ion fractions and electron densities), thermal balance (giving gas and dust temperatures), and radiation pressure. Photon wavelengths lie in the range 100 to \SI{e7}{\angstrom} and include the stellar radiation field as well as the diffuse field from gas/dust. The gas heating rates are from photoionisation heating of H and He, while the gas cooling rates involve recombination lines of H and He, collisionally excited metal forbidden lines, and free--free emission. Dust is heated by photon absorption and cooled by blackbody emission. Gas and dust temperatures are linked by a term accounting for collisional heating between the two. The included atomic species are \ion{H}{i--ii}, \ion{He}{i--iii}, \ion{C}{i--iv}, \ion{N}{i--iii}, \ion{O}{i--iii}, \ion{Ne}{i--iii}, and \ion{S}{i--iv} (the total abundance of each element remains constant throughout). The model also contains silicate dust grains with a median size of 0.12\mic -- dust can attenuate ionising photons, reducing the size of \hii regions compared to models without dust \citep{2015MNRAS.453.2277H}.

The initial condition is a spherical cloud with a uniform density inner core extending to half the sphere radius, with the density in the outer half going as $r^{-1.5}$. The density outside the sphere is 1\% of the density at the edge of the sphere. The sphere has a total mass $M= 1000$\,\msun, radius $R = 2.66$\,pc, and mean surface density $\Sigma = \SI{0.01}{\g\per\cm\squared}$. The 3D grid is Cartesian, uniform, and fixed with $256^3$ cells. The physical size of the grid is 15.5\,pc in each axis, yielding a resolution of 0.06\,pc per cell. The cloud evolves without stars under a seeded turbulent velocity field and self-gravity for 75\% of the mean free fall time of the initial sphere. This is the time at which \citet{Krumholz_2011} found the SFE to be 10\%. At this time, stars are added from a random sampling of the \citet{1955ApJ...121..161S} IMF, such that the cumulative stellar mass is 10\% of the cloud mass (100\,\msun), and at least one massive star is present. This results in a massive star of mass 33.7\,\msun, which is placed at the cloud's most massive clump, with 28 other stars placed according to a probability density function proportional to $\rho^{1.5}$. The distribution of stars at this stage is shown in Fig.\,\ref{fig:col_dens_stars}, overlaid on column density. The radiation field is then switched on and the simulation evolves until all of the mass leaves the grid. The stars evolve using \citet{1992A&AS...96..269S} tracks, with stellar spectra interpolated from atmospheric models by \citet{2003ApJS..146..417L} and \citet{1991ASIC..341..441K}. The initial mass, luminosity, radius, effective temperature and ionising photon rate of the  massive star are listed in Tab.\,\ref{tab:starparameters}. 

The ionising photon rate of the massive star is slightly larger than the mean of the normally distributed values of MAGPIS \hii regions considered in \citetalias{2018MNRAS.477.5486C} (log $N_{ly}$ of 48.87 and 48.40, respectively), but is well within the maximum observed log $N_{ly}$ value of 49.77. We also stated in \citetalias{2018MNRAS.477.5486C} that our estimates of the ionising photon rate are lower limits, due to only considering the mass within the \hii region shape boundary considered for analysis. In terms of ionising flux and mass, the numerically modelled \hii region from \citetalias{2018MNRAS.477.5422A} is therefore representative of an example \hii region from the MAGPIS sample considered for comparison in this work. Furthermore, the turbulence in the simulation results in a inhomogeneous density distribution and an off-centre massive star, meaning there could be differences in the shape of the ionised region when viewed from different projections, which will be investigated in detail in this work.

We note from Fig.\,3 in \citetalias{2018MNRAS.477.5422A} that from the onset of feedback ($t=0$) to \SI{0.6}\,Myr, there is a steady mass flow, with mass beginning to leave the simulation grid at $\sim$0.4\,Myr. After \SI{0.6}\,Myr, the overall mass flux begins to decrease. Spikes in the distribution correspond to removal of the clumps. The size of the spikes grows with time, as the densest clumps are the last to leave the grid. By $\sim$1.6\,Myr, or $0.74\,\langle t_\textrm{ff} \rangle$, all the mass has left the 15.5\,pc$^3$ grid. The peak value of ionised mass is 440\,\msun \,at 0.5\,Myr. The peak ionised mass fraction, which is just under 40\% of the total mass, is reached at 0.6\,Myr. At this time, the fraction of volume ionised is $\sim$80\%, showing that the majority of the neutral gas remains in the small, dense clumps, which resist the ionisation.

\begin{figure*}
	\centering
	\includegraphics[width=0.32\textwidth]{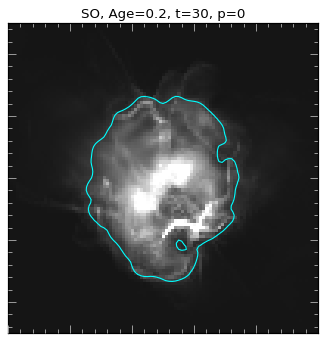}
	\includegraphics[width=0.32\textwidth]{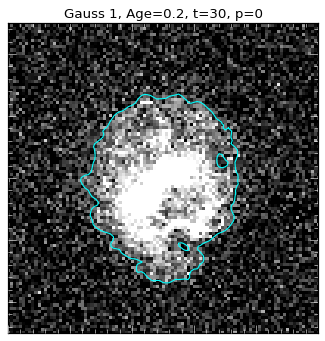} 
	\includegraphics[width=0.32\textwidth]{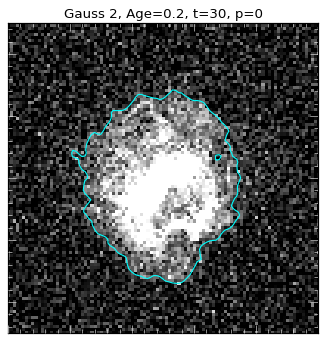}       
	\caption[Comparison of Original SO versus SO with Random Gaussian Noise]{Comparison images of example SO projection at 0.2\,Myr. Left: Original SO. Middle: Same SO with random Gaussian noise added to each pixel value. Right: As middle but with a different Gaussian distribution used. t and p are the $\theta$ and $\phi$ projection angles of the SO, respectively. The blue contours (for the Gaussian noise tiles) are the `edge' identified from the clipped mean pixel values plus 1$\sigma$. The contour in the original SO tile uses the threshold level from the middle tile.}
	\label{fig:gauss_comp}
\end{figure*}

For the synthetic observations in this paper, we use \textsc{torus} to produce free--free continuum images using the previously calculated temperature and density as inputs. The free--free emission coefficient in the radio regime is 
\begin{equation}
j_{\nu} = 5.4 \times 10^{-39} T^{-1/2} n_e^2 g_{\nu}
\end{equation}
\citep{1979rpa..book.....R} for gas temperature $T$ and electron density $n_e$, where the Gaunt factor $g_{\nu}$ is approximately
\begin{equation}
    g_{\nu} = \frac{\sqrt{3}}{\pi} \left[ \ln \left(\frac{T^{3/2}}{\nu}\right) + 17.7 \right]
\end{equation}
\citep{2006agna.book.....O}. We use $\nu=$1.4\,GHz ($\lambda=$20\,cm). 

Snapshots can be taken at given simulation ages and from any $\phi$ and $\theta$ spherical viewing angles. We chose to consider simulation ages of 0.1, 0.2, 0.4 and 0.6\,Myr. For each of these respective ages, 18, 22, 19 and 18 projections were included, resulting in 77 synthetic observations of the numerically modelled \hii region. The lower limit of the simulation ages was selected in order to produce a SO of a diffuse observed \hii region, rather than a compact or ultra-compact \hii region, which would be observed at earlier ages. We chose to include the 0.4 and 0.6\,Myr ages since they take place at key stages in the mass flow of the simulation grid, hence representing more evolved, late stage \hii regions.

Figure\,\ref{fig:raw_so} shows 12 example SOs of the 20\,cm radio continuum emission produced by the simulations. Three example projections are shown for each of the evolutionary stages (ages given in Myr). In this example, the $\phi$ angle is kept fixed (labelled $p$ in the figure headings) and the $\theta$ angle is 30, 60 and 90 $deg$ (labelled $t$ in the figure headings). The axes for each image are in parsecs. In the following section we detail how the shapes were extracted in a manner complementary to that carried out for the MAGPIS observational data in \citetalias{2018MNRAS.477.5486C}.

%%%%%%%%%%%%%%%%%%%%%%%
%%%%%%%%%%%%%%%%%%%%%%%%%%%%%%%%%%%%%%%%%%%%%%%%%
\section{Shape Extraction}
\label{sec:shape_ext}
The purpose of investigating the SOs in this work is to test the efficacy of the simulations by comparing them to the MAGPIS observations (from \citetalias{2018MNRAS.477.5486C}) and then see how different parameters may influence the shape. The extraction of the shape of the \hii regions hence needed to be performed in the same manner as it was for the MAGPIS observations. We therefore needed to process the `ideal' SOs into realistic SOs, with properties matching the observational MAGPIS sample. To achieve this, we converted the intensity units of the SOs from MJy/sr to those used in the MAGPIS observations, Jy/beam, whilst accounting for the different pixel scales. This was so that artificial noise could be added to the SOs in order to perform the contouring procedure to identify the `edge' of the \hii regions, exactly as was carried out in \citetalias{2018MNRAS.477.5486C}. 

\subsection{Gaussian Noise Profiles}
\label{ssec:gauss_np}

For the MAGPIS data, the shape of each \hii region was extracted using an image contouring procedure. After applying sigma clipping to all of the pixel values, which removed the signal from the ionised emission, the mean and standard deviation were taken from the remaining clipped values. This provided the contour value to apply to the original images to systematically define the \hii region boundaries. In order to carry out this procedure on the SOs, artificial noise was introduced to the SOs by taking a random value from a Gaussian distribution with mean, $\mu$, and standard deviation, $\sigma$, from one of the MAGPIS tiles, after sigma clipping. The standard probability density function for the Gaussian distribution was used:

\begin{equation}
	P(x) = \frac{1}{{\sigma \sqrt {2\pi } }}~e^{{{ - \left( {x - \mu } \right)^2 } \mathord{\left/ {\vphantom {{ - \left( {b - \bar{b} } \right)^2 } {2\sigma ^2 }}} \right. \kern-\nulldelimiterspace} {2\sigma ^2 }}} \label{eq:gauss}
\end{equation}

\begin{figure*}
    \centering
    %row 1
    \includegraphics[width=0.24\textwidth]{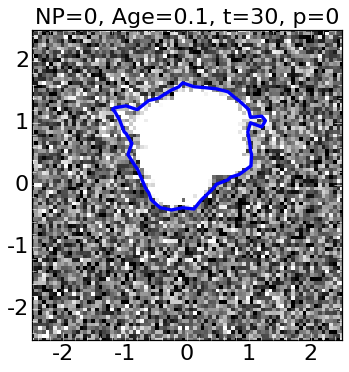} 
    \includegraphics[width=0.24\textwidth]{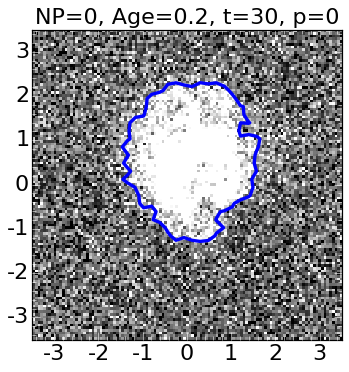} 
    \includegraphics[width=0.24\textwidth]{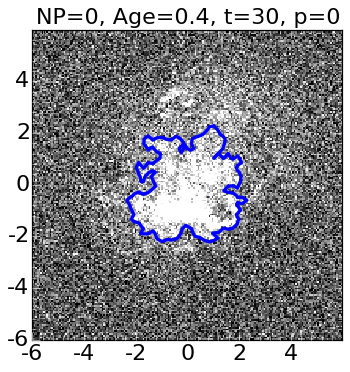} 
    \includegraphics[width=0.24\textwidth]{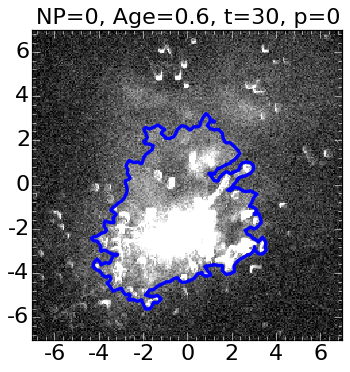} 

    %row 2
    \includegraphics[width=0.24\textwidth]{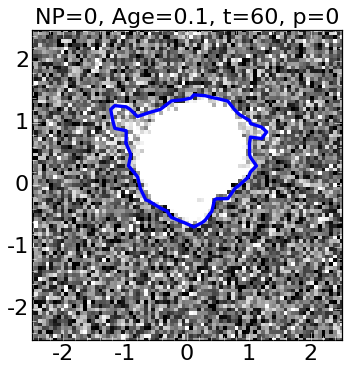}
    \includegraphics[width=0.24\textwidth]{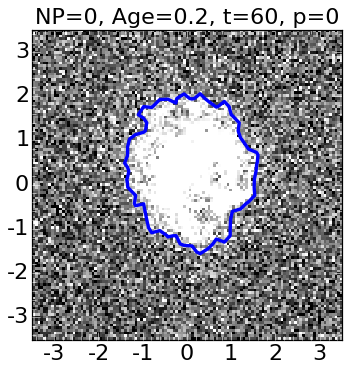}
    \includegraphics[width=0.24\textwidth]{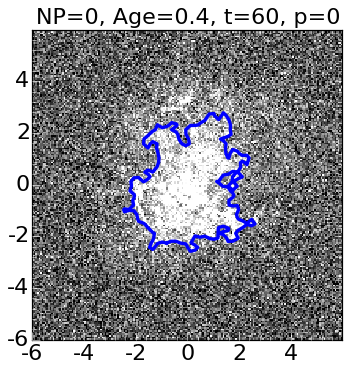}
    \includegraphics[width=0.24\textwidth]{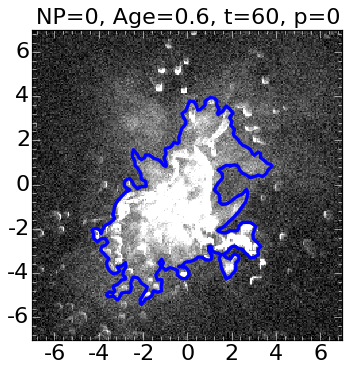}
    
    %row 3
    \includegraphics[width=0.24\textwidth]{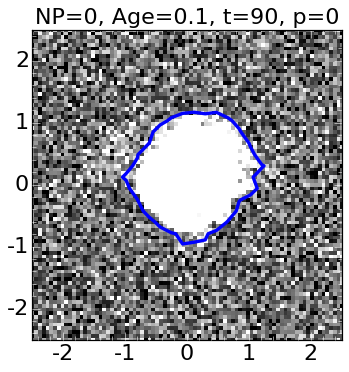}          
    \includegraphics[width=0.24\textwidth]{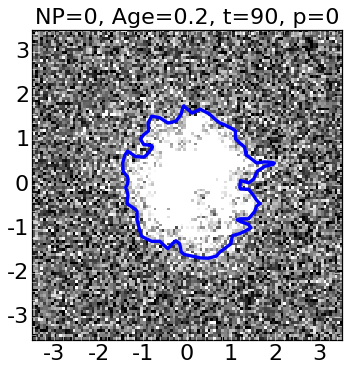}          
    \includegraphics[width=0.24\textwidth]{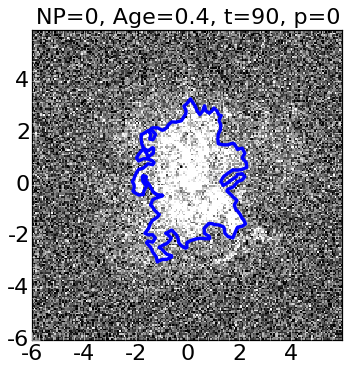}          
    \includegraphics[width=0.24\textwidth]{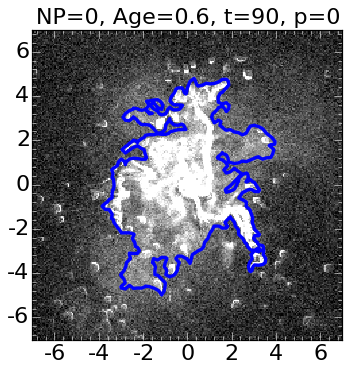}
                      
    \caption[Overview of the SOs with random Gaussian noise]{Overview of the addition of random Gaussian noise to the same SOs shown in Fig.\,\ref{fig:raw_so}. Contours are shown at a constant level of 1$\sigma$ plus the clipped mean of the values from the Gaussian distribution introduced. Coordinates are shown in pc scale.}
    \label{fig:gauss_so}
\end{figure*}

\begin{figure*}
	\centering
	\begin{subfigure}{0.24\textwidth}
		\includegraphics[width=\textwidth]{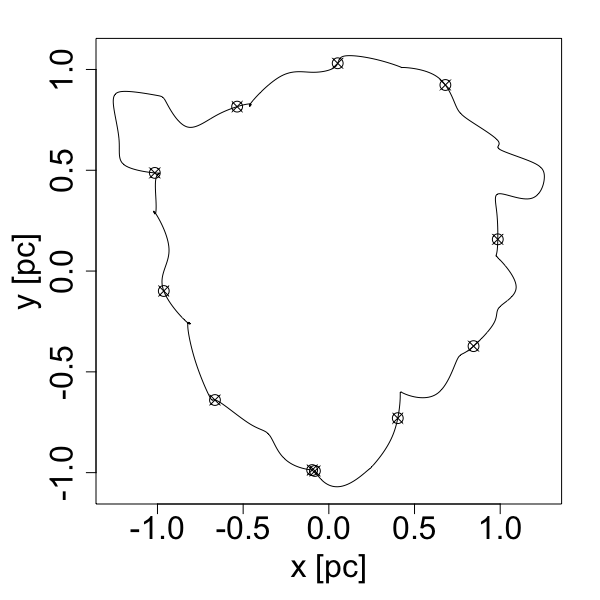}
		\caption{0.1\,Myr SO Shape}
	\end{subfigure}
	\begin{subfigure}{0.24\textwidth}
		\includegraphics[width=\textwidth]{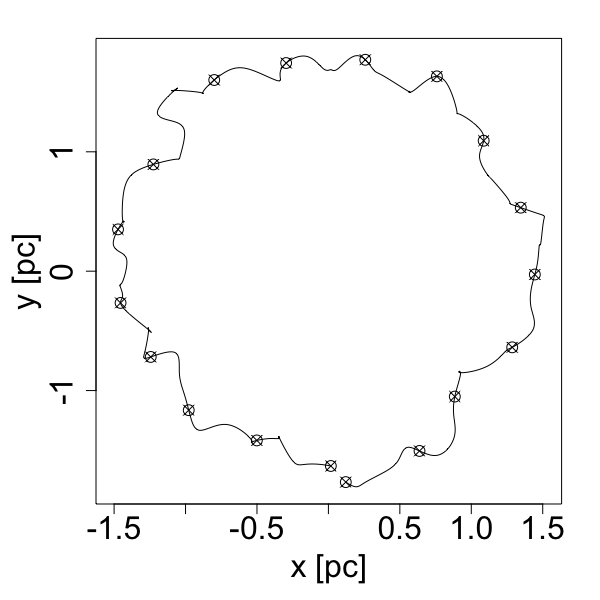}
		\caption{0.2\,Myr SO Shape}
	\end{subfigure}
	\begin{subfigure}{0.24\textwidth}
		\includegraphics[width=\textwidth]{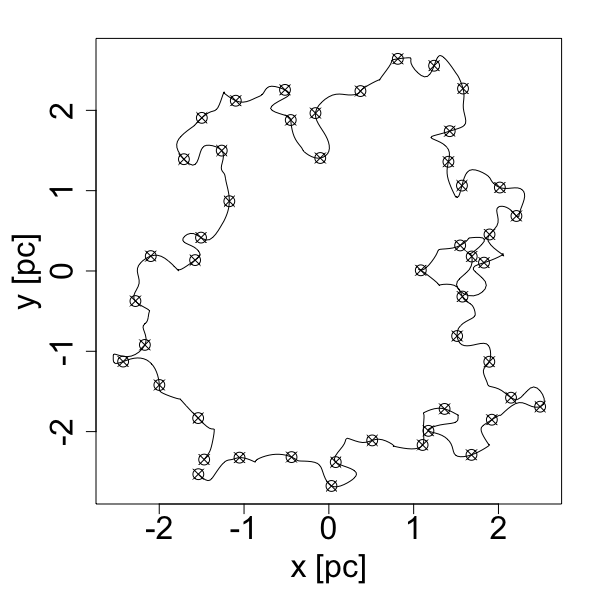}
		\caption{0.4\,Myr SO Shape}
	\end{subfigure}
	\begin{subfigure}{0.24\textwidth}
		\includegraphics[width=\textwidth]{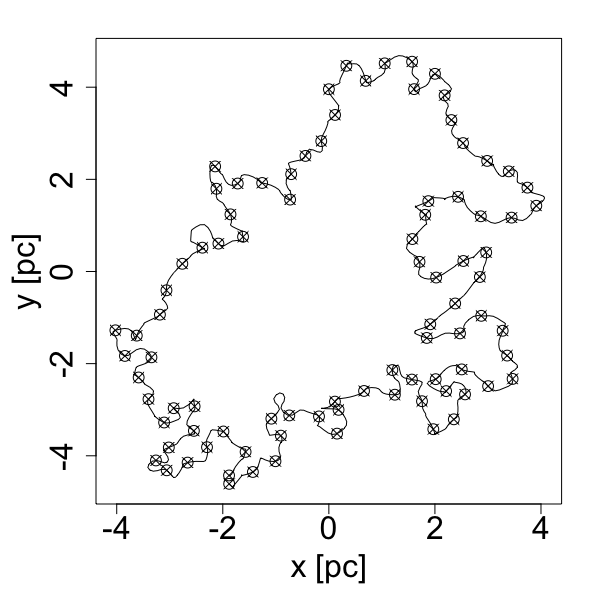}
		\caption{0.6\,Myr SO Shape}
	\end{subfigure}
	
	\caption[Shape Extraction from SOs]{Boundaries of an example synthetic observation \hii region for each of the four ages in the sample (those in the second row of Fig.\,\ref{fig:gauss_so}). Points signify the approximately equally spaced interpolation spline knots, where the curvature was calculated.}
	\label{fig:so_spline}
\end{figure*}

\begin{figure}
	\centering
	\includegraphics[width=\columnwidth]{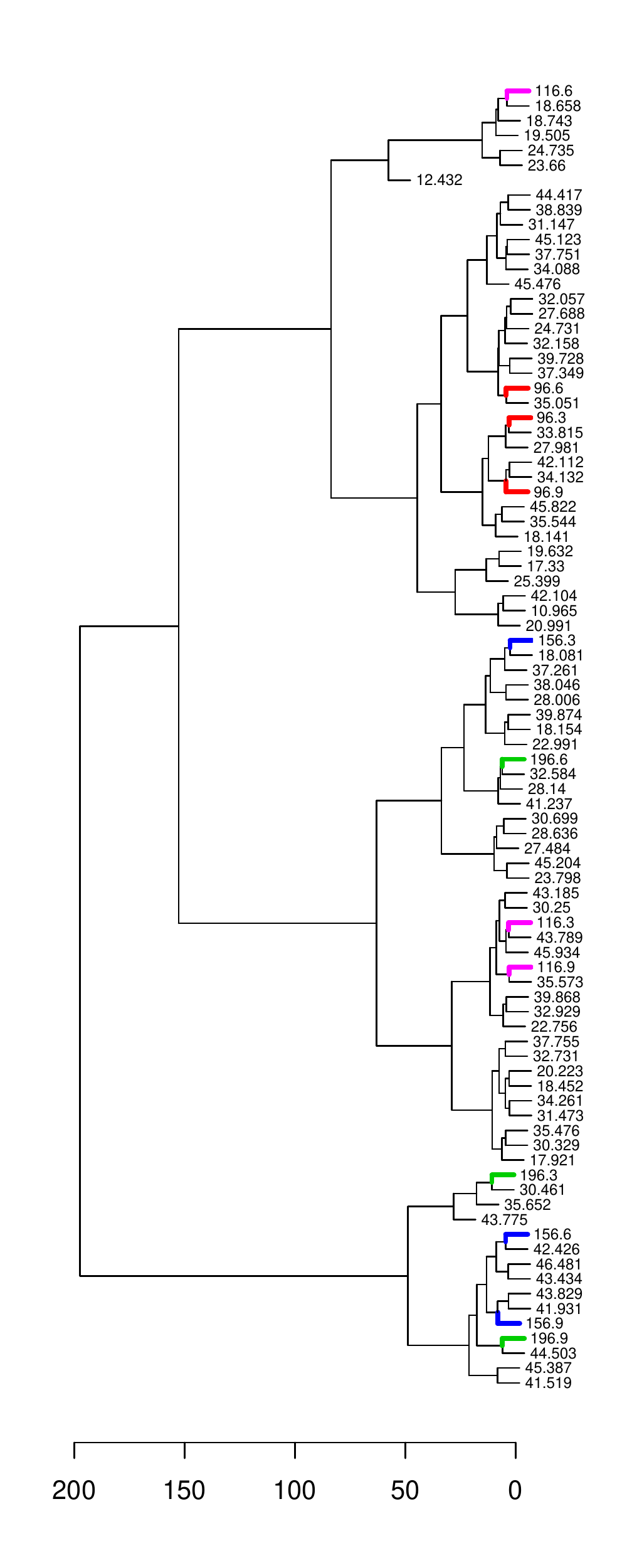}
	\caption[Dendrogram Comparing Shapes of Example SOs with Added Gaussian Noise to the MAGPIS Sample]{Dendrogram of the MAGPIS sample of \hii regions from \paperIt~ with the 12 example SOs with added Gaussian noise (those shown in Fig.\,\ref{fig:gauss_so}). The dendrogram represents the results from applying hierarchical clustering of the shape data of each \hii region. The branches of the \hii Regions are labelled by their Galactic longitude (for the MAGPIS sample) or an ID number (for the SO sample). by The branches of the SOs are coloured by their age: 0.1\,Myr in \textcolor{red}{red}, 0.2\,Myr in \textcolor{magenta}{pink}, 0.4\,Myr in \textcolor{blue}{blue} and 0.6\,Myr in \textcolor{green}{green}. The horizontal axis represents the height computed from the agglomerative clustering method.}
	\label{fig:gauss_mag_dend}
\end{figure}

Figure \,\ref{fig:gauss_comp} shows an example of one of the 0.2\,Myr SOs. The left panel shows the original SO, the middle and right panels show the same SO with the random Gaussian noise added to each pixel. The middle panel has its Gaussian profile taken from the MAGPIS tile centred at $l$ = 12.43\dg, $b$ = -0.04\dg , with $\mu$ = 0.959\,mJy/beam and $\sigma$ = 0.389\,mJy/beam. The right panel's Gaussian noise profile follows the MAGPIS tile centred at $l$ = 30.25\dg, $b$ = 0.24\dg, with $\mu$ = 0.830\,mJy/beam and $\sigma$ = 0.345\,mJy/beam. The contours for the two SOs with Gaussian noise are determined by taking the clipped mean plus 1$\sigma$, the same way in which each contour was calculated for each of the \hii regions in \paperIt. The level for the contour of the original SO in the left panel is the same as the middle panel, since the pixel vales of the SO had already been converted to match that of the MAGPIS data. There is not much visual difference between the three contours. The original SO has a much smoother contour as expected. Small perturbations along the boundaries of each of the Gaussian profile SOs are noted, they are approximately the same size. Whilst some extrusions along the boundary appear in the SOs with the Gaussian profiles, the intrusion on the original SO on the upper right side is smoothed out by the Gaussian noise. 

The conversion of MJy/sr to Jy/beam was suitable for the 0.1 and 0.2\,Myr SOs. However, for the 0.4 and 0.6\,Myr SOs, using the actual conversion factor resulted in the emission from the SOs being drowned out by the introduction of the Gaussian noise from the MAGPIS data distributions. This is explained by referring back to the bulk properties and the volume-average electron density of the simulations \citepalias[Figures.\,3 \& 9, respectively][]{2018MNRAS.477.5422A}, which shows that after 0.4\,Myr, mass starts to leave the grid of the simulation and the electron density decreases. Therefore, the integrated radio continuum intensity at 20\,cm obtained from the radiative transfer code, which is used to produce the SO, is less than should be expected. The intensity of these older SOs in Jy/beam was hence artificially boosted by a factor of $\sim 4$ for the 0.4\,Myr SOs and $\sim 70$ for the 0.6\,Myr SOs. These values were determined such that a single contoured central region, representative of the original SO before the noise was introduced, was obtained by the automated contouring procedure. These were minimum values for the single central region to be obtained and for the shape to not substantially change with further increase. Since the contour level is determined from the image tile, as the intensity boosting increases, the distribution of pixel values changes, therefore the final contour achieved is still consistent with further intensity boosting. We tested this for the 0.6\,Myr regions using a factor of $\sim 95$ and $\sim 135$, with the resulting amount of variation along the boundaries being well within the spatial sampling resolution of the shape analysis (c.f. Sec.\,\ref{ssec:comp_so_mag}).

Figure\,\ref{fig:gauss_so} shows a summary of the addition of a Gaussian noise profile to the same 12 example SOs shown in Fig.\,\ref{fig:raw_so}. The images for all 77 SOs with this Gaussian noise profile are shown in Appendix\,\ref{app:so_images}\footnote{Available in the online version of this paper.}. The noise profile used was that of the middle panel in Fig.\,\ref{fig:gauss_comp}, with the two younger SOs having the actual intensity conversion factor used and the older two have their emission boosted as described above. It is clear from this summary that there is a higher amount of perturbation along the \hii region boundaries as the age of the SOs increases. It appears that the contoured shape changes more with each projection for the 0.4 and 0.6\,Myr regions than for the 0.1 and 0.2\,Myr regions (however, larger perturbations are noted for some of the different projections for the earlier ages in the full overview in Appendix\,\ref{app:so_images}). The axes of the plots are given in spatial parsec scale, hence the effective radius of each \hii region boundary also increases with the age of the SO. For the remainder of this study, only one Gaussian noise profile was used for all of the 77 SOs. This was because the \hii region shape did not change by a substantial amount with the introduction of Gaussian noise profiles with different distributions from the MAGPIS data. This was numerically assessed with the shape comparison method utilised in this work and \paperIt, resulting in lower pairwise test scores than for any other comparison thus far. Having the same noise profile also meant that the contour level applied to each SO was consistent, since this is calculated from the noise itself - distinguishing the signal of the \hii region from the background noise.

\subsection{Shape comparison: Gaussian SO \& MAGPIS}
\label{ssec:comp_so_mag}

The first test of the SOs was to directly compare them to the MAGPIS observational sample from \paperIt. The further steps to extract and compare the shapes of the regions were carried out in the same manner as before. To summarise: interpolation splines were fitted to the region boundaries identified from the contouring procedure, with interpolation knot intervals of $\sim 0.54$\,pc (see Fig.\,\ref{fig:so_spline}); then the local curvature values were calculated at each knot. The empirical distribution functions (EDFs) of curvature values were then statistically compared pairwise, using the two-sided Anderson-Darling (A-D) test statistic \citep{anderson1952,pettitt1976two}. The A-D test returns a dissimilarity measure between the pair of samples, whereby the null hypothesis that the samples are drawn from the same parent population is rejected for large test result scores. After applying a Euclidean distance transformation to the A-D test scores, hierarchical clustering was performed on the distance matrix of \hii region shape distances using Ward's agglomerative method \citep{ward1963hierarchical,murtagh2014ward}. The resulting hierarchical structure was then investigated using the dendrogram graphical representation.

In this primary investigation of how the shapes of the SO \hii regions compare to those of the MAGPIS \hii regions, the 12 example SOs from Fig.\,\ref{fig:gauss_so} were considered, along with the 76 MAGPIS \hii regions from the \paperIt. Figure\,\ref{fig:gauss_mag_dend} shows the resulting dendrogram from the hierarchical clustering. The branches of the SOs are highlighted by colours that correspond to the age of the SO: 0.1\,Myr in red, 0.2\,Myr in pink, 0.4\,Myr in blue and 0.6\,Myr in green. The first clear result is that the shapes of the SO \hii regions are grouped in amongst the MAGPIS \hii regions, with none appearing as outliers. This result confirms that these numerical simulations are producing \hii regions that are representative of what we observe in our Galaxy; since we ensured that the shape of the SO \hii regions was extracted and quantified in the exact same way as the MAGPIS sample.

The next point to note from Fig.\,\ref{fig:gauss_mag_dend} is that the three different projections considered for each SO age are not all grouped together in the dendrogram. The 0.1\,Myr projections are close to each other on the dendrogram, however, one of them belongs to a different parent group than the other two. Similarly, for the 0.2 and 0.4\,Myr  projections, two each belong to the same parent group, with the third positioned a few groups away, respectively. The most spread in group allocation is seen for the 0.6\,Myr projections. These results show that for each of the SO projections, there is a MAGPIS \hii region that shares the most similar shape, such that the projections are having a significant influence on the identified shape of the region. This will be further investigated and discussed in section\, \ref{sec:so_discussion}. 

Since the dendrogram represents the results of the hierarchical procedure, which is a bottom-up approach to forming groups, excluding part of the sample does not change the resulting groupings. This was tested by excluding the 0.4 and 0.6\,Myr regions in Fig.\,\ref{fig:gauss_mag_dend} and rerunning the clustering procedure, the resulting structure and groups match those presented here. The group structure that would be obtained from cutting the dendrogram at a height that intersects seven branches (i.e., seven groups) is 95\% concurrent with the groupings identified in \paperIt, with a notable exception being the outlier \hii region labelled `12.432' being joined to a different group here.

\begin{figure}
	\centering
	\includegraphics[width=\columnwidth]{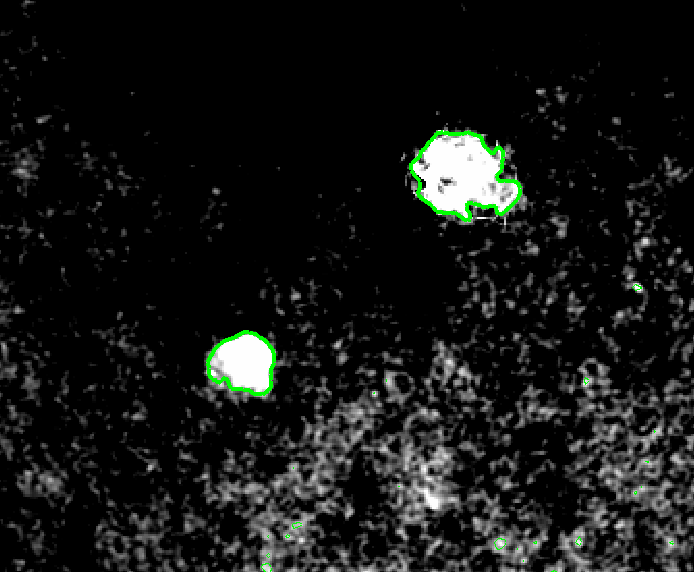}
	\caption[Example MAGPIS Tile \& \hii Region with SO Inserted]{Example MAGPIS image tile and \hii region G030.252$+$00.053 (lower-left). One of the 0.2\,Myr SOs has been inserted to the tile (top-right). Contours shown are that obtained from the image tile using the clipped mean plus one sigma.}
	\label{fig:so_mag_comp}
\end{figure}
 
\subsection{MAGPIS Noise Profiles}
\label{ssec:mag_np}

Whilst the results of the previous subsection show that the numerical simulations are producing well representative SOs, the introduction of noise to the SOs in order to extract the shape can be developed further. The distribution of noise from radio interferometry images does follow a Gaussian distribution, however, it is not completely homogeneous across the image tiles. Artefacts from the reduction process and emission from fore- and background sources each contribute to the non-homogeneity of the noise distributions. We stated  earlier in this section that small changes to the Gaussian mean and standard deviation of the noise profile had minimal influence on the shape of the SO \hii region, however, we did propose in \paperIt~ that observational noise may be a significant contribution to the observed and extracted shape we obtain using our methods.

In order to see how much of an effect the observational noise distributions from observations have on the shape of the SOs, we inserted the SO data directly into the MAGPIS tiles used in \paperIt. The intensity converted SO pixel values were added to those in an area of the image tile deemed to contain only image noise and away from the signal of the \hii region. An example of this is shown in Fig.\,\ref{fig:so_mag_comp}, where one of the 0.2\,Myr projections (upper-right) is inserted to the MAGPIS tile containing \hii region G030.252$+$00.053 (lower-left). The non-homogeneity of the image noise can be seen clearly by the gradient of noise distribution across the tile. G030.252$+$00.053 is at a distance of 4.5\,kpc, has an effective spatial radius of 0.8\,pc and a dynamical age of 0.1\,Myr \citep{2014ApJS..212....1A,2018MNRAS.477.5486C}. The spatial pixel scale of the SOs is 0.06\,pc per pixel. If we assume that the \hii region in the SO is at a distance of 6.2\,kpc, then this corresponds to an angular pixel scale of 2" per pixel, the same as the MAGPIS tiles. This meant that the smoothing factor applied to the contours was uniform for both the MAGPIS and the SO \hii regions. As before, the contours shown are calculated from the clipped mean plus one standard deviation of the entire tile, which identifies both \hii regions well.

\begin{table}
	\centering
	\caption[Summary of Noise Profile Properties]{Summary of the different MAGPIS Noise Profile (NP) properties that the SOs were inserted into.}
	\label{tab:NP_info}
	\begin{tabular}{rrrrr}
		\hline
		NP & \textit{l} [\dg] & \textit{b} [\dg] & $\mu$ [mJy/beam] & $\sigma$ [mJy/beam] \\ 
		\hline
		1 & 20.99 & 0.09 & 0.120 & 0.242 \\
		2 & 12.43 & -0.04 & 0.945 & 0.367 \\
		3 & 43.76 & 0.06 & 0.105 & 0.263 \\
		4 & 20.22 & 0.11 & 0.329 & 0.222 \\
		5 & 30.25 & 0.24 & 0.829 & 0.343 \\
		6 & 41.93 & 0.04 & 0.144 & 0.265 \\
		\hline
	\end{tabular}
\end{table}

\begin{figure*}
    \centering
    
    \begin{subfigure}{0.48\textwidth}
       \centering
        \includegraphics[width=0.32\textwidth]{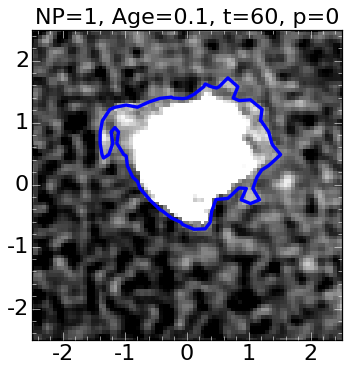} 
        \includegraphics[width=0.32\textwidth]{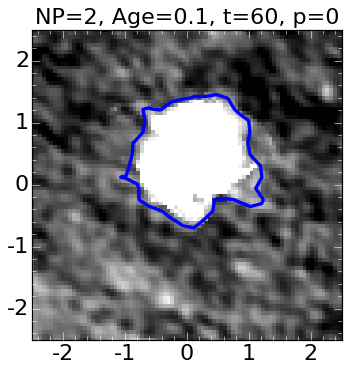}
        \includegraphics[width=0.32\textwidth]{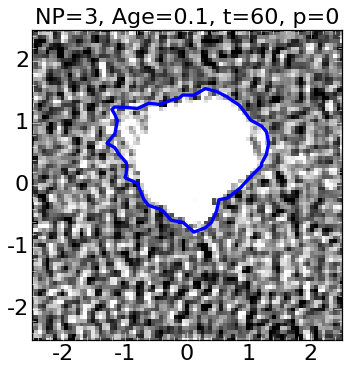}
                
        \includegraphics[width=0.32\textwidth]{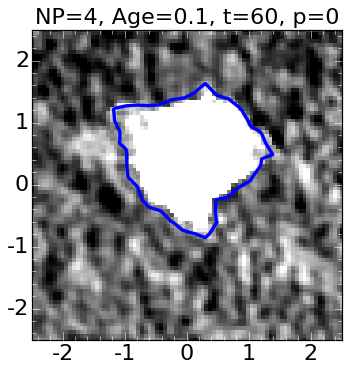}          
        \includegraphics[width=0.32\textwidth]{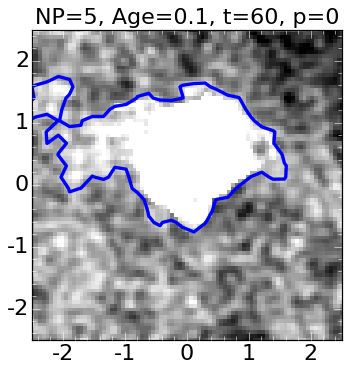}       
        \includegraphics[width=0.32\textwidth]{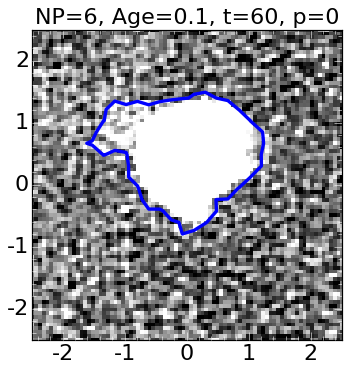}
           
        %\hspace{5cm}
        \caption{0.1\,Myr SOs} \label{fig:magpis_so_096}
        %\vspace{0.5cm}
    \end{subfigure}   	
	\begin{subfigure}{0.48\textwidth}
    \centering
        \includegraphics[width=0.32\textwidth]{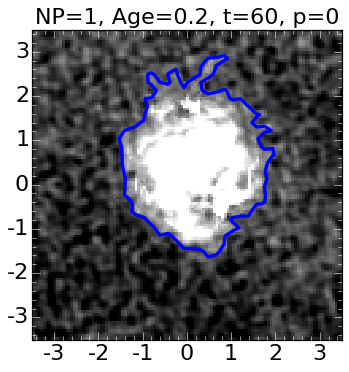} 
        \includegraphics[width=0.32\textwidth]{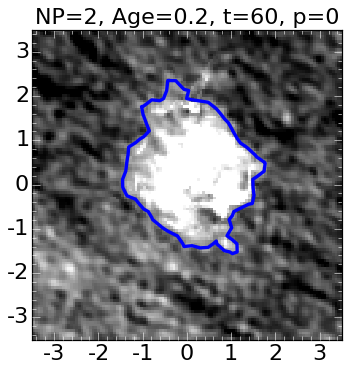}
        \includegraphics[width=0.32\textwidth]{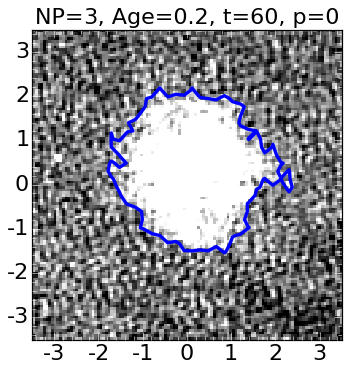} 
             
        \includegraphics[width=0.32\textwidth]{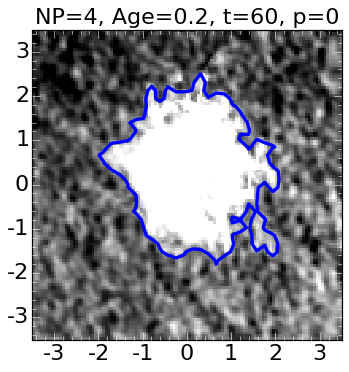}   
        \includegraphics[width=0.32\textwidth]{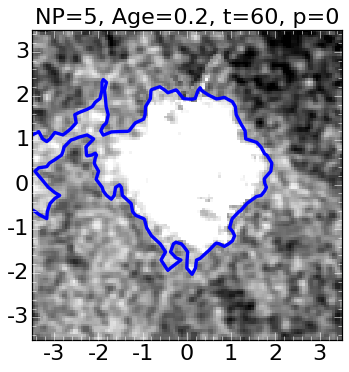}    
        \includegraphics[width=0.32\textwidth]{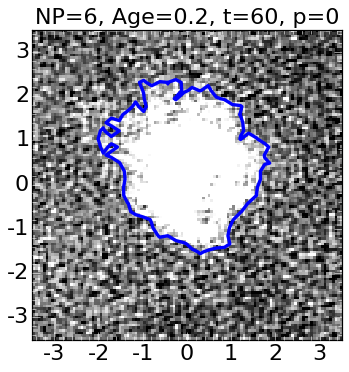}
                  
        %\hspace{5cm}
     \caption{0.2\,Myr SOs} \label{fig:magpis_so_116}
    \end{subfigure}  
    
    \begin{subfigure}{0.48\textwidth}
       \centering
        \includegraphics[width=0.32\textwidth]{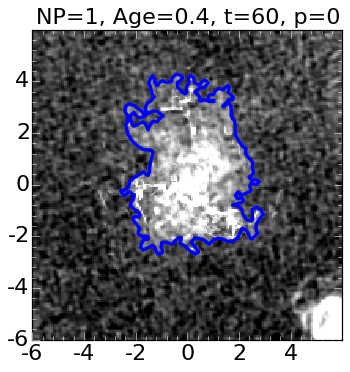} 
        \includegraphics[width=0.32\textwidth]{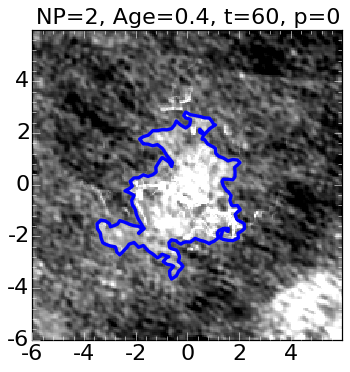}  
        \includegraphics[width=0.32\textwidth]{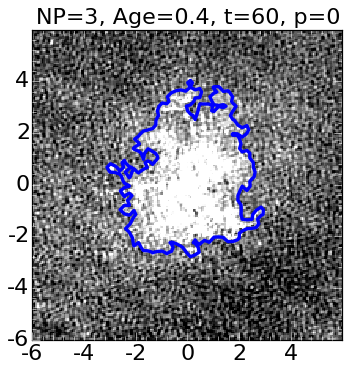}      
        
        \includegraphics[width=0.32\textwidth]{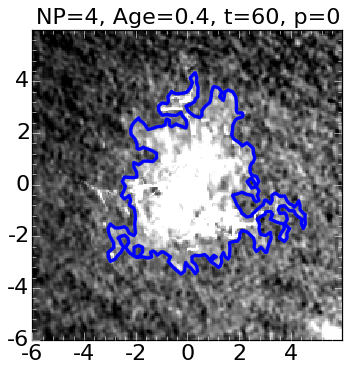}     
        \includegraphics[width=0.32\textwidth]{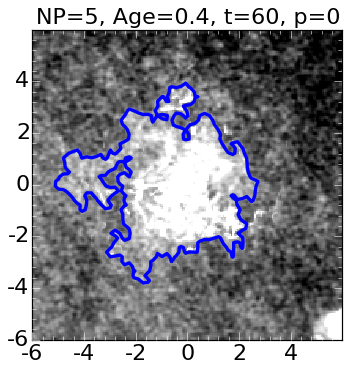}      
        \includegraphics[width=0.32\textwidth]{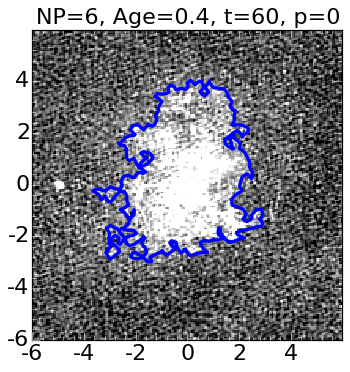}
         
        %\hspace{5cm}
        \caption{0.4\,Myr SOs} \label{fig:magpis_so_156}
        %\vspace{0.5cm}
    \end{subfigure}   
	\begin{subfigure}{0.48\textwidth}
    \centering
        \includegraphics[width=0.32\textwidth]{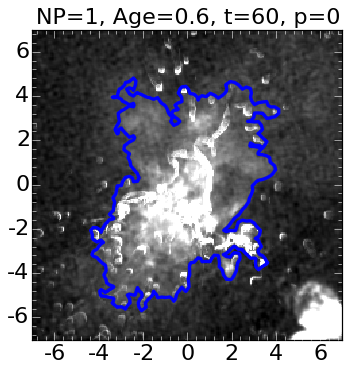} 
        \includegraphics[width=0.32\textwidth]{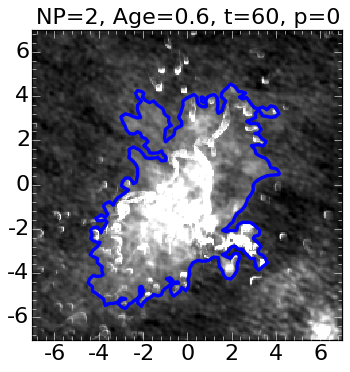}   
        \includegraphics[width=0.32\textwidth]{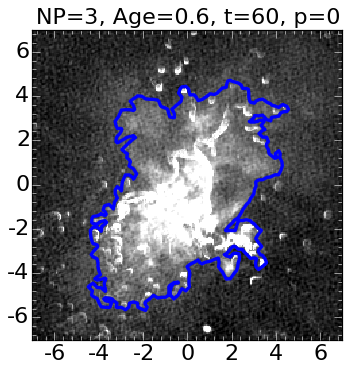} 
        
        \includegraphics[width=0.32\textwidth]{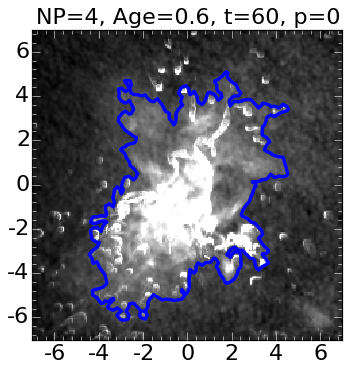}         
        \includegraphics[width=0.32\textwidth]{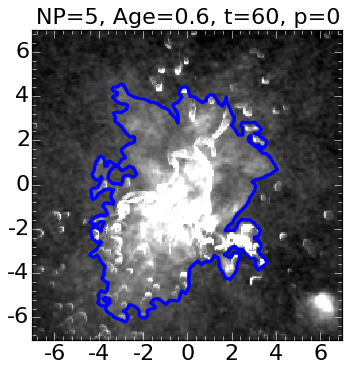}    
        \includegraphics[width=0.32\textwidth]{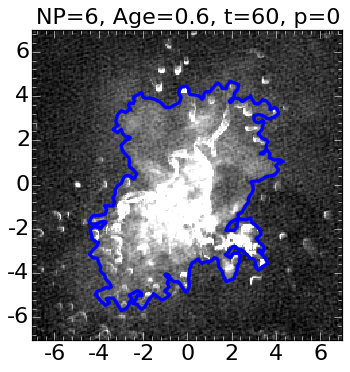}
                 
        %\hspace{5cm}
     \caption{0.6\,Myr SOs} \label{fig:magpis_so_196}
    \end{subfigure}         

    \caption[Overview of the SOs Inserted into Six Different MAGPIS Tiles]{SOs inserted into six different MAGPIS tiles to show how different noise profiles influences extracted shape. For each age, the same projection is shown, along with the 1$\sigma$ plus clipped mean contours. Axes are in pc, with a pixel scale of 0.06\,pc/pixel. The corresponding angular size of the cutouts for increasing SO are are 2.8$^\prime$, 3.9$^\prime$, 6.7$^\prime$ \& 7.8$^\prime$, respectively. Linear intensity scaling with 95\% maximum is used for each tile, to highlight the differences in noise structure.}
    \label{fig:magpis_so}
\end{figure*}

Figure\,\ref{fig:magpis_so} shows an overview of an example SO projection for each of the four ages that has been inserted into six different MAGPIS tiles. The example SOs used for each age are those in the second row of Fig.\,\ref{fig:raw_so} (the same ones used in Fig\,\ref{fig:so_spline} to exemplify the interpolation spline fitting to the boundary contours). The MAGPIS tiles that the SO were inserted into will be referred to as the noise profile (NP). Details of which MAGPIS tile each NP represents is given in Tab.\,\ref{tab:NP_info}. The means and standard deviations were calculated from the sigma clipped MAGPIS tiles, with the contour level applied in each of the images then taken as the mean plus one standard deviation, as before. The example NPs are each from MAGPIS \hii region tiles that were sorted into different groups by the shape analysis clustering in \paperIt. We surmised there that the observational noise may be influencing the shape obtained, hence the reasoning for selecting tiles from different resulting groups. From the 76 \hii region tiles in our preceding study, the range of mean intensity values were 0.1 -- 0.945\,mJy. The range of standard deviations in the tiles were 0.2 -- 0.4\,mJy, which concurs with the variation in the RMS noise of the overall MAGPIS survey of $\sim$0.25\,mJy \citep{2006AJ....131.2525H}. The example NPs selected here therefore cover the range of Gaussian noise distributions of the observational data we are considering for comparison. 

 In the example images shown in Fig.\,\ref{fig:magpis_so}, for NPs 1, 2, 4 \& 5 at 0.4 and 0.6\,Myr, part of the MAGPIS \hii region can be seen in the bottom right corner. Also, as with the Gaussian noise examples shown previously, the 0.4 and 0.6\,Myr SOs have had their emission boosted by the same amount as before (factor of $\sim$4 and $\sim$70, respectively). As previously explained, this enabled the central part of the \hii region to be contoured appropriately, accounting for the lower mass and electron density remaining in the simulation grid. 

The change in the noise structure and intensity is apparent from Fig.\,\ref{fig:magpis_so}, where the image tiles are each linearly scaled with a maximum intensity limited to 95\%, to visually highlight the differences between the NPs. NPs 3 and 6 appear to have the most `salt and pepper' like noise, similar to that seen in the random Gaussian distributions used previously. The examples shown for the 0.6\,Myr SOs appear to have the least dissimilarities by NP, although this may be because the regions themselves are larger and it is harder to visually discern the differences along the boundaries. It is worth remembering here that whilst these contours represent the systematically defined region boundaries, the shapes that are compared in the subsequent analyses are quantified from the shape landmarks, which are given by the interpolation spline fitting (see Fig.\,\ref{fig:so_spline}). Since we are still interested in the direct comparison of these SOs to the MAGPIS \hii regions, the spline sampling remained at $\sim 0.54$\,pc. Therefore, each of the boundaries shown here were under-sampled, with small perturbations along the contour smoothed out by the splines, with the level of smoothing proportional to the spatial size of the regions. Nevertheless, the differences in the contoured \hii region shapes seen here, resulting from the different NPs, will enable us to investigate how such changes in NP effect the final comparison of the shape that we perform in the analysis. This will be discussed in detail in Sec.\,\ref{sec:so_discussion}.

\begin{figure}
	\centering
    \includegraphics[width=\columnwidth]{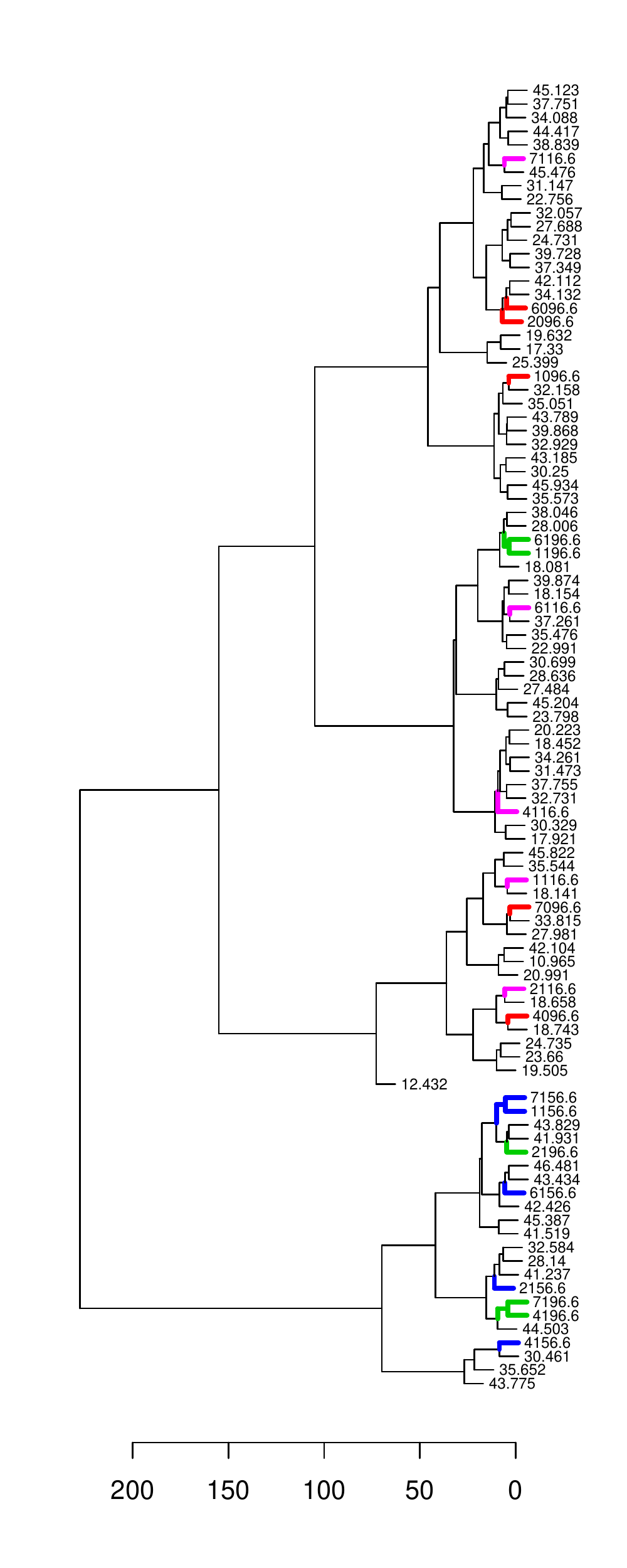}
    \caption[Dendrogram Comparing Shapes of Example SOs inserted into different MAGPIS NPs to the MAGPIS Sample]{Dendrogram of the MAGPIS sample of \hii regions from \paperIt~ with the 20 example SOs that were inserted into the MAGPIS NPs (shown in Fig.\,\ref{fig:magpis_so}, excluding NP 5). The dendrogram represents the results from applying hierarchical clustering of the shape data of each \hii region. The branches of the 20 SOs are coloured by their age: 0.1\,Myr in \textcolor{red}{red}, 0.2\,Myr in \textcolor{magenta}{pink}, 0.4\,Myr in \textcolor{blue}{blue} and 0.6\,Myr in \textcolor{green}{green}.}
    \label{fig:magpis_np_dend}
\end{figure}

For the first three ages, NP 5 results in a spurious extrusion on the left side of the \hii region. This is an example of where there may be underlying signal in what was thought to be only background noise, and thus it is having a clear effect on the identified shape.  We are only able to identify this since we have prior knowledge of the original SO data, along with the other NPs for comparison. If this were a real observation it would be systematically included as is, with a larger image tile used to ensure the contour is closed. For the rest of this analysis and the purposes of testing, however, we chose to exclude NP 5 and consider the remaining five profiles as our representative observational noise.

Figure\,\ref{fig:magpis_np_dend} shows the resulting dendrogram from applying the clustering algorithm to the 76 MAGPIS \hii regions, along with the 20 example SO \hii regions (from Fig.\,\ref{fig:magpis_so}), whose shapes were extracted from the five NPs detailed here. As with the previous dendrogram of the Gaussian noise profiles, the branches of the SO \hii regions are coloured by their age: 0.1\,Myr in red, 0.2\,Myr in pink, 0.4\,Myr in blue and 0.6\,Myr in green. Figure\,\ref{fig:magpis_np_dend} shows that all of the 0.4\,Myr SO \hii regions belong to the same parent branch, along with three of the five 0.6\,Myr SO regions. The other two 0.6\,Myr regions are paired together in a separate group. There are only a few examples of where SOs of the same age are paired together in this manner. As with the Gaussian noise profiles, most of the SO \hii regions are being paired with one of the MAGPIS \hii regions. There is also a 97\% concurrence between the groups identified from this dendrogram and those identified from the MAGPIS sample in \paperIt. This further confirms the efficacy of the simulations for producing representative SOs.

\begin{figure*}
	\centering
	\includegraphics[width=0.7\textwidth]{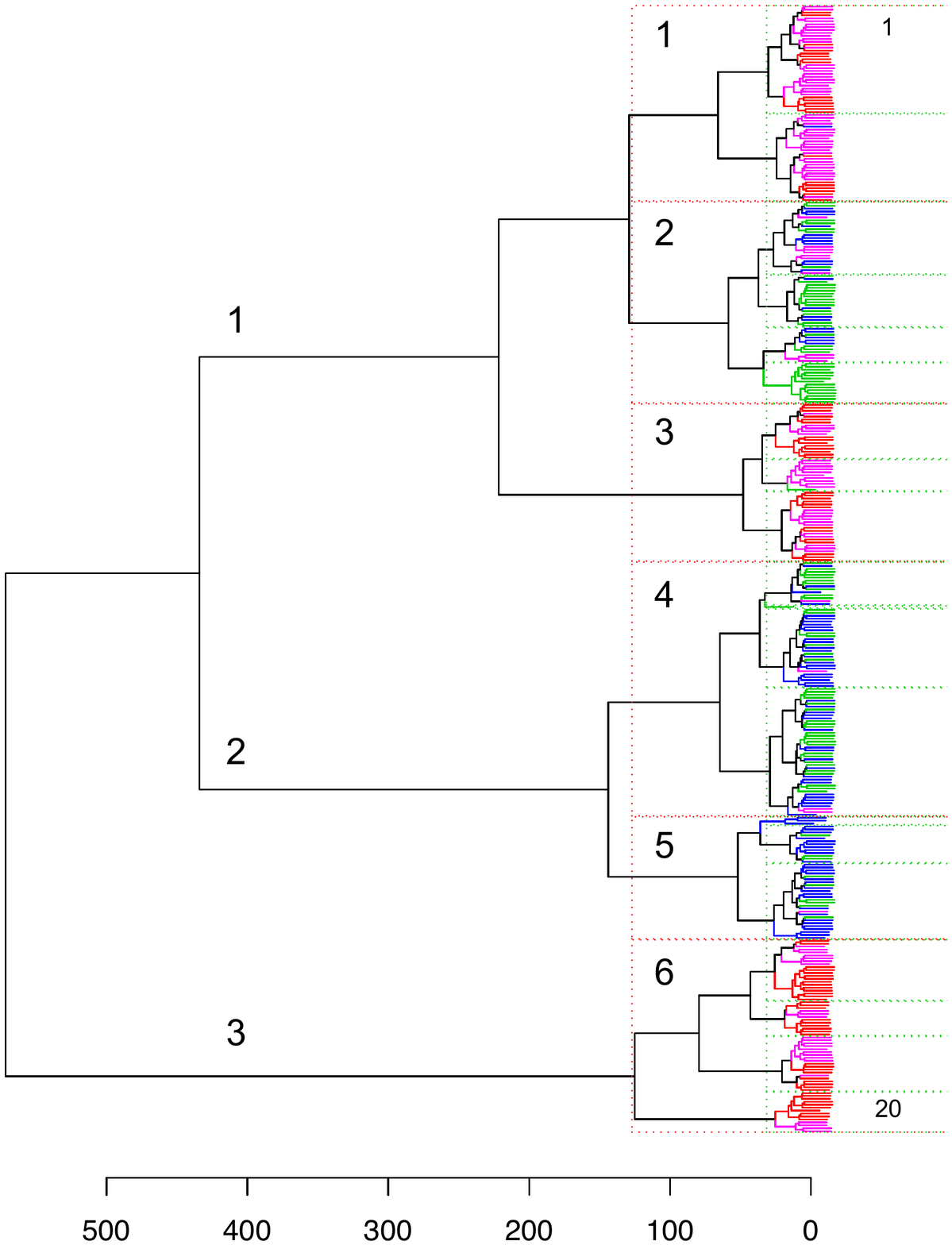}
	\caption[Dendrogram of the Shapes of the Full SOs Sample Inserted into Different MAGPIS NPs]{Dendrogram resulting from applying hierarchical clustering to the shape data of the sample of 385 SO \hii regions inserted into MAGPIS NPs. As with the previous dendrograms, the branches are coloured by their age: 0.1\,Myr in \textcolor{red}{red}, 0.2\,Myr in \textcolor{magenta}{pink}, 0.4\,Myr in \textcolor{blue}{blue} and 0.6\,Myr in \textcolor{green}{green}. The three groups displayed in Fig.\,\ref{fig:age_so_5np_3groups} are indicated by the respective group numbers at the first split into three. Six groups are delineated by the dashed red boxes and labelled 1 through 6. 20 further groups as seen in Fig.\,\ref{fig:radius_so_5np} are delineated by the green boxes with the first and last labelled 1 and 20, respectively.}
	\label{fig:dend_so_5np}
\end{figure*}

\begin{figure}
	\centering
	\includegraphics[width=\columnwidth]{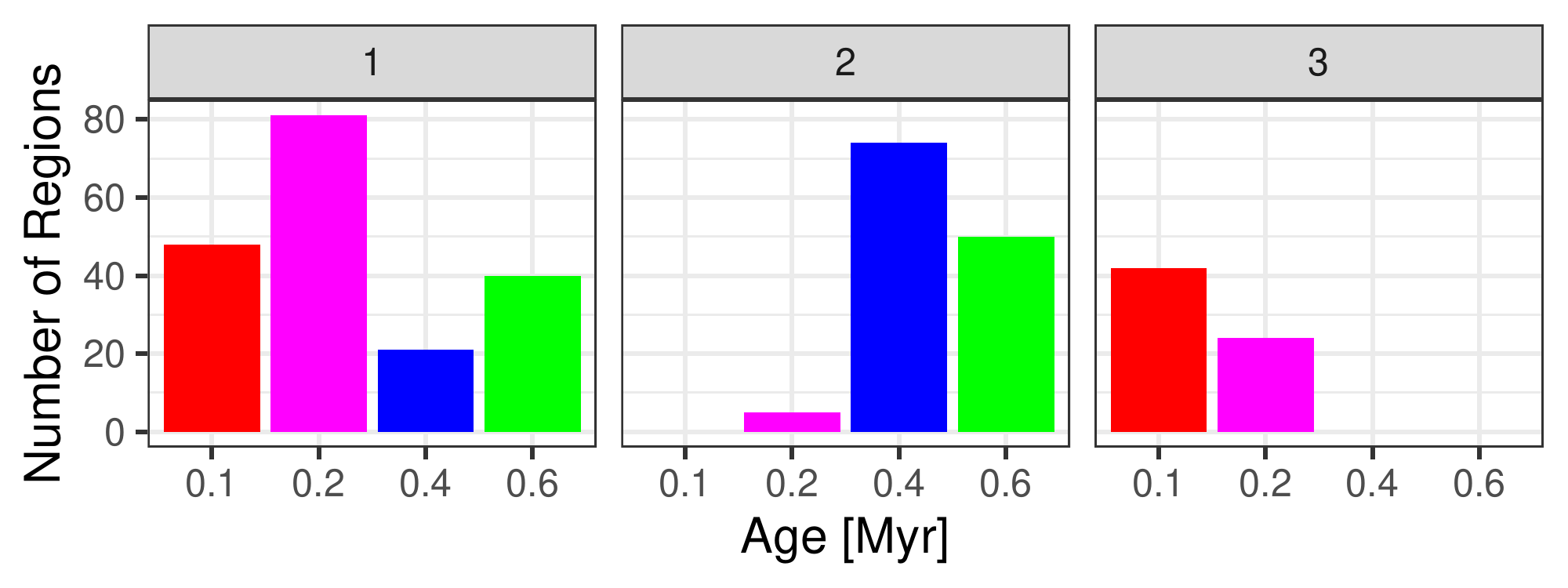}
	\caption[Summary of the Ages of the SOs by Group]{Bar chart showing the ages of the SOs that have been assigned to the first three groups in the dendrogram in Fig.\,\ref{fig:dend_so_5np}, at a cut height of e.g. 300.}
	\label{fig:age_so_5np_3groups}
\end{figure}

%%%%%%%%%%%%%%%%%%%%%%%
%%%%%%%%%%%%%%%%%%%%%%%
\section{Discussion}
\label{sec:so_discussion}
Having shown that the insertion of the SOs into different NPs from the MAGPIS data leads to \hii region shapes that are mathematically similar to what we observe in the MAGPIS sample, we can now further investigate the properties of the SOs and how they may influence the obtained shapes. Whilst we are only considering SOs from one set of initial conditions, i.e., star cluster mass, ambient density, etc., introduction of the MAGPIS NPs essentially expands our number of observations at each simulation time-step. For the 77 projections, across the four ages, with five NPs, we arrive at a sample of 385 individual \hii regions. These variables are hence the parameters that we will discuss further in this section. In addition to showing that shape analysis can be used as a tool to confirm the reliability of synthetic data, a further aim of this investigation is to see whether the SOs could be used as a training set for supervised classification of \hii regions via their shapes. We therefore need to understand how each of these parameters is affecting the obtained shapes and defining the groups. 

A potential variable that we are deferring to future work is distance. In this investigation, all SO \hii regions are assumed to be at the same distance from the observer, ensuring that the shape could be extracted in exactly the same manner as the MAGPIS \hii regions from \paperIt; with the corresponding pixel scales matching those used for the contouring parameters. We found in \paperIt~ that using an angular sampling scale for the shape analysis resulted in biased results, with nearby regions that possess higher angular resolution data being grouped together by the procedure. Furthermore, we discussed there how variations in the determined distance (due to the many errors associated with kinematic distance estimations) is directly related to variations in the spatial sampling scale used for the shape extraction. We are planning a detailed investigation in to how distance influences the shape analysis methodology and feasibility for future applications. Brief details of this and other planned future work is discussed at the end of this section.

\subsection{Hierarchical Clustering of the SO Sample}
\label{ssec:hclust_np_so}

Figure\,\ref{fig:dend_so_5np} shows the dendrogram resulting from applying the hierarchical clustering method to the shape distances of the 385 SO \hii regions `observations' across the different projections and NPs. As with the previous dendrograms, the branches of the SOs are coloured by their age: 0.1\,Myr in red, 0.2\,Myr in pink, 0.4\,Myr in blue and 0.6\,Myr in green. It is clear from this dendrogram that there is a divide between groups containing the 0.1 and 0.2\,Myr regions (early-type regions) and those containing the 0.4 and 0.6\,Myr regions (late-type regions). This is displayed concisely in Fig.\,\ref{fig:age_so_5np_3groups} by a bar chart of number of regions of a given age for the first three distinct groups. These groups are obtained by cutting the dendrogram at a height of e.g. 400. In this situation, group 1 contains mostly early-type regions with some late-type regions. Group 3 hosts exclusively early-type regions and group 2 hosts mainly late-type regions with some 0.2\,Myr regions being included. In terms of grouping the synthetic \hii regions purely by age, the dendrogram in Fig.\,\ref{fig:dend_so_5np} shows that only a few of the smaller groups are all the same colour, that is, regions of the same age. A cut on the dendrogram, resulting in six groups is shown by the dashed red boxes. This is the maximum height (i.e., fewest number of groups) required to produce groups that each contain either entirely late- or entirely early-type regions, with only a few exceptions of mixing. If the cut were slightly higher, groups 1 and 2 would be the first to merge. We can investigate the division of ages between groups further by taking a lower cut with more resulting groups, this is represented on Fig\,\ref{fig:dend_so_5np} by the green boxes.

\begin{figure}
	\centering
    \includegraphics[width=\columnwidth]{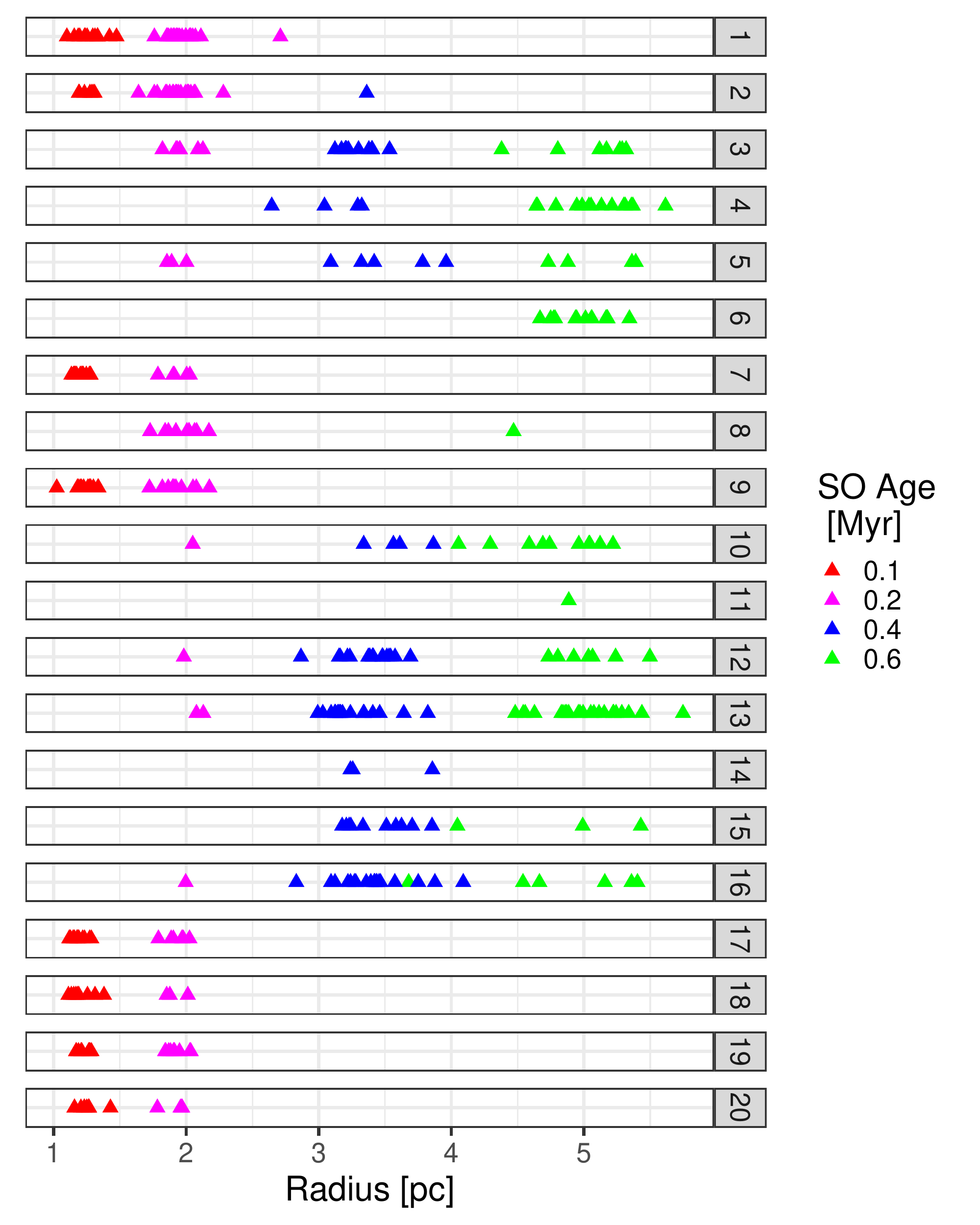}
    \caption[Radii and Ages of SOs by Group]{Distribution of SO effective radii and age for 20 groups from the hierarchical clustering of the shape data.}
    \label{fig:radius_so_5np}
\end{figure}

Figure \,\ref{fig:radius_so_5np} shows the ages and effective radii of the SO \hii regions across 20 groups from the dendrogram. Here, the mean number of region per group is $\sim$20. It is clear that even with this many groups, it is still most common for there to be a mix of both early-type regions and both late-type regions belonging to each group. The few exceptions are group 8, with mostly 0.2\,Myr regions and one 0.6\,Myr region; group 14 has only 0.4\,Myr regions; and groups 6 and 11 contain only 0.6\,Myr regions. Increasing the number of groups further (to e.g. 40 groups), results in the same pattern of the grouping of early- and late-type regions, with a few more instances of exclusively differentiating the respective individual ages. This result shows that there is a lot of similarity between the shapes of the early-type and late-type SO \hii regions, an observation we can also make from considering the interpolation splines which define the shape in Fig.\,\ref{fig:so_spline}. 

A noteworthy point here is that the late-type regions are those which had their emission artificially boosted before being inserted to the MAGPIS tiles, to account for the loss of mass and lower density within the simulation grid. Whilst this could be the defining factor for the differentiation between these regions shapes, we do still see associations between some of the early-type regions and these late-type regions. Referring to the overview of \hii region shapes from the purely Gaussian distributions in Appendix\,\ref{app:so_images}, we see that even with the more uniform noise, certain projections from the 0.2\,Myr SOs feature boundaries with more perturbations. Such examples are those likely to be grouped with the late-type regions, based upon what we already know from how the grouping procedure works \paperIp. Furthermore, we do still see from Fig.\,\ref{fig:radius_so_5np}, both the cross over and distinction in the obtained groups between the 0.4 and 0.6\,Myr shapes. For the purpose of further investigations, we will maintain that the late-type regions are thus representative of their Galactic counterparts. We will, however, return to this in future work when considering simulations from a larger grid. 

Figure\,\ref{fig:radius_so_5np} also shows that there is not a clear distinction in region radii by group, apart from that seen in the main split in early and late type regions (possessing small and large radii, respectively). In fact, some of the groups that host both early- and late-type regions display a large spread in region radii. The results from the MAGPIS data in \paperIt~ showed that one of the identified groups was exclusively small regions. The fact that this result is not seen here could reaffirm the notion that those regions from the previous work had similar shapes because of their young ages, and not purely their small sizes.

\begin{figure}
	\centering
    \includegraphics[width=\columnwidth]{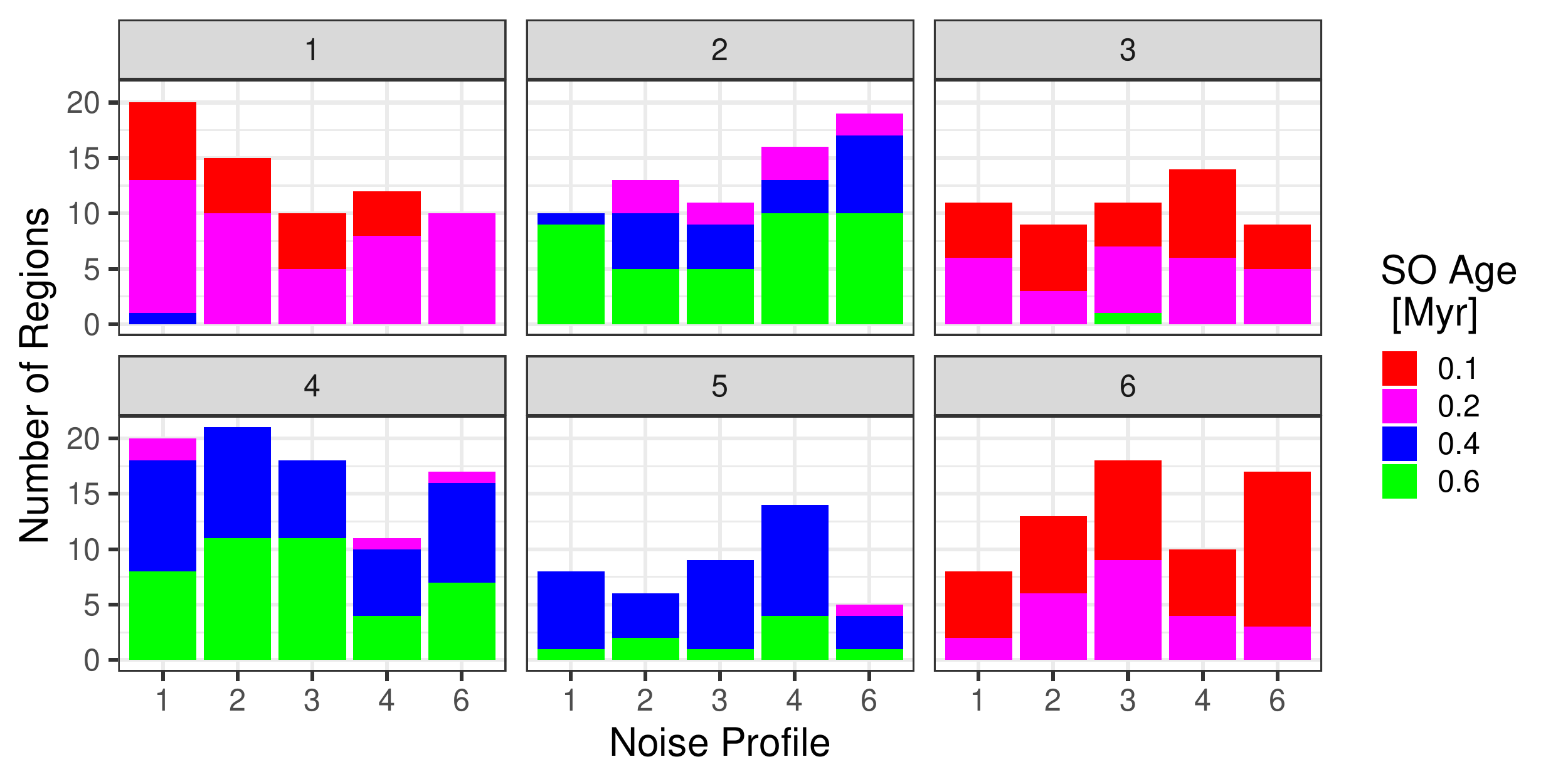}
    \caption[Distribution of NPs by Group]{Bar chart showing the distribution of MAGPIS noise profiles hosting the 385 SOs, for the six groups delineated in Fig.\,\ref{fig:dend_so_5np}.}
    \label{fig:np_by_group}
\end{figure}

The remaining parameters to investigate if and how they influence the \hii region shapes are the noise profile and projection. Figure\,\ref{fig:np_by_group} shows the distribution of NPs across the six groups identified by the red boxes on the dendrogram. It appears that there is no clear preference for regions belonging to a given NP to be placed in a particular group. Group 1 has slightly more regions from NP 1 than the other NPs. Group 4 has fewer regions from NP 4 than any of the other NPs, whilst group 5 shows the opposite result. Considering the ages along with the NPs, the regions from NP 6 that appear in group 1 are only 0.2\,Myr old, and all but one of the regions from NP 1 in group 2 are 0.6\,Myr old. The majority of NPs in each group, however, are associated with at least two of the ages, again, split by early- and late-type ages. Similar results to these are obtained by considering each of the $\theta$ and $\phi$ projection angles. There is no clear preference for a given observation angle to result in regions being assigned to the same group. 

\begin{figure}
	\centering
    \includegraphics[width=\columnwidth]{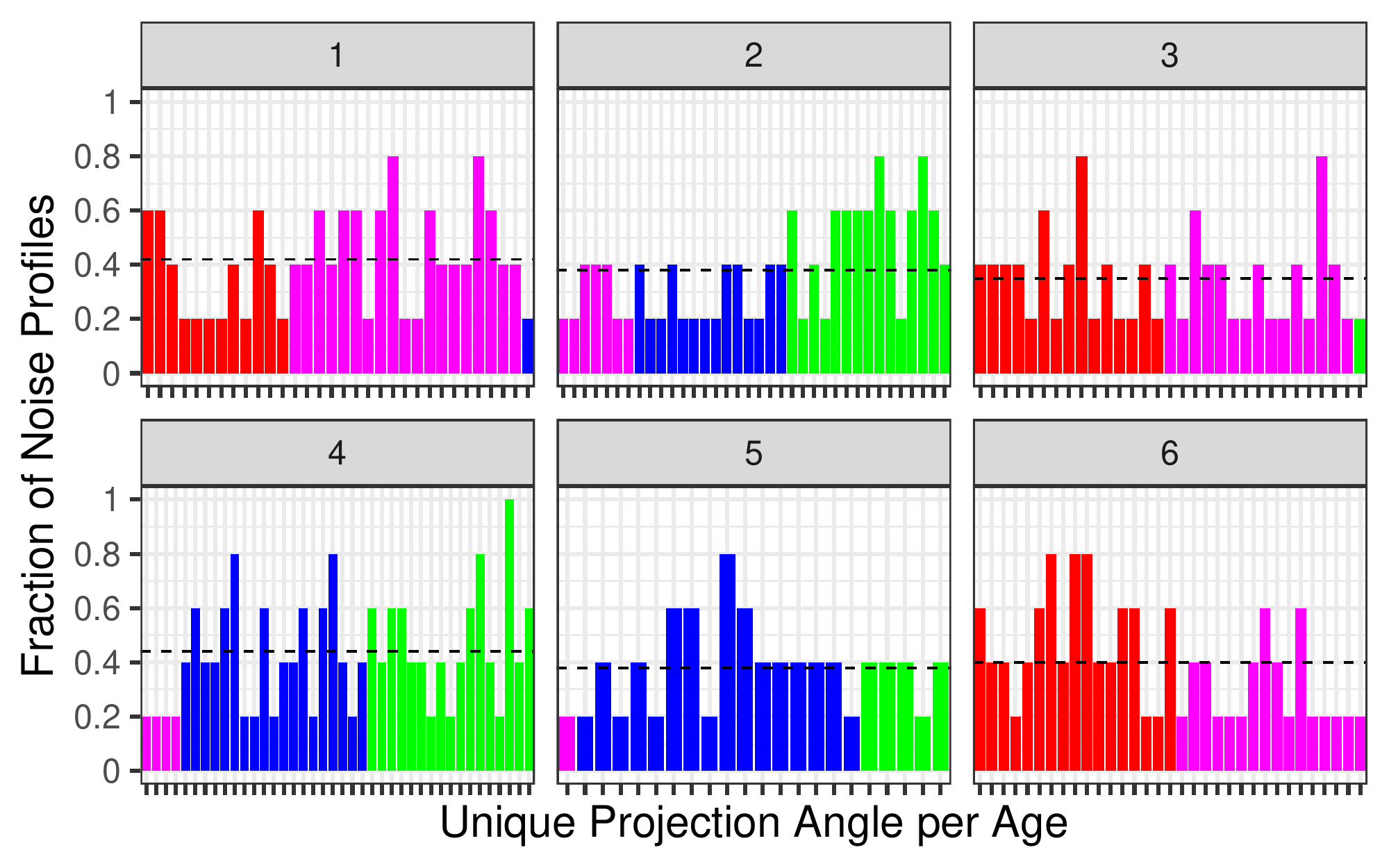}
    \includegraphics[width=\columnwidth]{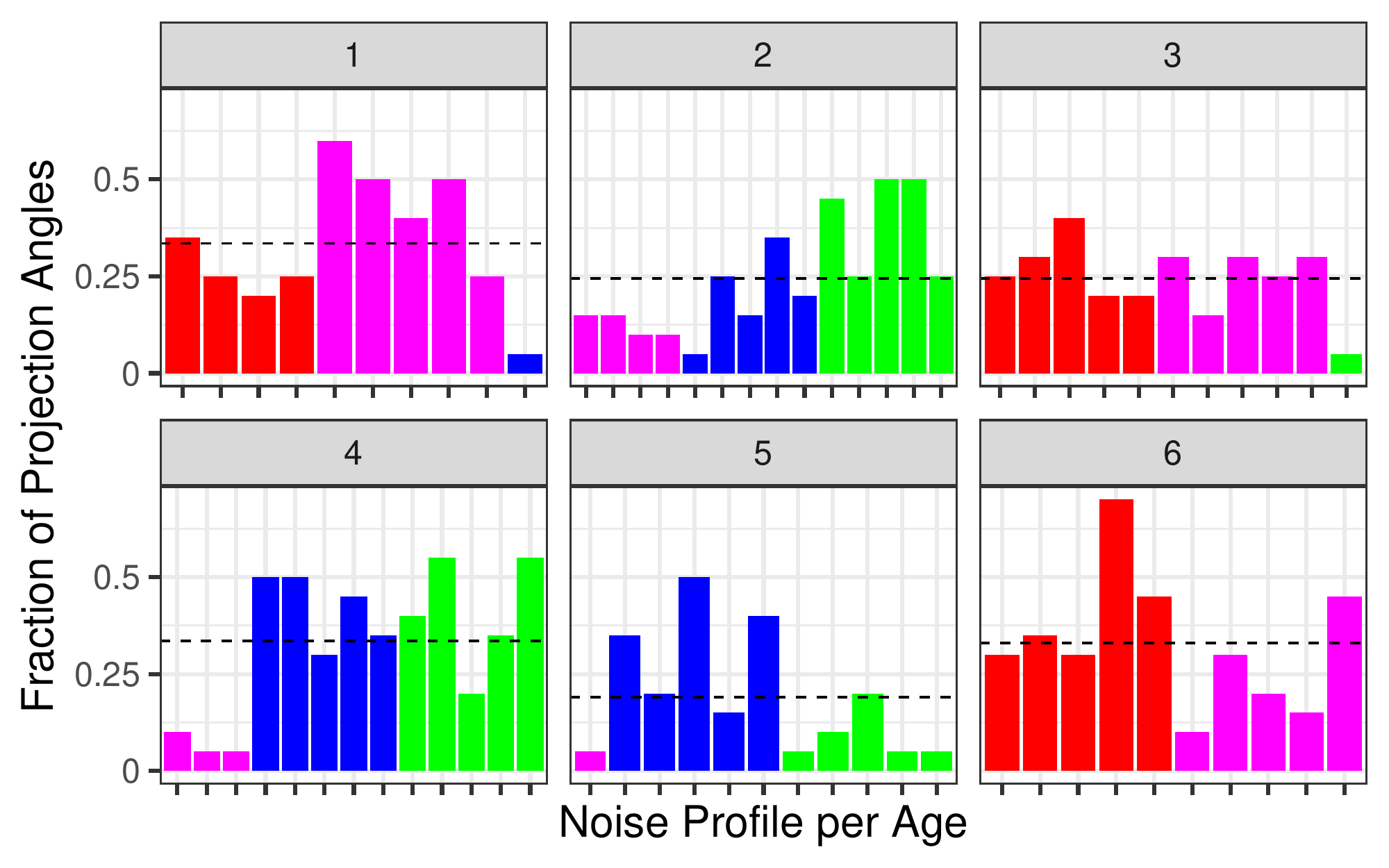}
    \caption[Overview of NP and Projection Influence on Group]{Bar charts showing the respective influence of changing the NP or the projection angle of the SO on resulting group. Ages are as with previous plots: 0.1\,Myr in \textcolor{red}{red}, 0.2\,Myr in \textcolor{magenta}{pink}, 0.4\,Myr in \textcolor{blue}{blue} and 0.6\,Myr in \textcolor{green}{green}. Top: Distribution of different NPs for a fixed age and projection. Each bar represents a given SO projection angle and age, with the fraction of NPs belonging to the respective group shown. Bottom: Distribution of different projection angles for a fixed age and NP. Each bar represents a given SO age and NP, with the fraction of projection angles belonging to the respective group shown. Mean values in each group are shown by the dashed black lines.}
    \label{fig:proj_np_comp}
\end{figure}

Another way to discern whether the NP or the projection angle has more of an influence on the shapes and obtained groups, is to consider for a given NP, how many projections of each snapshot age are grouped together. Conversely, for a given projection angle, how many of the five NPs for each age are grouped together. These results are shown in Fig.\,\ref{fig:proj_np_comp}. For the six groups described previously, the top panel shows for a given projection angle and age, what fraction of the five NPs are placed in a given group. The bottom panel shows for a given NP and age, what fraction of the 18-22 different projection angle SOs are grouped together. The respective x-axes have not been labelled since it is only the relative distributions we are interested in. Mean values for each group are indicated by the dashed black line. 

The top panel of Fig.\,\ref{fig:proj_np_comp} shows that it is most likely for two of the NPs for a given age and projection to be grouped together, suggesting that the NP is having a large influence on the extracted shape of the \hii region. For the examples where three or four of the NPs are grouped together, there is no preference for this to occur in a given group. There is only one situation where all five of the NPs are grouped together, that is for one of the 0.6\,Myr projections in group 4. The bottom panel shows some slight differences for the fraction of projections for a given NP and age that are grouped together. Group 1 has on average, 33.5\% of the total number of projections for a given NP grouped together, with a slight preference for up to half of the 0.2\,Myr projections to be grouped together. In group 6, the average is again 33.5\% projections, however, NP 6 for the 0.1\,Myr regions includes 70\% of the different projections. For the late-type regions, group 4 shows the highest average number of projections per NP, followed by the 0.4\,Myr regions in group 1 and the 0.6\,Myr regions in group 2. The average number of given projections per group is only slightly higher for the early-type regions. This is an interesting result since the early-type regions are much more spherically symmetric and less perturbed than the late-type regions. This again shows that the curvature method for representing the region shapes is robustly quantifying the regions based on their boundaries. 

The result that, in some groups, $\sim 50\%$ of the different projections of a given age and NP are grouped together shows that changes in the viewing angle may not influence the shape of the \hii region substantially. Whilst this may be due to the initial condition of spherical symmetry throughout the numerical simulations, this is still a result that can only be achieved via study of the SOs. These are also likely the regions that remain the most spherically symmetric across the different projection angles, i.e., those that do not possess and intruding or protruding features. Previous investigations of inferring the projection angle of \hii regions relative to the observer has focused on such `cometary' and `champagne flow' features, to see whether environmental properties that influence these inhomogeneities can be determined from the \hii regions themselves \citep[e.g.][]{1983A&A...127..313Y,2017MNRAS.466.4573S}. Since we do only have the one vantage point for our observational data, the requirement for classification is still to be able to carry this out in conjunction with the information we can collect. Therefore, in the present work, having the different projection angles of the SOs essentially gives us different observations of \hii regions that share the same initial conditions but can be thought of as evolving differently due to differences in ambient density or ISM structure along our line of sight. However, investigating further how the noise structure of our observations influences the shape we observe is an important aspect towards refining an observational morphological classifier.

The manner by which the \hii region shapes were extracted from both the MAGPIS tiles and the SOs was by analysing the background noise from the radio continuum images. By removing the radio signal and setting a threshold level that was above the remaining noise profile, this enabled the boundary of each region to be defined by the contouring procedure. This led to a systematically defined data sample, whereby the \hii regions were each extracted, such that their signal levels should be consistent across the Galactic Plane. Nevertheless, one could argue that if you took one of the \hii regions and placed it in a different area of the Galaxy, defining the boundary from the background radio noise in the vicinity could lead to the shape being different. This was thus what the SOs allowed us to test by doing exactly that. We have seen here that these different NPs, defined from the MAGPIS tiles, are influencing the shape. The results also show, however, that this is not a systematic influence, and SO \hii region shapes from given noise profiles are not concurrently being paired together by the hierarchical clustering. We carried out some further investigation into a systematic influence by the different NPs using the ordinance visualisation of multi-dimensional scaling (MDS, detailed in Appendix\,\ref{app:so_noise_shape}), but reach the same conclusions as we do in this subsection; that the influence is not systematic. This is discussed further in Appendix\,\ref{app:so_noise_shape} and Sec.\,\ref{ssec:future_work}, where we suggest how we may overcome issues with noise in our future work.

\begin{figure}
	\centering
	\includegraphics[width=\columnwidth]{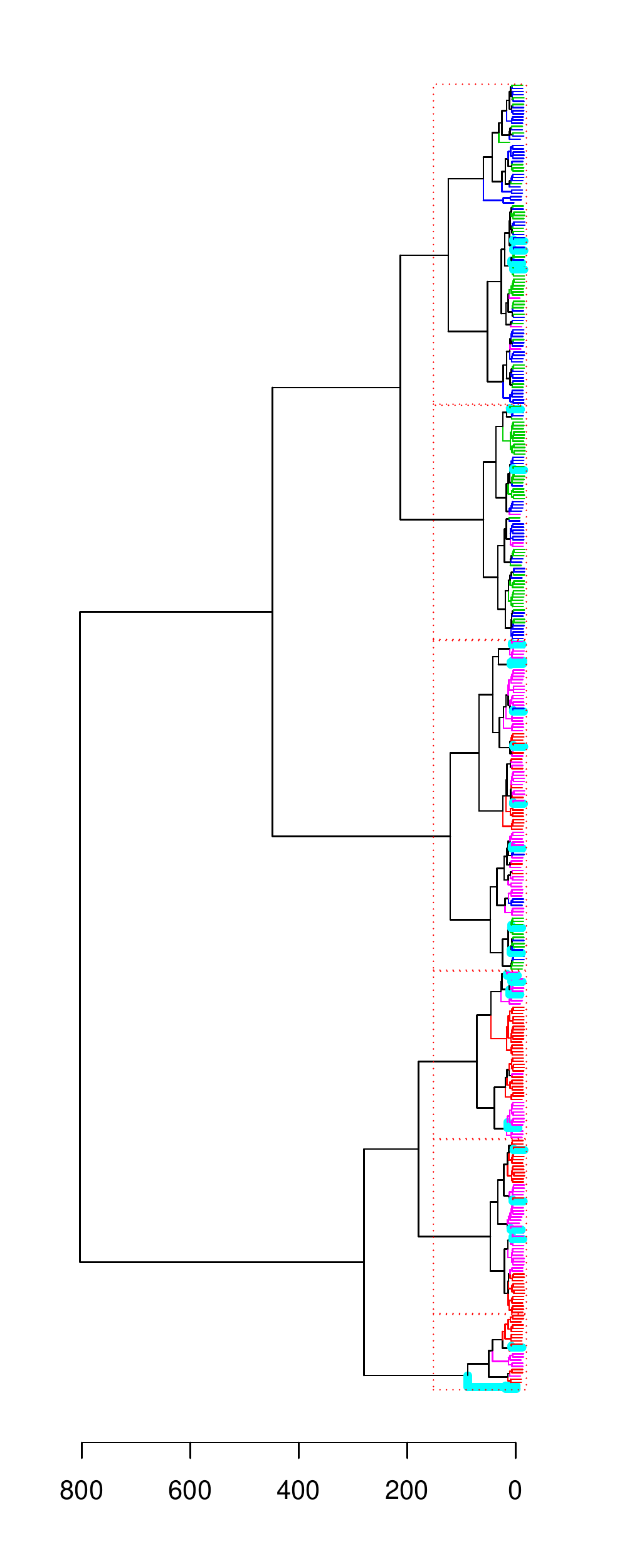}
	\caption[Overview Dendrogram of SOs Sample as a Training Set for MAGPIS Subsample]{Dendrogram resulting from applying hierarchical clustering to the shape data of the sample of 385 SO \hii regions along with 26 of the MAGPIS \hii regions. The branches are coloured by their age for the SO data: 0.1\,Myr in \textcolor{red}{red}, 0.2\,Myr in \textcolor{magenta}{pink}, 0.4\,Myr in \textcolor{blue}{blue} and 0.6\,Myr in \textcolor{green}{green}.; and the MAGPIS \hii regions are in \textcolor{cyan}{cyan}.}    
	\label{fig:training_dend}
\end{figure}

\subsection{Using the SO Sample as a Training Set}
\label{ssec:training}

Although we may not be able to fully quantify a systematic effect that the different noise profiles or projections are having on the SO \hii region shapes, the SOs still have much better constraints than the Galactic observations in terms of initial conditions and properties. We have also shown here, via the shape analysis method, that they are explicitly representative of the Galactic \hii regions. Considering the sample of 385 SOs, we can say with certainty, which evolutionary stage each SO is at. The different projections and noise profiles essentially give us many more observations of the given age. This is similar to what we observe along the Galactic plane, with no preference for \hii regions of a given age to be at a given place along the Plane \citep{2014ApJS..212....1A}. Ultimately, for a training set of SOs, we would also require SOs of differing initial conditions. Using the data we have analysed in this work, however, we can test whether the SO data sample can be used to infer the ages of the MAGPIS observational sample. 

The initial conditions of the numerical simulations, which produce the SOs analysed here, involve ionisation and radiation pressure feedback from a star of mass 33.7\,\msun, with an ionising photon rate of N$_{\rm{ly}} =$ 7.36 $\times 10^{48}$ s$^{-1}$, or log N$_{\rm{ly}}$ = 48.86 \citepalias{2018MNRAS.477.5422A}. Since the observational sample we considered in the \paperIt~ cover a range of masses ($\sim$ 17 - 45\,\msun), we have selected a mass limited sample from the MAGPIS \hii regions to use in the following test. The limit used was 48.3 $<$ log N$_{\rm{ly}}$ $<$ 48.8. This corresponds to a mass range between 23 and 34\,\msun\, \citep{2010A&A...524A..98W}, and was around the mean of the normally distributed values for the MAGPIS sample (Fig.\,8, \paperIt). The lower limit was taken to account for photon leakage out of the system, along with the mass leaving the simulation grid and decreasing density. The resulting SOs may hence be representative of regions of lower masses. This resulted in a subsample of 26 MAGPIS regions, with ages ranging between 0.1\,Myr and 1.9\,Myr. Whilst this age range of the MAGPIS subsample covers a considerably larger range than the SOs considered in this work, there is not a one to one correspondence between the two ages used. The ages from the SO snapshots begin with $t = 0$ when feedback starts in the simulation. Whereas, the ages calculated for the MAGPIS sample are only an estimation of the dynamical ages. These estimates involve assumptions regarding the surrounding ISM and require accurate distances to the regions. The dynamical age then considers the observed expansion with respect to the theoretical Str{\"o}mgren radius. Therefore, for the purpose of this test, we can consider the respective age ranges and distributions of both the SOs and MAGPIS sample, to see how they compare.

\begin{figure}
	\centering
	\includegraphics[width=\columnwidth]{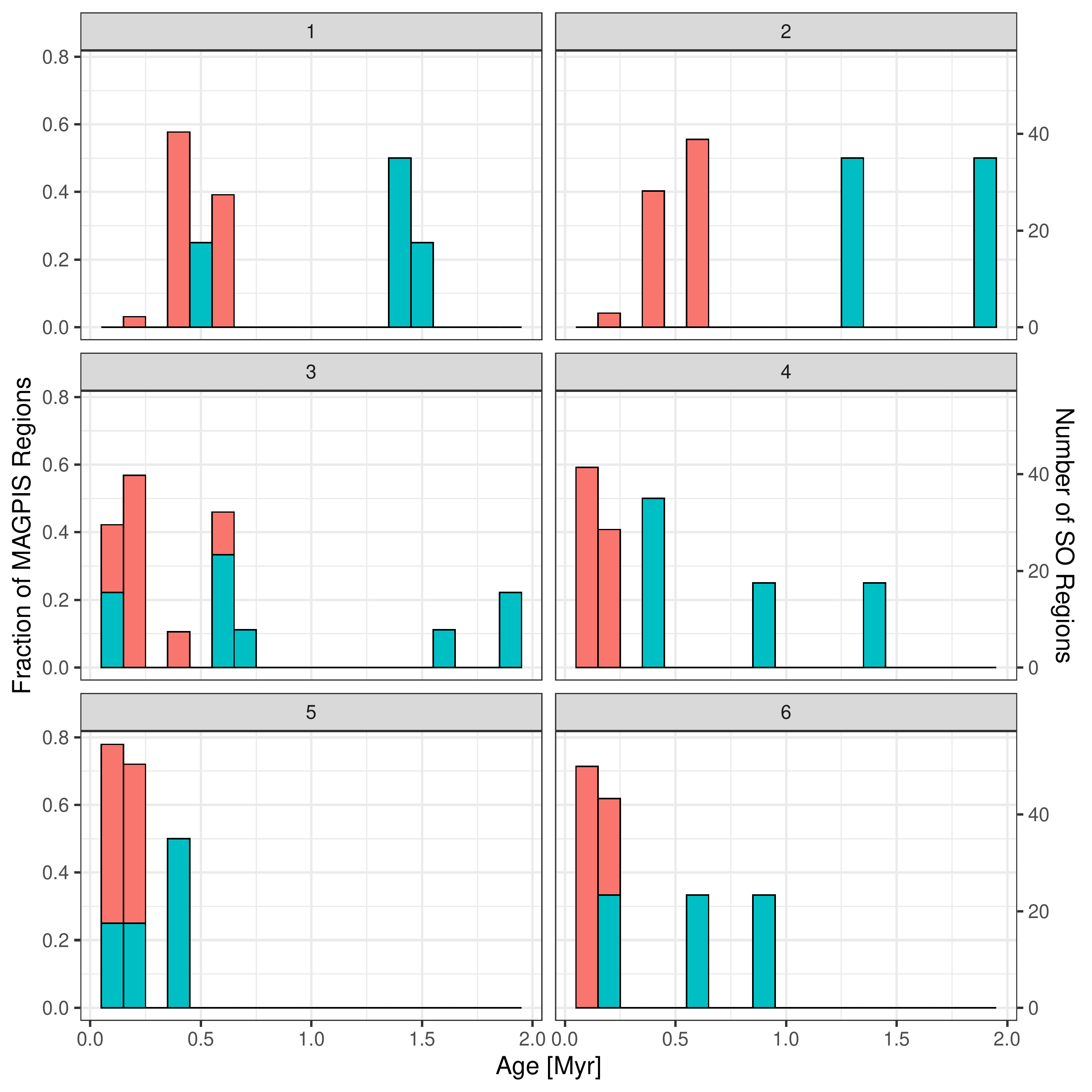}
    \caption[Summary of Ages of the Training Set Groups]{\hii Region ages for six groups identified from Fig.\,\ref{fig:training_dend}. Groups are numbered 1 through 6 corresponding to top down on the dendrogram, as with previous figures. Number of SO \hii regions in each group is shown by the orange bars. For comparison, the fraction of MAGPIS \hii regions per group is shown by the cyan bars.}
    \label{fig:training_ages}
\end{figure}

Using the data set of 385 SOs described in Sec.\,\ref{ssec:hclust_np_so}, we inserted the 26 mass limited MAGPIS \hii region shapes to the SO sample; and looked to see where the MAGPIS \hii regions would be placed in relation to the SOs in the resulting group structure. Figure\,\ref{fig:training_dend} shows the dendrogram resulting from the hierarchical clustering procedure for the SO training data and the MAGPIS target data. Branches are coloured as before - 0.1\,Myr in red, 0.2\,Myr in pink, 0.4\,Myr in blue and 0.6\,Myr in green - with the addition of the MAGPIS \hii regions in cyan. The introduction of the MAGPIS regions has changed the ordering of the six delineated groups, with mostly late-type regions shown in the top two groups and mostly early-type regions in the rest. The ordering of the final groups is arbitrary. It is the resulting group associations and hierarchy that matter. The MAGPIS \hii regions appear slotted in to the SO \hii regions. The same result of good correspondence that was seen with the data the other way around in Sec.\,\ref{ssec:comp_so_mag}. There are two MAGPIS regions, paired together, joined to the bottom group, These two may represent slight outliers since they join to the adjacent group at a substantial height. We will discuss this is more detail later in the section. 

Figure\,\ref{fig:training_ages} shows the respective ages of the SO and MAGPIS \hii regions that are grouped into the six groups in Fig.\,\ref{fig:training_dend}. In order to compare the respective age distributions of the data, the number of SO \hii regions are shown in orange, with the fraction of MAGPIS \hii region per group shown in cyan. As seen previously with the SO data, there is a clear division between early- and late-type regions, with group 3 showing the most cross over between ages. Following from the respective age discrepancies, mentioned previously, we can split the MAGPIS \hii regions into early- and late-types by considering those with age less than 1\,Myr to be early-type and those greater than 1\,Myr to be late type. 

Group 4 hosts exclusively early-type SOs with a mean age of 0.14\,Myr. 75\% of the MAGPIS regions assigned to this group are also early-type. Group 3 has the largest mix of early- and -late type SOs, but with majority 0.2\,Myr SOs and a mean age of 0.25\,Myr. Two thirds of the MAGPIS regions in group 3 are also early-type, with the remaining late type, showing a good agreement in spread with the SO data. Group 5 has exclusively early-type regions for both the SO and MAGPIS data. The same result is seen for group 6. Group 2 hosts majority late-type SOs, with a mean age of 0.5\,Myr. The two MAGPIS regions assigned to this group are also late-type. 75\% of the MAGPIS regions in group 1 are late-type, in good agreement with the late-type assignment of the SOs. In terms of this distribution of relative ages for each sample, these results are promising for the prospect of using the SOs as a training set for supervised classification. We see here that even with only one parameter of investigation, the evolutionary stage of the regions, we have good agreement between the SOs and MAGPIS observed sample. We do not, however, suggest that one set of initial conditions substantially represents the entirety of the age distribution of the observational sample. The identified groups from \paperIt\, were shown to have a spread of mass ranges, which is why we used the mass limited sample here. 

In addition to these results, Appendix \ref{ap:mag_images} shows an overview of the mass-limited MAGPIS \hii region sample images and shapes that were assigned to each SO group. We can see from the figures in Appendix \ref{ap:mag_images} that the MAGPIS regions sorted into each of the test groups appear to share similar morphological features. This reaffirms the notion that the shape analysis and statistical methods employed here are performing as expected. Group 3 is the largest group, showing the most visual differences between MAGPIS regions. This would be the first group to split if the cut was made lower on the dendrogram in Fig.\,\ref{fig:training_dend}, which would separate the more uniform regions from the more perturbed. The first two MAGPIS regions shown in the images for group 6 are those located at the edge of the dendrogram. They appear to each host at least one tight point of inflection, which would result in a large outlier in the curvature distributions. This is the likely cause for why they are slightly apart from the rest of the data. The comparatively smooth sections along the rest of these region's boundaries are likely why they were grouped to the other MAGPIS region, and the corresponding SO regions in group 6.  

The notion of using the SOs as a training set in supervised morphological classification of \hii regions would require an input sample with many varied initial conditions and known parameters. We have shown throughout this investigation that the SOs produced in \citetalias{2018MNRAS.477.5422A} are well representative of their Galactic counterparts. Furthermore, we can see here that the SOs of different ages can be used to suggest whether a Galactic \hii region is early- or late-stage. With different initial masses and ambient densities, we could further refine the parameters of each training set group, and repeat this investigation with a correspondingly larger sample of Galactic \hii regions.

\subsection{Future Applications}
\label{ssec:future_work}
The foremost proposed application of this work is to increase the number of parameters investigated by using SOs from numerical simulations of differing initial masses, ambient densities, environmental conditions and temperatures. If similar groupings that are shown in this investigation are seen for the larger sample set, we will be one step closer to a thorough morphological classification scheme. For example, we might find that the groups from the hierarchical procedure still differentiate between early- and late-type \hii regions, but also by intermediate mass and high mass ionising sources, a result that was indicated for the Galactic regions in \paperIt. Furthermore, with a different initial condition set-up, and a larger simulation grid, we would not be required to boost the emission of the regions in the SOs at later times, as we did in this investigation. We plan to extend this work to the simulations of \citet{2019MNRAS.487.4890A}, which features a 10$^4$\,\Msolar cloud. This could lead to regions that are even more representative of the observed sample. We also have further simulations currently in preparation that feature clouds of different metallicity environments, together with differing ambient densities. Another factor to consider for future SOs used for comparisons is the discretisation of parameters. A more continuous sample of ages is well within reason and is only limited by the simulation time-steps. Varying parameters such as initial mass or electron density would be more computationally expensive, but as the data becomes available, it will be useful to have a shape analysis tool ready for the analysis and comparison.

In terms of the observational sample of \hii regions, a further application would be to compare samples from different radio continuum surveys. The work carried out here regarding a systematic effect of the noise profiles on the \hii regions was non-conclusive, i.e., if there is a systematic effect, it as not revealed by our investigation. However, comparing different observational data with different noise profiles may lead to progress in this area. Similar to how we considered the same SO with a different noise profile, we could consider the same Galactic \hii region from different surveys in the same manner. A survey with complementary coverage to the MAGPIS survey that could be used is The HI/OH/Recombination line survey of the inner Milky Way (THOR) \citep{2016A&A...595A..32B}. In addition to this, whilst we have only considered radio continuum images of \hii regions in this work, a full classification scheme based on their morphologies should also consider data from further wavelength ranges, such as MIR. This would be the first step towards evolving this method into a multi-variate classifier.

The unsupervised clustering analysis utilised herein from \paperIt~ has potential to be adapted to a machine learning (ML) algorithm. One requirement of ML is a large training set, so that the algorithm can learn the classes. This work has shown good potential for SOs of corresponding astrophysical data to be used as the training set for future applications. If the hierarchical clustering procedure was able to decide whether to continue with certain groups or reject them based upon predefined criteria, the resulting training set could be more accurate for investigating parameters of the observed samples. For example, late-type regions being assigned to known early-type groups could be excluded from the procedure and reassessed. The ML process could also consider the MDS investigation (Appendix \ref{app:so_noise_shape}) of the different NPs on the fly, rejecting any data with large ordination differences. The clustering procedure and future ML adaptations would also be useful for comparing different methods of shape extraction and description. With future clustering and classification methods, we could use the observed properties of the \hii regions (such as ionised mass) together with quantised shape to build a multi-variate descriptor to replace the shape-distance scores, which can then be compared and clustered \citep[we applied such methods as a multi-variate descriptor of light-curve variations in][]{2018MNRAS.478.5091F}. This could be the natural evolution of this work, after the next steps of increasing the sample size of both the SO and observational samples. 

Only one shape descriptor is used in this work so far, the curvature distributions obtained from the contoured boundaries of the \hii regions. Whilst we have shown the potential associations between morphology of \hii regions and physical properties, there is no clear one to one correspondence between the extracted shapes and the properties considered thus far. Our original reasoning for selecting the contouring/curvature descriptor was for automation and bias reduction. However, given that we have seen here and in our previous work, potential for shape to be used as an indicative measure, future work should also further investigate the differences between different shape descriptors. We should also consider how to properly utilise angular resolution scales, when distances are unknown. This avenue could have potential applications for reducing kinematic distance errors in observational data. Different image analysis/quantisation techniques considered for future work include convolutional/artificial neural networks and self organising maps. The comparison of such methods with the shape analysis methods used so far would be useful for confirming the validity of which method to adopt for future classification requirements, potentially based on synthetic data as the training set.

There is also further potential for this shape analysis and clustering method to be applied to other astronomical phenomena. We have quantitatively shown here that modern SOs are producing well representative \hii regions; we could therefore begin to investigate how different observations and simulations perform. For example, \hii regions could be compared to supernovae remnants, which are visually similar in the radio continuum images \citep[e.g.][]{2009MNRAS.399..177G}. It would be interesting to discover whether our shape analysis method could tell the two nebulae apart, and if so, how it compares to established methods such as spectral energy distribution fitting. Looking further ahead, as telescope and imaging techniques continue to improve, alongside computational power, there will be even more high resolution data to analyse and characterise. Statistical shape analysis could prove to be a useful tool in the era of big data. 

%%%%%%%%%%%%%%%%%%%%%%%
\section{Conclusions}
\label{sec:conc}

The synthetic observations of an \hii region produced in the numerical simulations of \citet{2018MNRAS.477.5422A} were analysed using the shape analysis and statistical clustering methodology developed in \citet{2018MNRAS.477.5486C}. The numerical \hii region was the result of photoionisation and radiation pressure feedback of a 34\,\msun\, star, in a 1000\,\msun\, cloud. 77 SOs were considered, comprising four evolutionary snapshots (0.1, 0.2, 0.4 and 0.6\,Myr), and multiple viewing projection angles. After the addition of artificial Gaussian noise, following the distribution of observational noise from one of the MAGPIS tiles, the shapes of the SO \hii regions were extracted in the same manner as they were for the MAGPIS observational sample in \paperIt. The shape analysis results provided confirmation of the efficacy of the numerical simulations, such that they are quantifiably consistent, in terms of their shape, with the real observational counterparts. When considering the 76 MAGPIS \hii regions from \paperIt~ and 12 representative SO \hii regions, across the four ages, the SO \hii regions were placed in amongst the MAGPIS \hii regions, in the resulting dendrogram from the hierarchical clustering procedure. 

This result was also found when directly inserting the SO regions to different MAGPIS tiles, to represent realistic noise for the simulation images. By using five MAGPIS noise profiles for the 77 SOs, we essentially had 385 distinct observations of the numerical \hii region, at the given age snapshots and projections. As with the shapes of the \hii regions using the artificial Gaussian noise distribution, those from the MAGPIS NPs were grouped in amongst the MAGPIS \hii regions, with the majority of synthetic regions paired with one of the MAGPIS regions. This suggested that the different projection angles and noise profiles were having a significant impact on the regions shapes. Such that the SO \hii regions of the same age were not exclusively grouped with each other in the hierarchical clustering. 

When considering the hierarchical clustering of the 385 SOs that had been inserted into the MAGPIS tiles, the following results were obtained:
\begin{itemize}
  \item The determined hierarchy showed a clear divide between early- (0.1 and 0.2\,Myr) and late-type (0.4 and 0.6\,Myr) regions. This divide was not exclusive by age, with a low cut on the dendrogram (resulting in many groups) still producing groups that possessed a mix of both early- and late-type regions. Whilst this may be due to how the late-type region's emission had to be artificially boosted to account for mass leaving the simulation grid, the results for the late-type regions still show the same cross over as the early-type regions. Furthermore, these late-type regions were still shown to be representative of the MAGPIS observational sample, even with this boosting. However, this is a point to return to in future work with simulations from a larger grid. 
  \item There was no further association between the identified groupings and SO region radii, apart from the main split between the early- (small radius) and late-type (large radius) regions. This suggests that the result obtained in \paperIt, pertaining to one group hosting exclusively small regions could in fact be due to those regions all being young \hii regions. 
  \item There was no strong preference for SO regions from a given noise profile, nor given projection angle, to be assigned to specific groups. In terms of which of these parameters has more of an effect on the shape -- for a given SO age and projection angle, on average two of the five NPs were grouped together in the hierarchical clustering. For a given age and NP, on average $\sim 30\%$ of the different projections were grouped together. This result was consistent for both the early- and late-type SO \hii regions, even though the early-type regions appeared to be more spherically symmetric. 
  \item No systematic effect due to the different NPs was found by the analysis. The hierarchical grouping of the pairwise shape distance measures revealed no preference for SOs from a given NP to be grouped together. Further investigation using multidimensional scaling ordinance also revealed no such systematic influence.
  \item The MDS did show systematic effects for how the shape of the \hii regions is extracted. Different initial contour levels (for identifying the boundary of the \hii regions) changed the ordination in the MDS axes, but not the relative scores along the axes. Whilst different spatial shape resolutions changed the scores along the axes but not the relative ordination positions. Higher resolutions corresponded to a larger spread in MDS scores, showing that as more detail is considered, the variances in shape as a result of the different NPs is more profound. 
\end{itemize}

\noindent The results in this work have shown that the realistic SOs considered here are conclusively morphologically representative of the Galactic \hii regions we observe in the MAGPIS radio continuum survey. To determine whether the SOs could potentially be used to construct a training set for supervised classification of \hii regions, via their shapes, a mass limited sample of the MAGPIS \hii regions were considered along with the 385 SOs. These results showed that there was good correspondence between respective early- and late-type \hii regions from each sample. This suggests that there is a lot of potential for the utilisation of the SOs to construct such a training set. For this SO sample, we only investigated whether there was correspondence between the ages, since we only considered SOs from one set of initial conditions. For a larger training set, of varying masses and ambient densities, across the different evolutionary stages, the results shown here suggest that we should be able to make predictions of the physical nature of the Galactic \hii regions, based upon how their shapes compare to those of the model simulations.

\section*{Acknowledgements}
We thank the referee for their careful review of this manuscript. We thank S. E Ragan for organising the Cardiff Galactic Star Formation Workshop, which facilitated this collaboration. We also thank J. S. Urquhart and M. A. Thompson for the thorough discussion that led to multiple improvements in this work. JCW acknowledges the studentship provided by the University of Kent. AAA acknowledges funding from the European Research Council for the Horizon 2020 ERC consolidator grant project ICYBOB (grant number 818940), and a Science and Technology Facilities Council (STFC) studentship. The synthetic observations were calculated on DiRAC Complexity at the University of Leicester, and the DiRAC@Durham facility managed by the Institute for Computational Cosmology. These form part of the STFC DiRAC HPC Facility (www.dirac.ac.uk). Complexity was funded by BIS National E-Infrastructure capital grant ST/K000373/1 and  STFC DiRAC Operations grant ST/K0003259/1. DiRAC@Durham was funded by BEIS capital funding via STFC capital grants ST/P002293/1, ST/R002371/1 and ST/S002502/1, Durham University and STFC operations grant ST/R000832/1. DiRAC is part of the National e-Infrastructure.

%%%%%%%%%%%%%%%%%%%%%%%%%%%%%%%%%%%%%%%%%%%%%%%%%%



\bibliographystyle{mnras}
\bibliography{paper_bib.bib} % if your bibtex file is called example.bib

%\clearpage
%\newpage

%%%%%%%%%%%%%%%%%%%%%%%%%%%%%%%%%%%%%%%%%%%%%%%%%%

%%%%%%%%%%%%%%%%% APPENDICES %%%%%%%%%%%%%%%%%%%%%

\appendix

\section{Further Details on Noise Profiles and Selection Choices}
\label{app:so_noise_shape}

In an attempt to determine whether we can quantify the influence of the MAGPIS noise profiles on the shape of the SO \hii reigons, we return to the ordinance technique of multi-dimensional scaling (MDS) that was used in \paperIt~ to check that the hierarchical clustering was properly defining groups based upon the regions shapes. To recap: MDS reduces the dimensionality of an input distance matrix to a number of orthogonal principal coordinates. The eigenvectors of which, give the ordination and the eigenvalues give the relative importance of that axis for representing the data variation. In \paperIt, we saw that there was a correspondence between the amount of high curvature points along the region boundaries and the scores along axis 1 of the MDS ordination, and surmised that the variation along axis 2 was also directly resulting from features of the curvature distributions. We also showed that the ordinations on the MDS plots corresponded well with the groups from the hierarchical clustering. Here, we can apply MDS to the distance matrix of \hii region shapes for a given SO across the different NPs, to both visually and quantitatively see how the resulting mathematical shapes compare. 

\begin{figure}
	\centering
    \includegraphics[width=0.48\columnwidth]{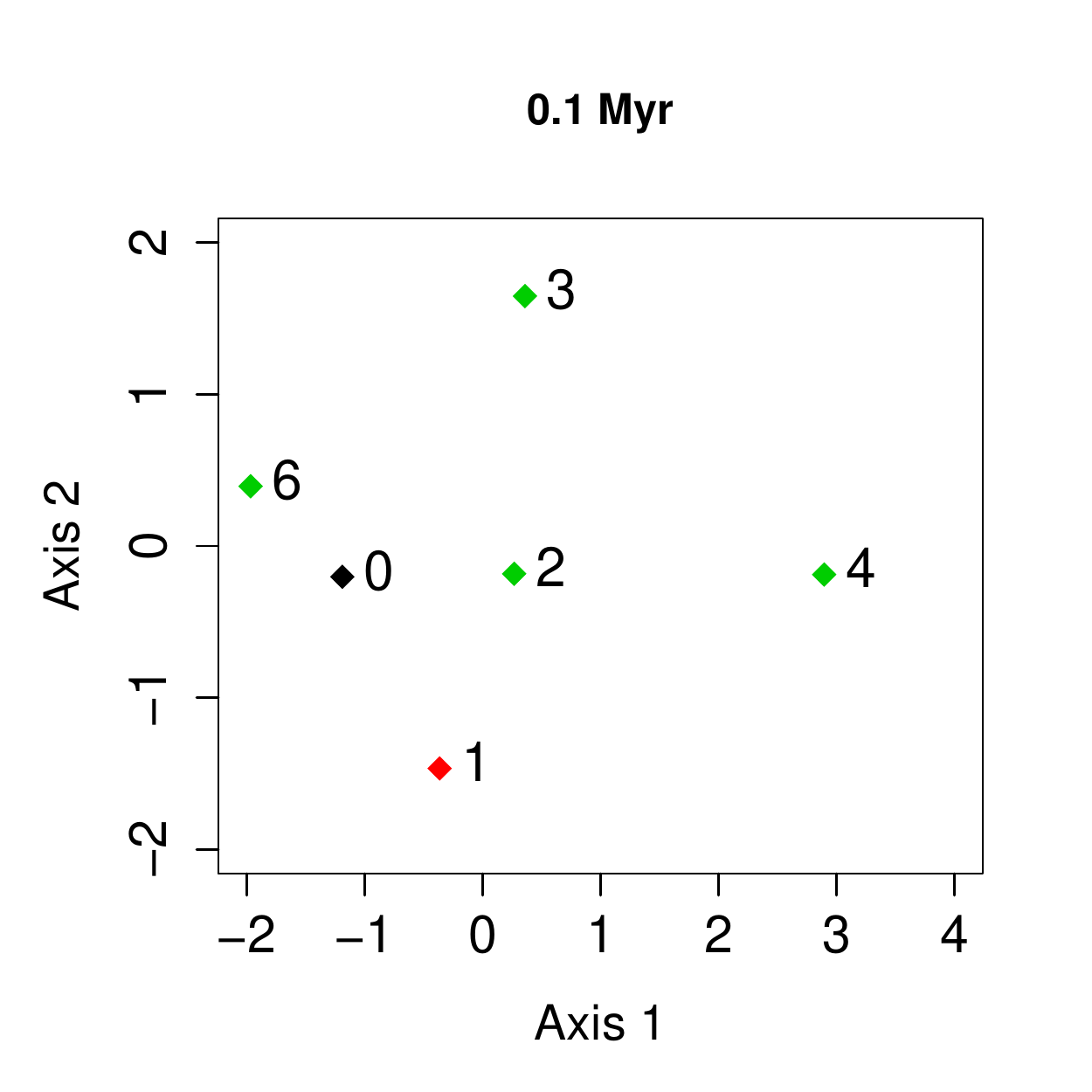}
    \includegraphics[width=0.48\columnwidth]{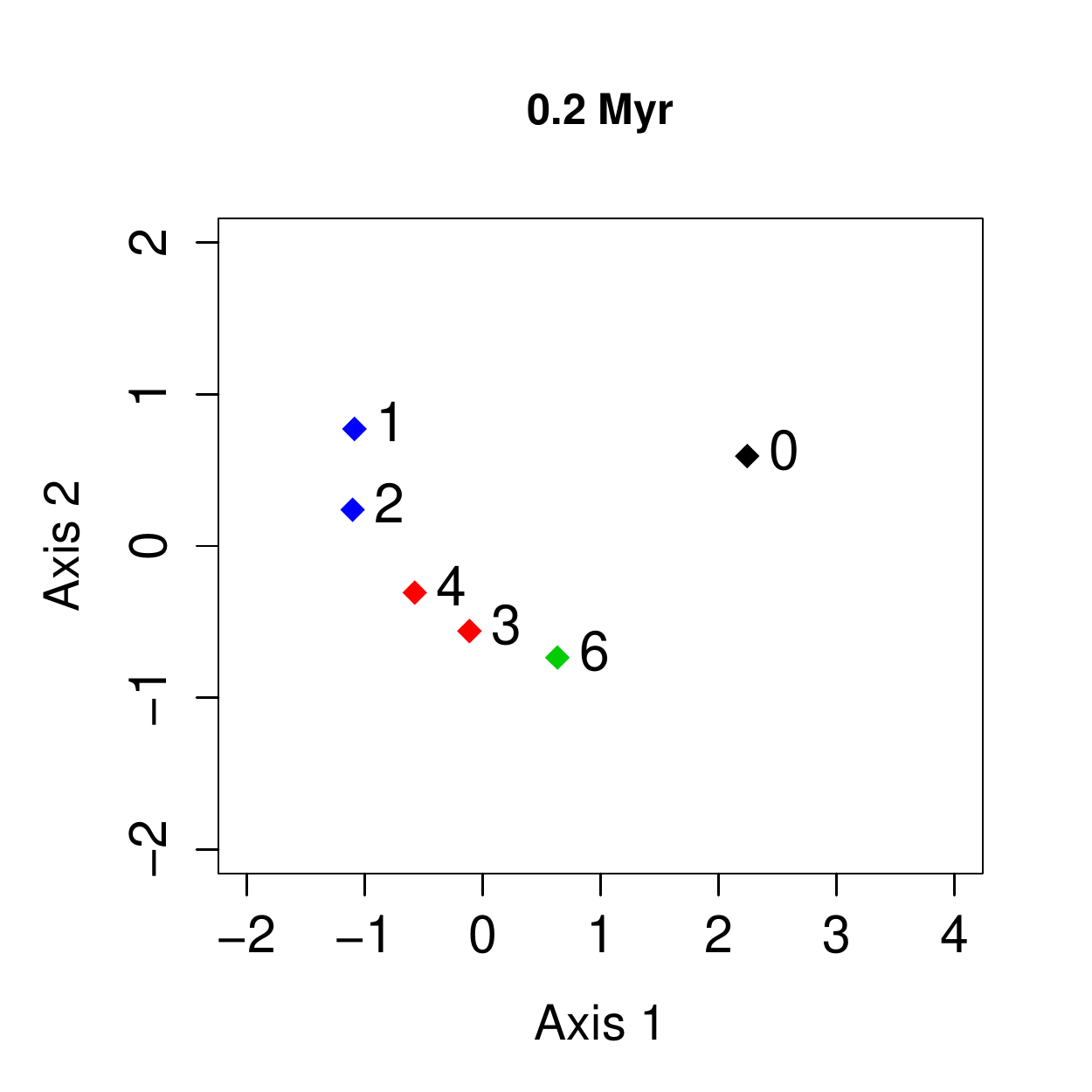}
    \includegraphics[width=0.48\columnwidth]{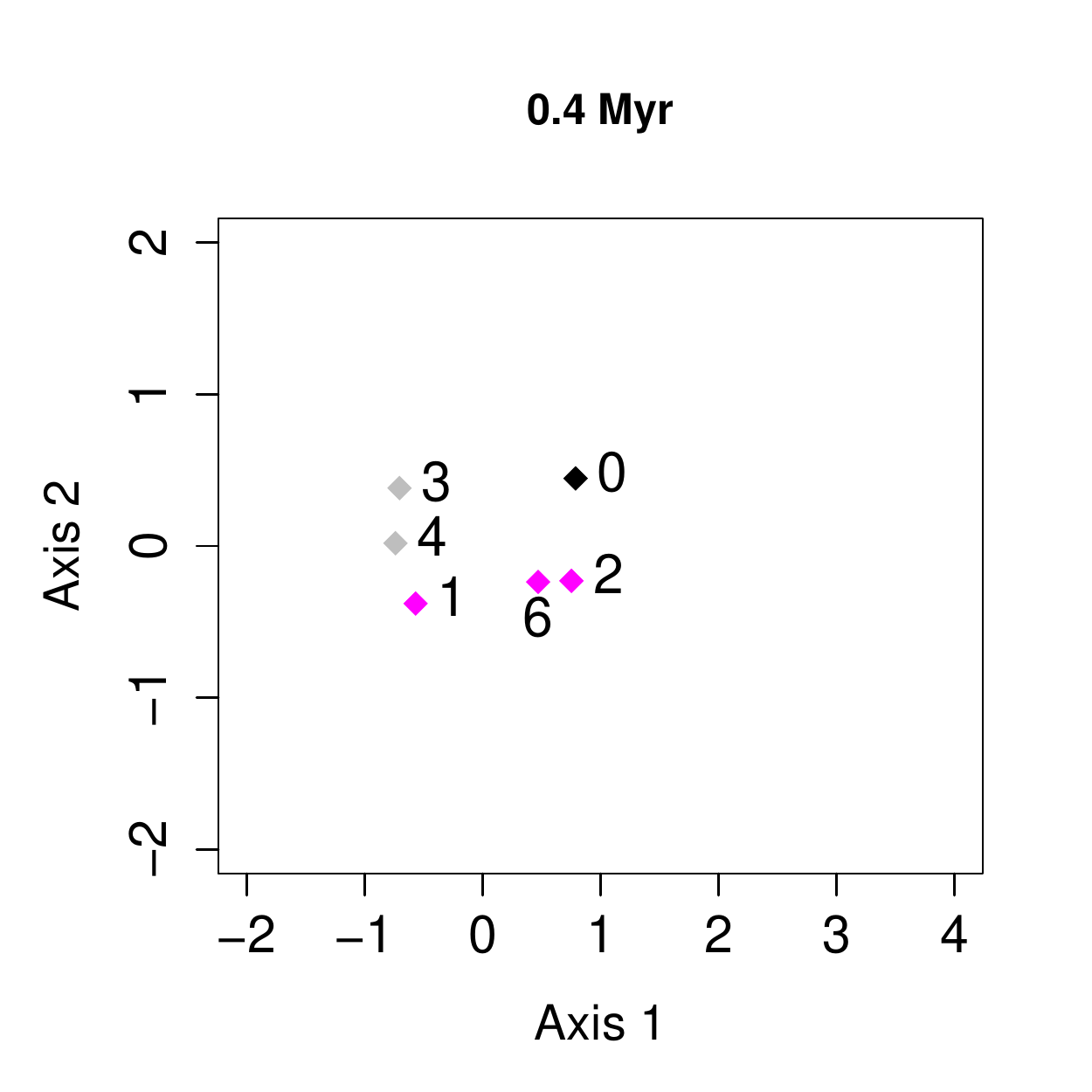}
    \includegraphics[width=0.48\columnwidth]{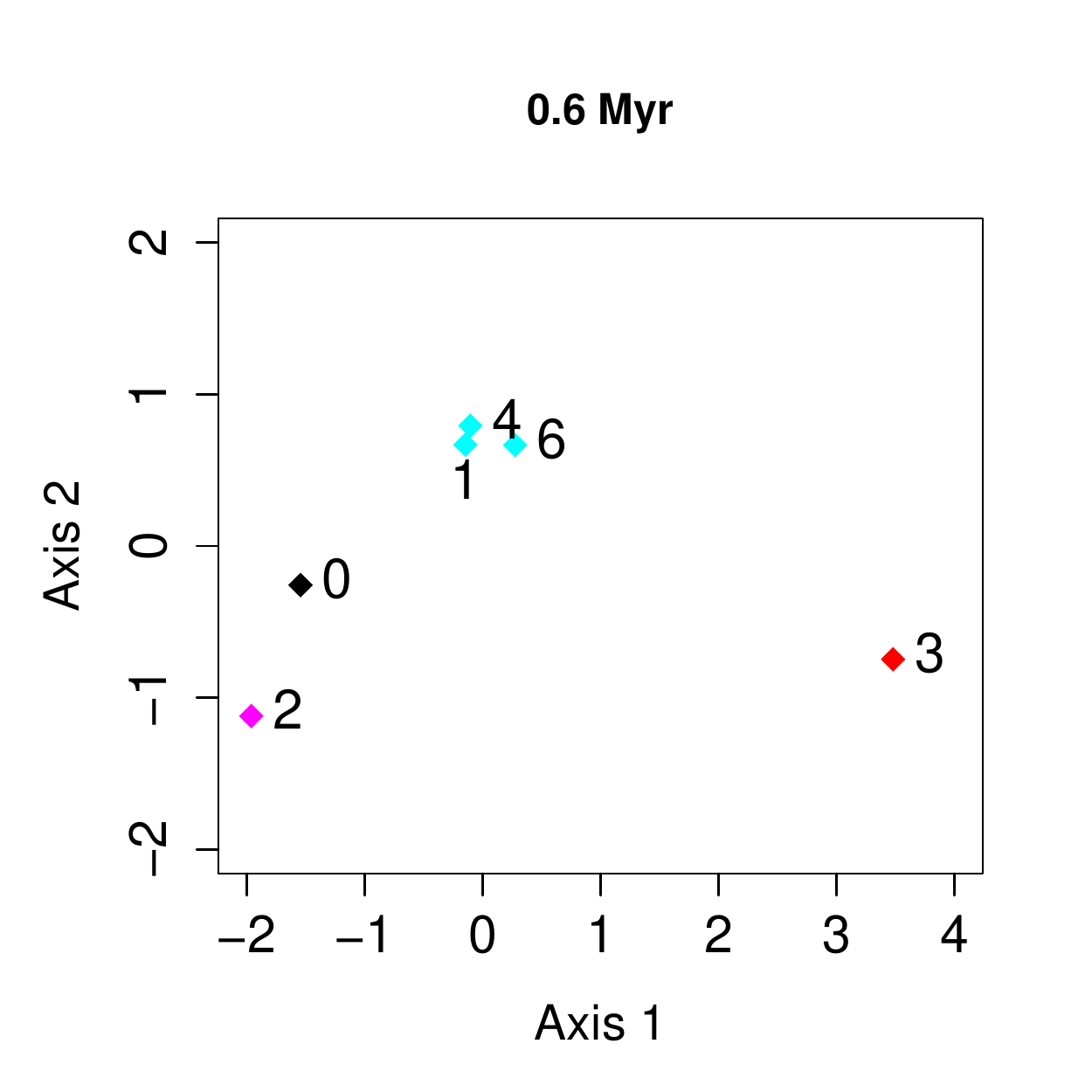}
    \caption[MDS Ordinations for Four Example SOs Showing NP Differences]{Multidimensional Scaling ordination plots for the shape distances of four example SO ages and projections. Labels signify NP, with 0 representing the shape obtained from the random Gaussian distribution. Point colours correspond to which group the SO \hii region shapes were allocated to in the hierarchical clustering of the full dataset.}
    \label{fig:mds_np_comp_app}
\end{figure}

Figure\,\ref{fig:mds_np_comp_app} shows the MDS ordinations for four of the SOs, one at each age. The numbering of the \hii region shape's points on the graphs corresponds to the NP of the SO (from Sec.\,\ref{ssec:mag_np}), with NP = 0 corresponding to the artificial Gaussian distribution (from Sec.\,\ref{ssec:gauss_np}). In each of the MDS plots, only the six \hii region shapes shown were compared pairwise using the A-D test. These ordination results are hence only showing the differences the NPs have on the shapes. In each of the four instances, for increasing SO age, axis 1 of the MDS accounts for 64, 76, 73 and 80\% of the shape variability, respectively. Axis 2 accounts for 23, 18, 16 and 15\%, respectively. Therefore, these two axes are sufficient for investigating the shape differences accordingly. 

In each of the four plots, the Gaussian noise profile 0 shape is ordinated away from the other five NPs. For 0.1\,Myr, all of the shapes are spread over the ordination plot. For 0.2\,Myr, NP 0 is ordinated away from the origin, at approximately an equal distance from each of the other NPs. A similar looking distribution is seen for the 0.4\,Myr data, with a tighter association. For the 0.6\,Myr data, NPs 1, 4 and 6 are ordinated very close to one another, with NPs 0, 2 and 3 ordinated away. In each case, the points are coloured by which of the six groups the shape was sorted into from the dendrogram in Sec.\,\ref{sec:so_discussion}. As expected, those ordinated close together are assigned to the same group in the larger data set. NP 3 in the 0.6\,Myr data is the one example of that age assigned to a group comprising otherwise only early-type regions. As we can see here, it has the furthest distance from the other points. Another interesting note is that, for the 0.1\,Myr shapes, the shape from NP 2 is ordinated close to the origin of the coordinate system. This means that this shape represents the `average' of the sample and the other shapes are all differing with respect to this shape. 

These plots were produced for a number of given projections to determine whether the respective positions in the MDS space for each NPs was systematic and reproducible. Unfortunately, the effect the different NPs have on the underlying shape of the SO \hii region is not a systematic across the ages. We do not see each of the respective NPs by age behaving in the same manner in each of the MDS plots; neither with respect to the other MAGPIS NPs, nor the Gaussian noise only shapes. The example shown for the 0.1\,Myr SO shows a large spread in the MDS ordination, yet this is an example where four of the five NPs are put in the same of the six groups from the hierarchical clustering of the entire data-set. On the other hand, NP 1 of the 0.4\,Myr data is ordinated close to NPs 4 and 3 but is placed into the same group as NPs 2 and 6. We must remember here, however, that the hierarchical groupings are for a much larger sample, such that many of the other projections will be influencing the groupings due to the agglomerative procedure. 

An explanation for this non-systematic effect of the NPs is that the noise from the radio continuum images is not exactly Gaussian, i.e., it has inherent structures within, which was a motivation for using them in the manner we have here. This means that the different sizes of the SOs at each age will be influenced by the inherent noise in the continuum observations by varying amounts. An example can be seen in the results where NPs 3 and 4 are ordinated close to one another in the 0.2 and 0.4\,Myr plots, but not in the other two. Whilst the MDS investigation is useful for visualising the spread in the shape data, we do not gather any summary results from this investigation for the data in its entirety. The MDS approach does, however, show the specific influence of each NP when compared to a reference or standardised version of the shape, which may be useful in future machine learning approaches.

Another application of the MDS ordination we performed was to further investigate how the different selection choices for defining the \hii region shapes affect the resulting spread in the shape data. That is, the initial sigma level used when extracting the boundaries and the spline knot spacings for controlling the spatial resolution of the shape landmarks. So far, for all of the SO shape data, we have only considered the 1 sigma contour level, along with the 0.54\,pc spline knot spacing. This was for the purpose of directly comparing the SOs to the previous results from the MAGPIS data. However, we suggested in \paperIt~ that the SOs could provide a better test set for determining how these two selection variables influence the resulting shape. When rerunning the MDS of the shape data using contours with 0.8 and 1.2 sigma above the mean value, the positions of the ordinations changes, but the overall spread in the data remains. This suggests that, as with the NPs, varying the initial sigma level is not having a systematic effect on the shape data. When rerunning the MDS with different spline resolutions, decreasing the spline interval (hence increasing the spatial resolution) results in a larger spread along the MDS axes. This was expected as more features and points for comparison are captured with a higher resolution. Furthermore, we have already shown that the amount of high curvature points corresponds to the score along axis 1 of the MDS. Increasing the interval (decreasing the resolution) results in a smaller spread in the MDS ordination. Whilst this may seem like a favourable result, the level of smoothing along the boundaries is significantly increased, resulting in fewer features along the curves. Increasing this smoothing amount by too much thus becomes redundant for the smaller diameter regions. As found with the previous tests of this nature in \paperIt, the 1 sigma, 0.54\,pc interval seems to be a good intermediary between extremes. The most important aspect here (and for future work of this nature) is that the shape extraction and quantification remains consistent for the sample. 

%id proj 0.1600, np 1 in group 1, np 2, 4, 6, 7 in group 3, spread in mds scores

%id proj 0.200, np 1 and 2 in group 2, np 4 and 7 group 1, np np 6 group 3, mds ordination confirms this, all far from gauss shape

%id proj 0.43030, np 1, 2 and 6 in group 4, np 4 and 7 in group 5, again confirmed by MDS, all far away from gauss, lower scores than previous example

%id proj 0.660120, np 1, 4 and 6 in group 6, np 2 in group 4, np 7 in group 1, only one of the 0.6myr data to not be in group 4 5 or 6 and is ordinated far away on MDS plot

\section{Mass-Limited MAGPIS Sample Images}
\label{ap:mag_images}

The following figures show the images of MAGPIS \hii Region assigned to each of the training groups in Fig.\,\ref{fig:training_dend}. Groups are numbered top down from the dendrogoram. Image tiles are in angular degrees scale with Galactic longitude and latitudes shown. The contoured outlines are those obtained from the shape extraction process, with the corresponding spline interpolation points used to obtain the curvature distributions indicated by the open squares. 

\begin{figure}
		\centering
		\includegraphics[width=0.2\textwidth]{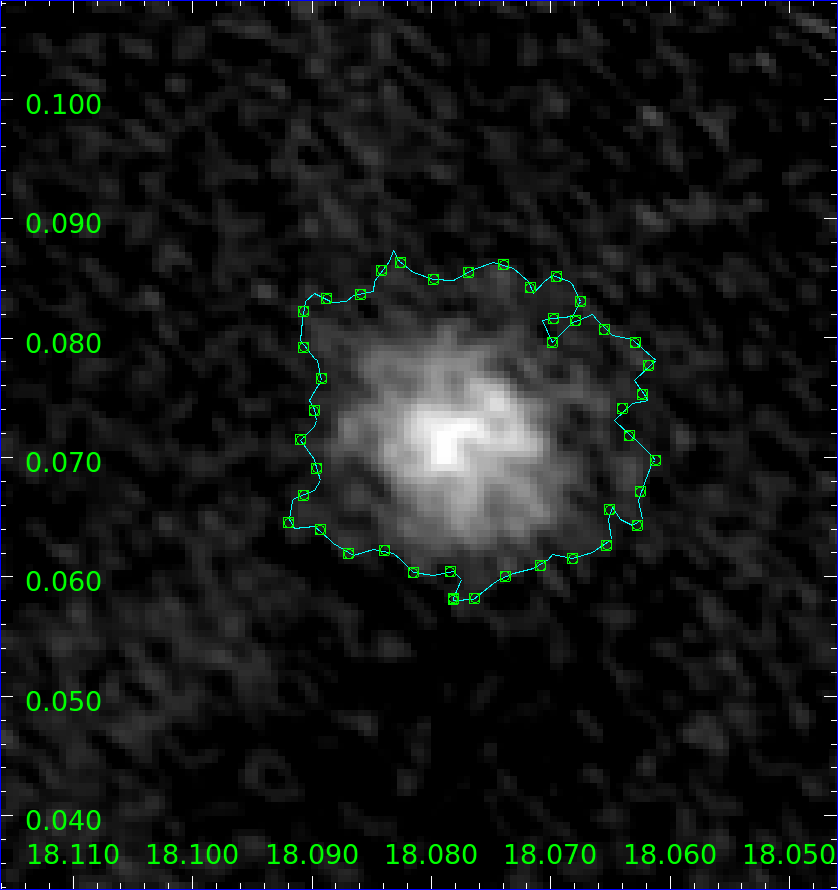} 
		\includegraphics[width=0.2\textwidth]{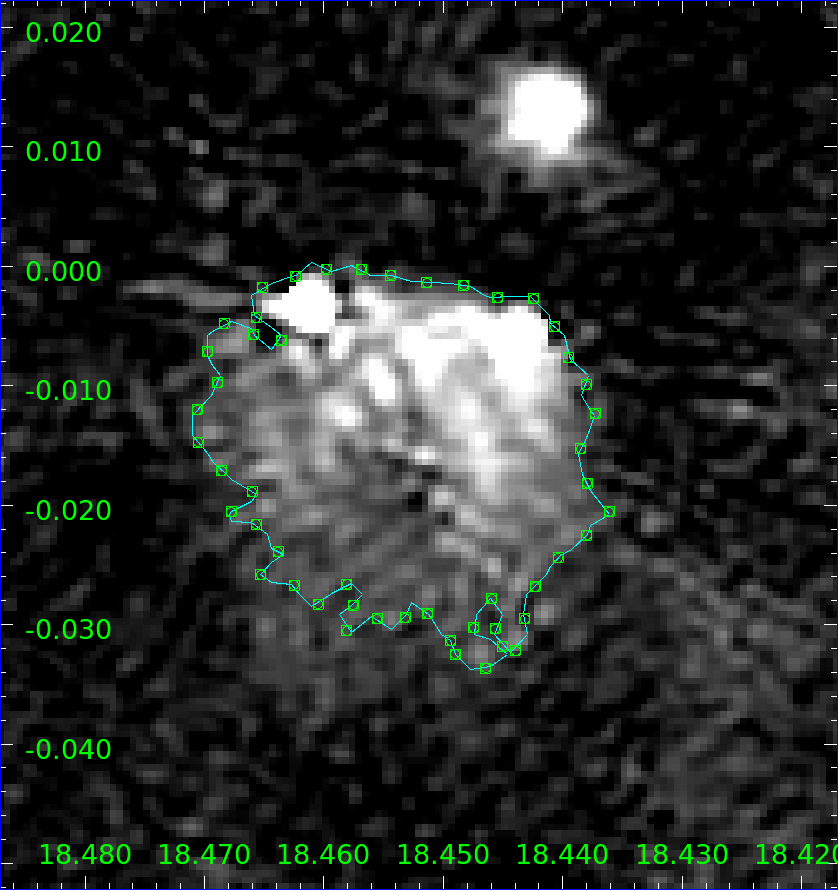}
		\includegraphics[width=0.2\textwidth]{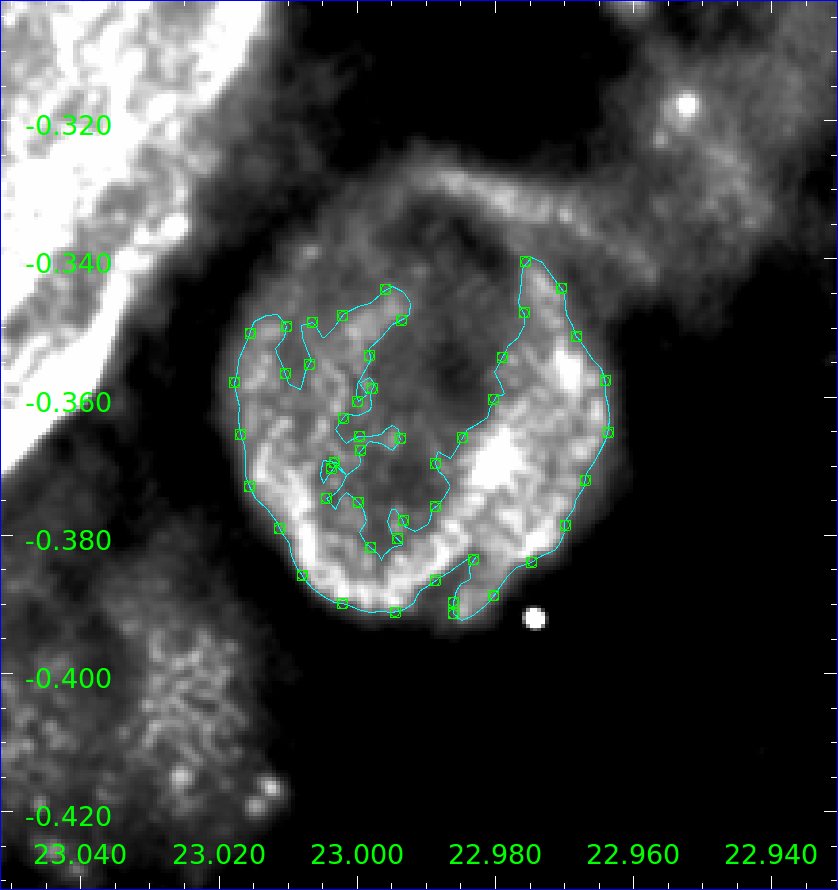}
		\includegraphics[width=0.2\textwidth]{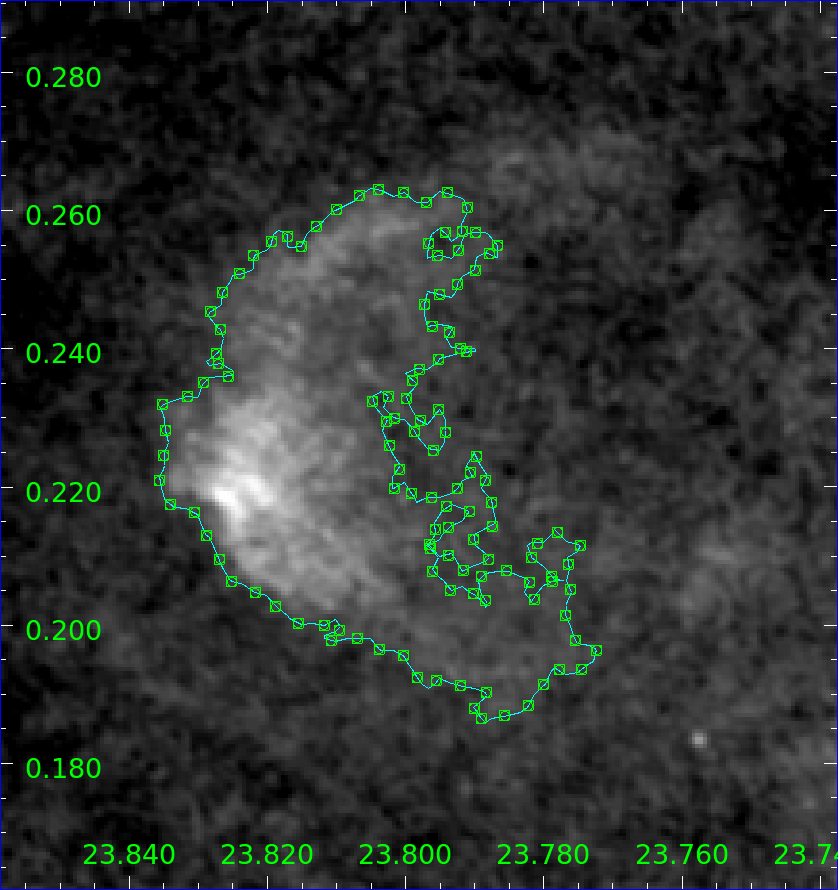}
		
		\includegraphics[width=0.2\textwidth]{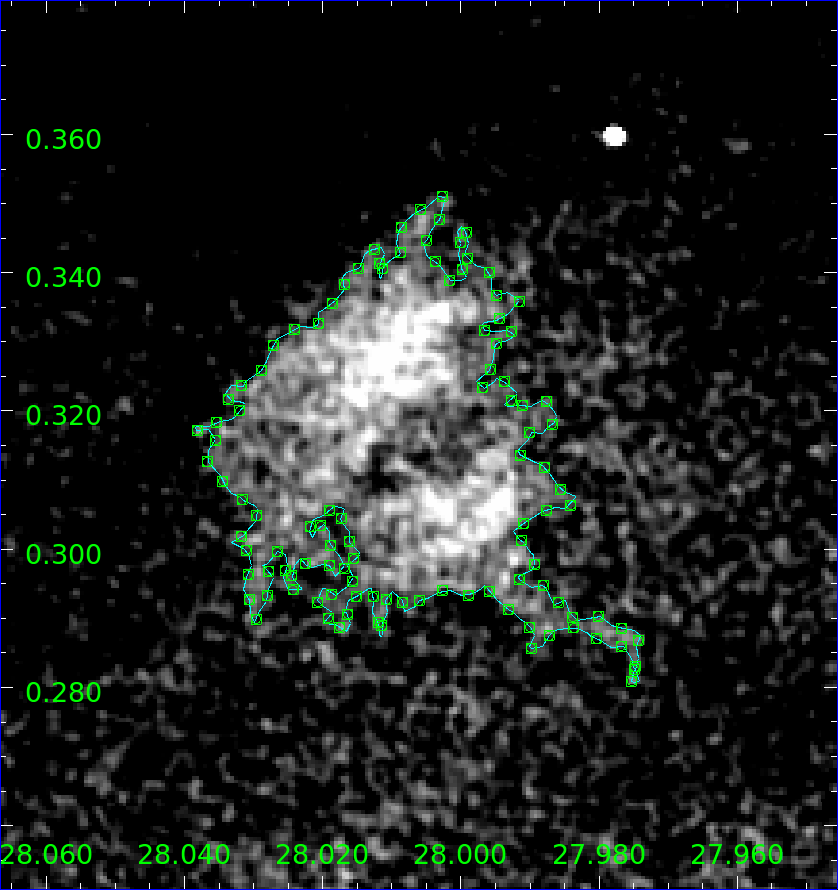}
		\includegraphics[width=0.2\textwidth]{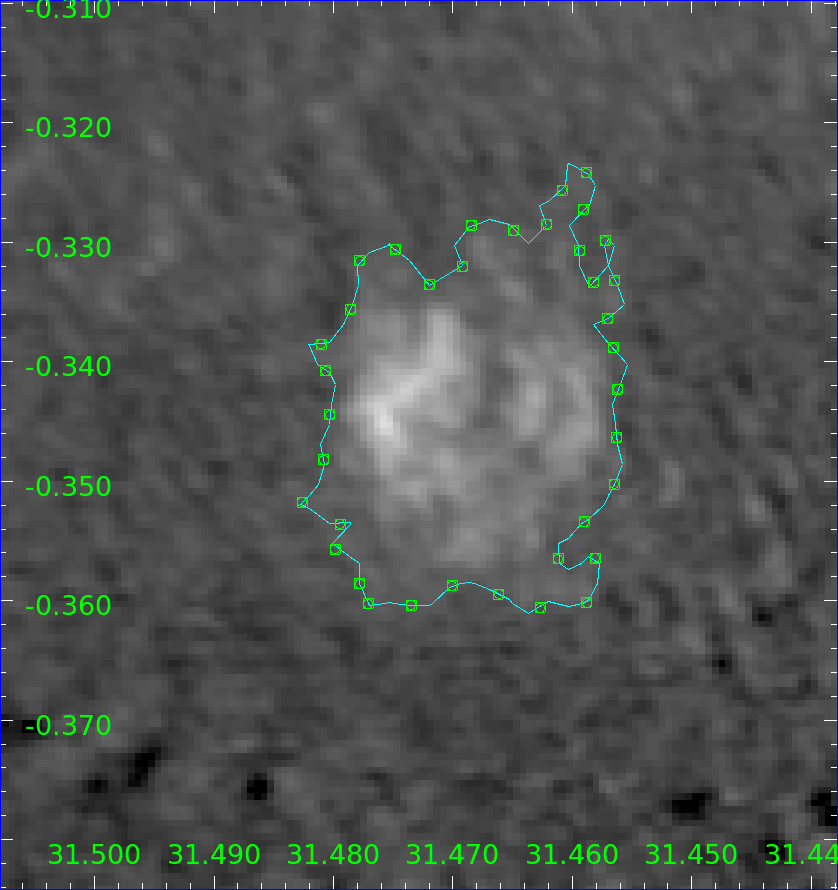}
		\includegraphics[width=0.2\textwidth]{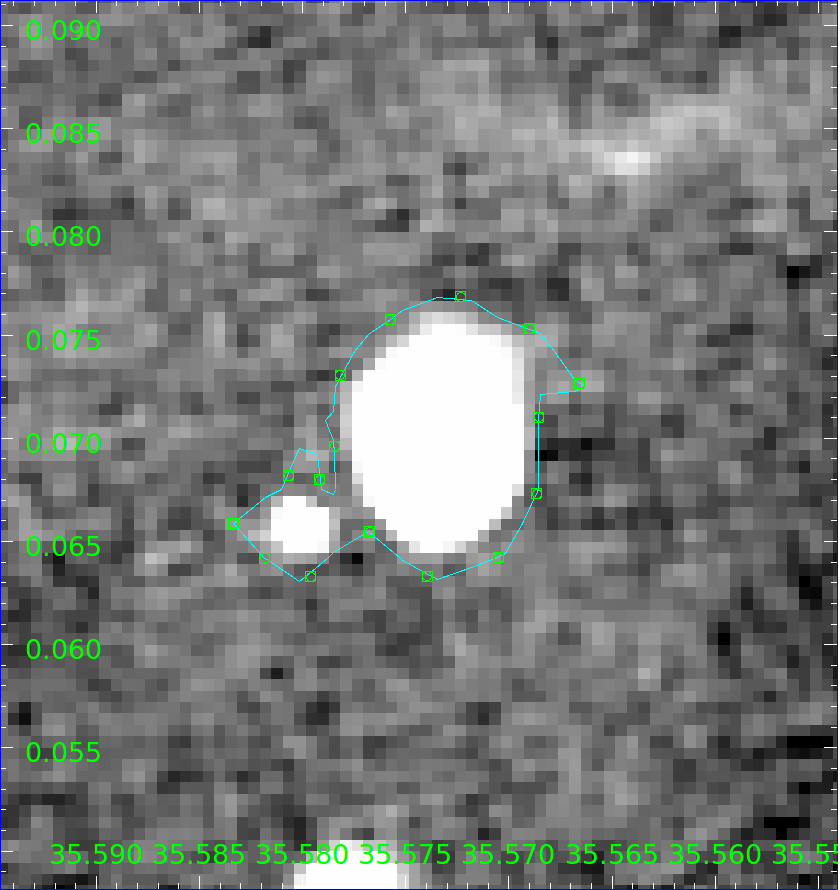}
		\includegraphics[width=0.2\textwidth]{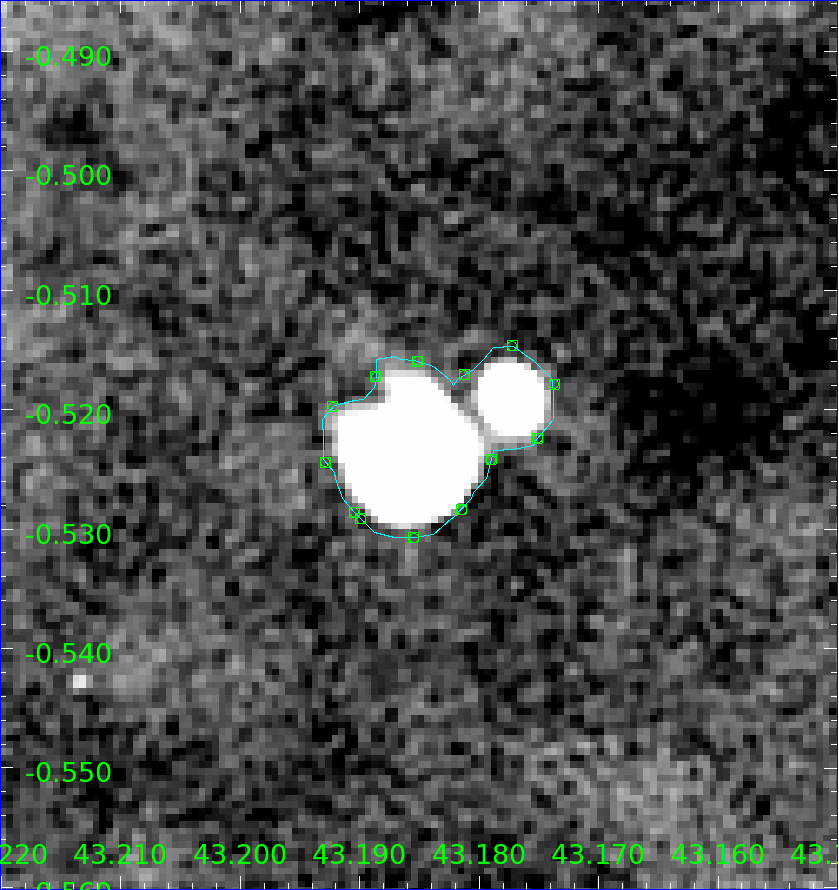}
		%\hspace{5cm}
		\caption{Group 3, note that region G045.204+00.744 is not shown} \label{fig:mag_train2}
\end{figure}  

\begin{figure}
		\centering
		\includegraphics[width=0.2\textwidth]{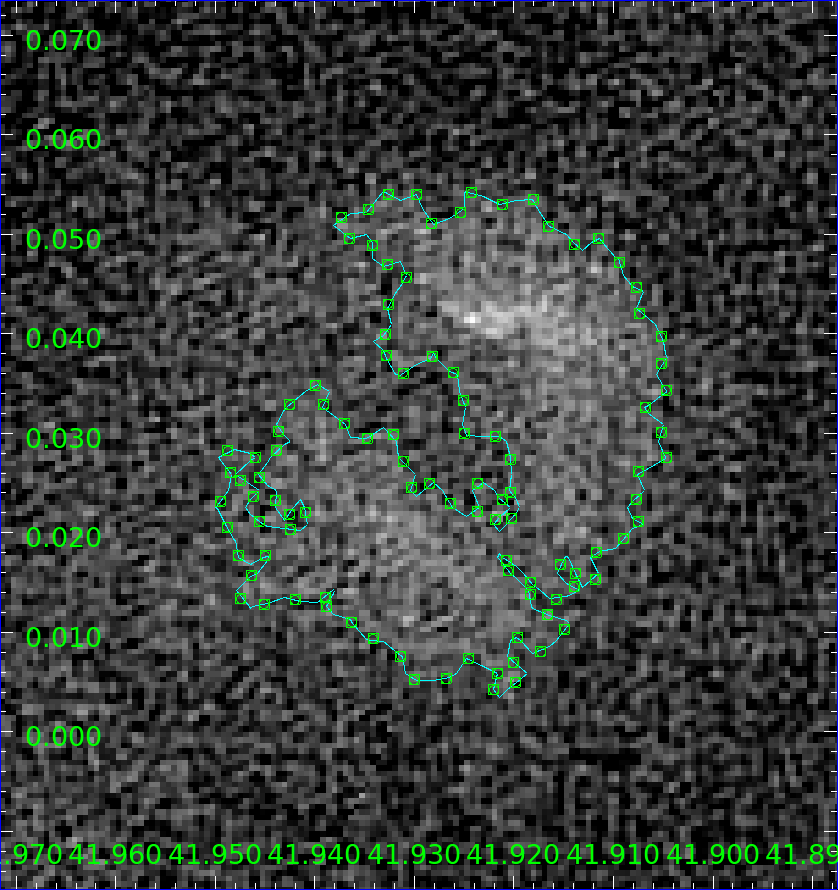} 
		\includegraphics[width=0.2\textwidth]{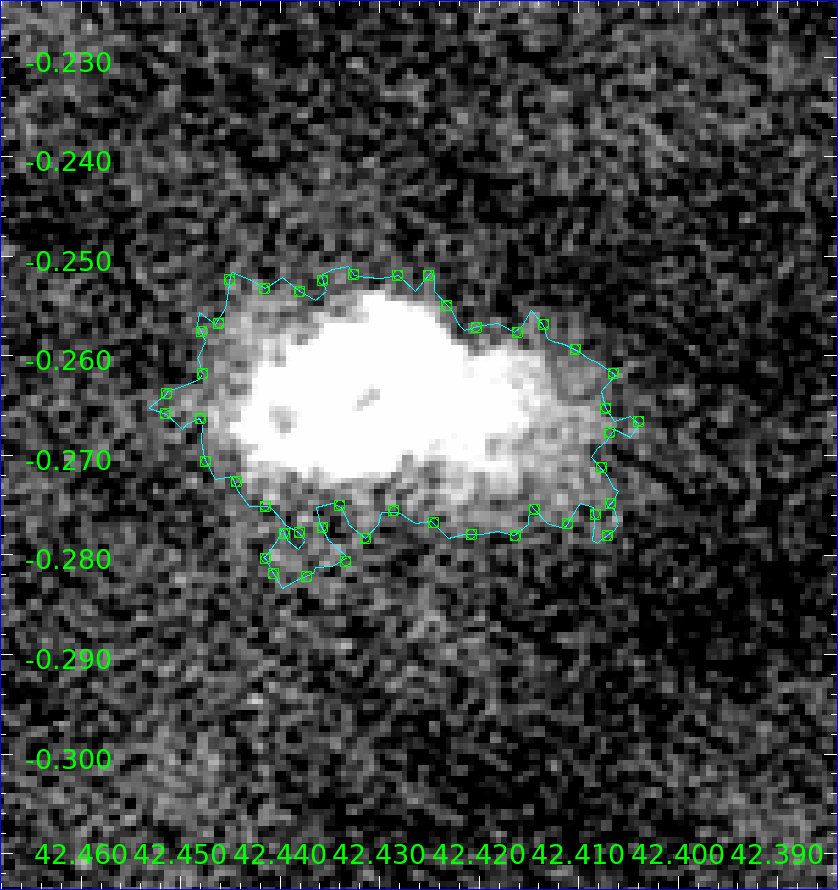}
		\includegraphics[width=0.2\textwidth]{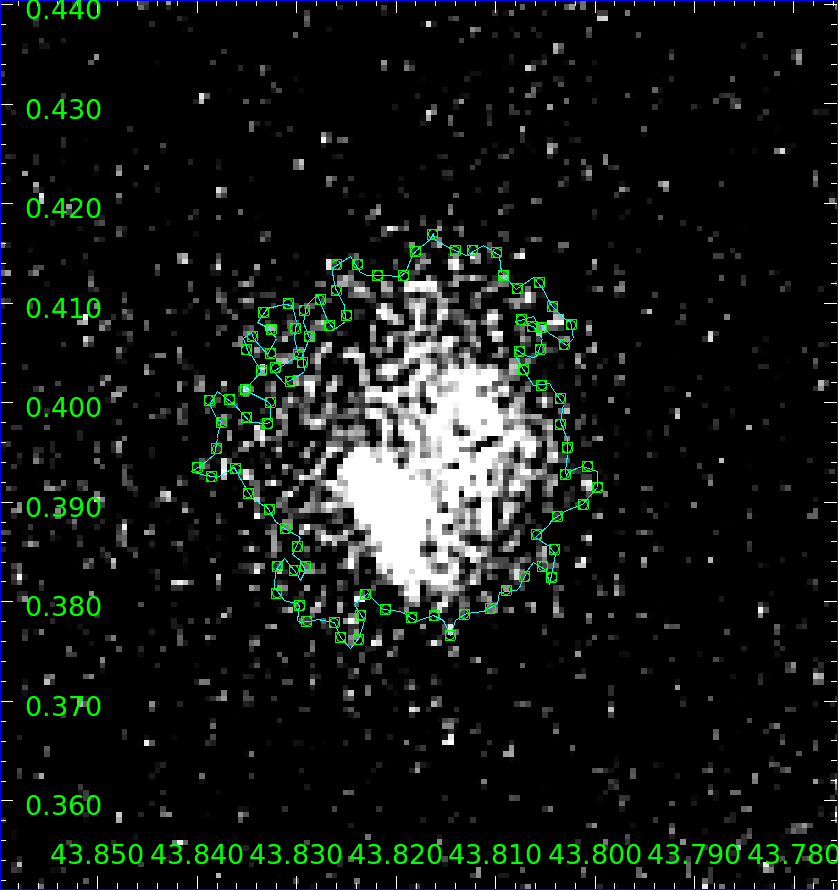}  
		\includegraphics[width=0.2\textwidth]{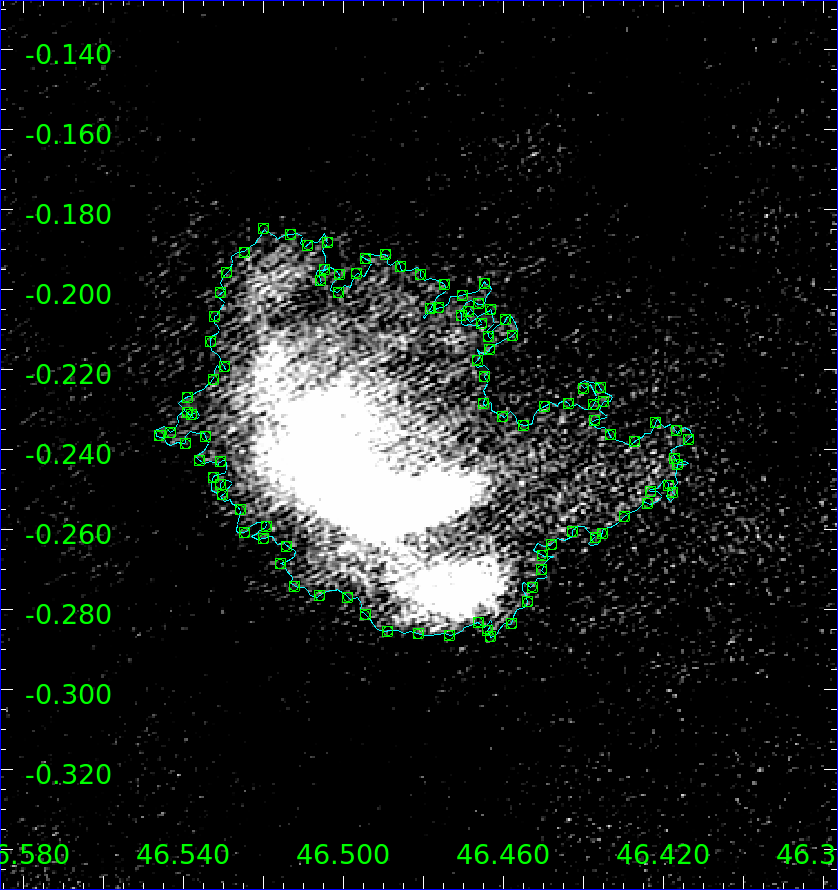}                    
		%\hspace{5cm}
		\caption{Group 1} \label{fig:mag_train6}
\end{figure} 

\begin{figure}
		\centering
		\includegraphics[width=0.2\textwidth]{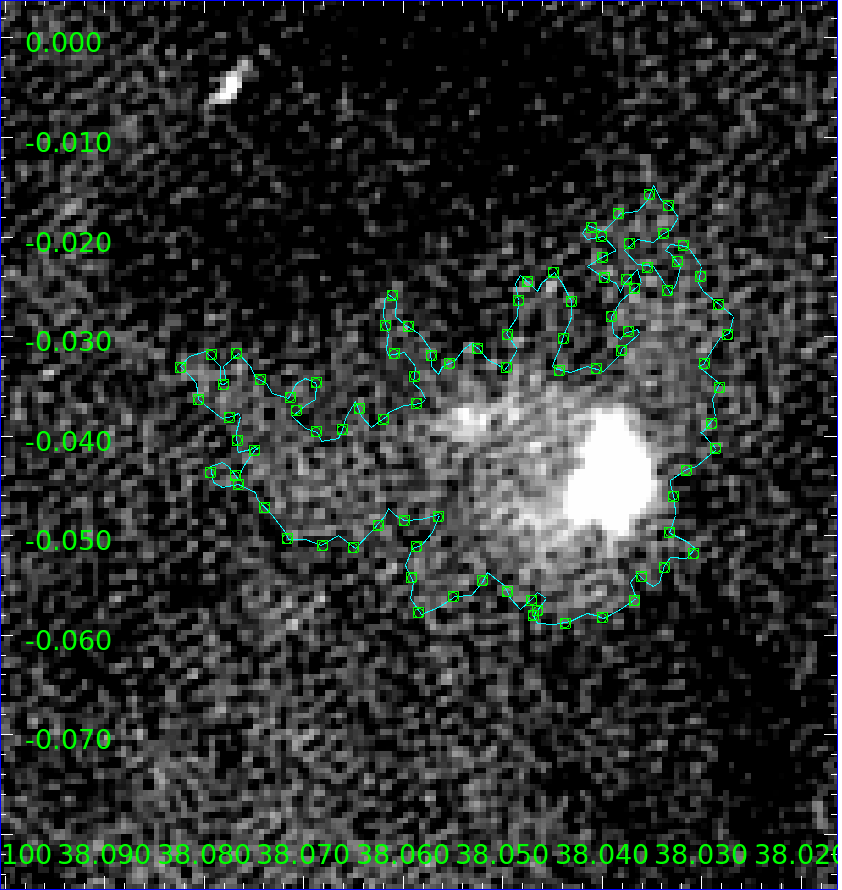} 
		\includegraphics[width=0.2\textwidth]{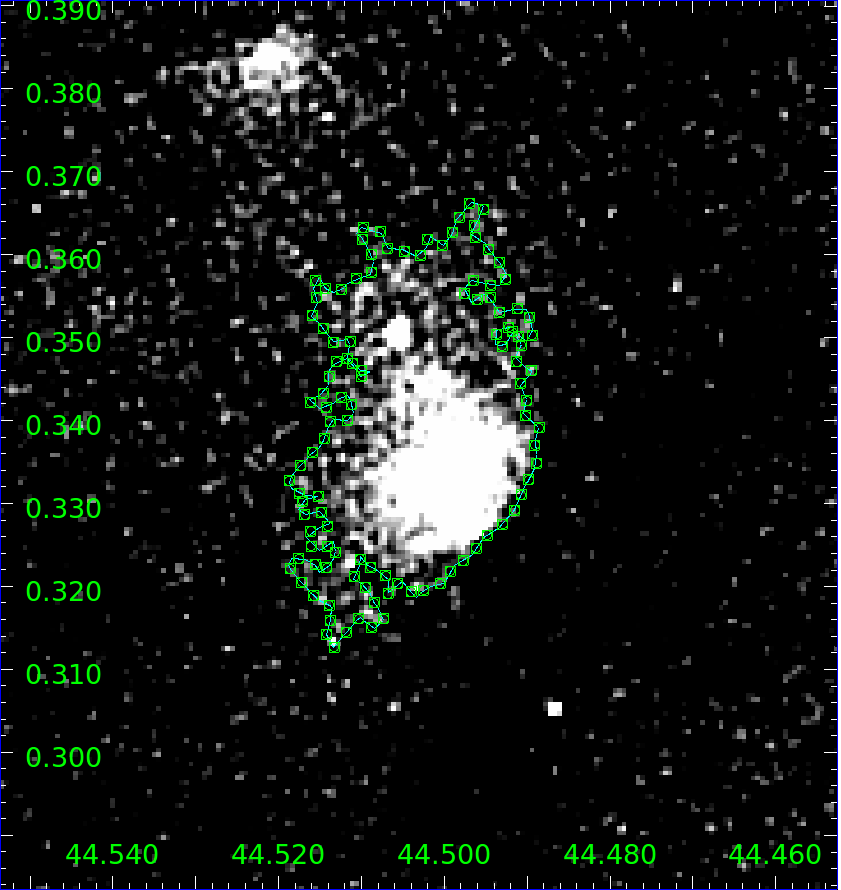}             
		%\hspace{5cm}
		\caption{Group 2} \label{fig:mag_train5}
\end{figure}  

\begin{figure}
		\centering
		\includegraphics[width=0.2\textwidth]{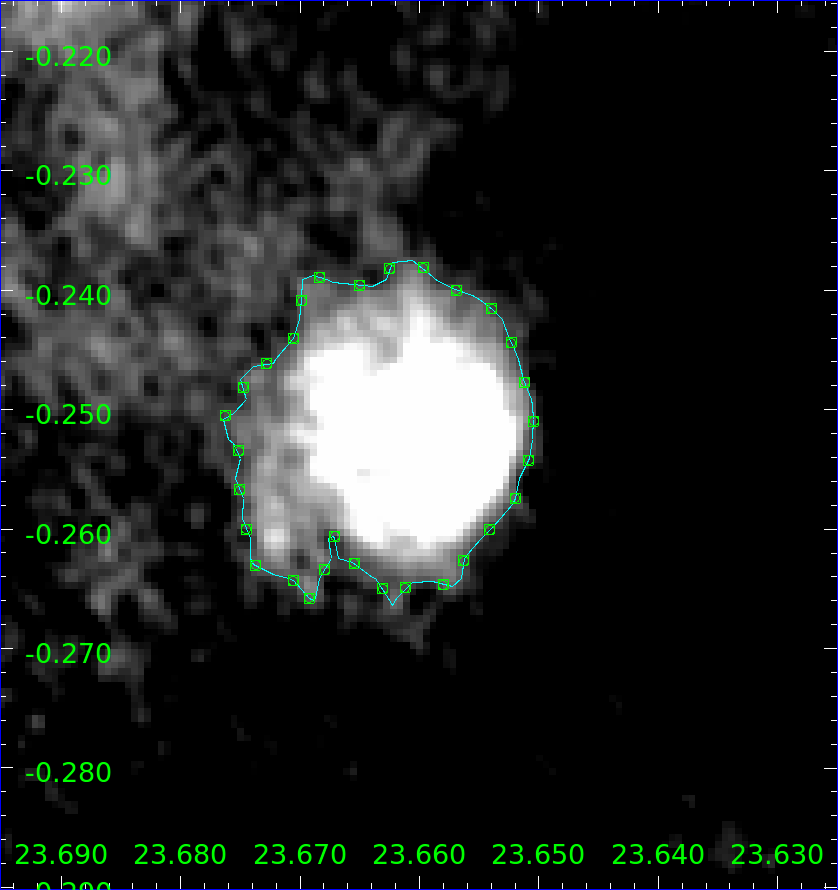} 
		\includegraphics[width=0.2\textwidth]{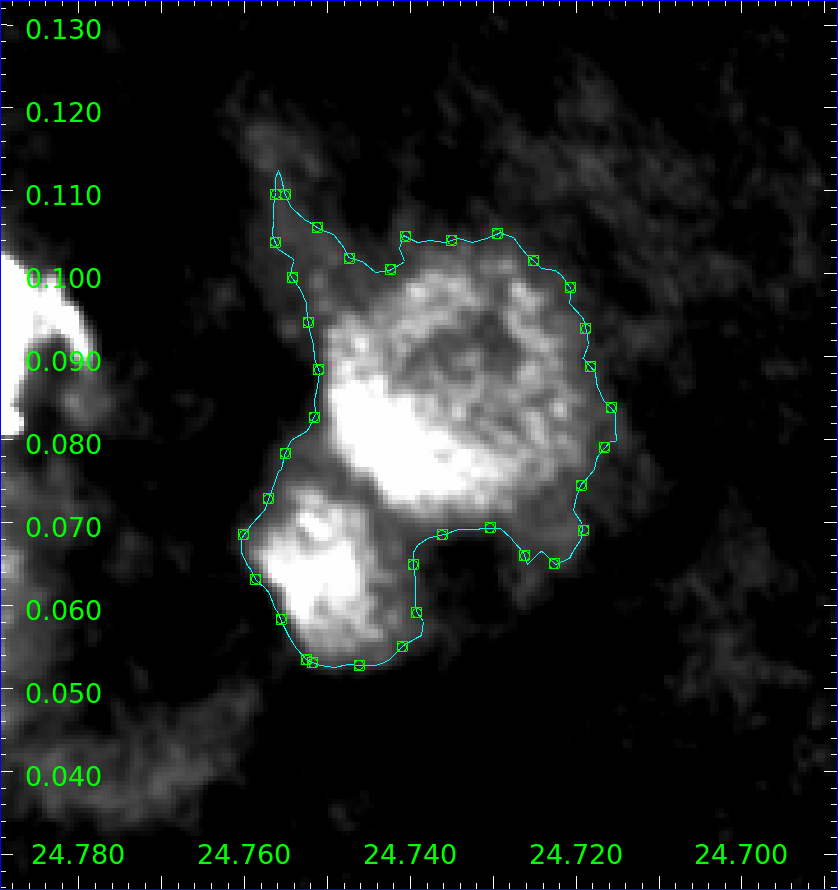}
		\includegraphics[width=0.2\textwidth]{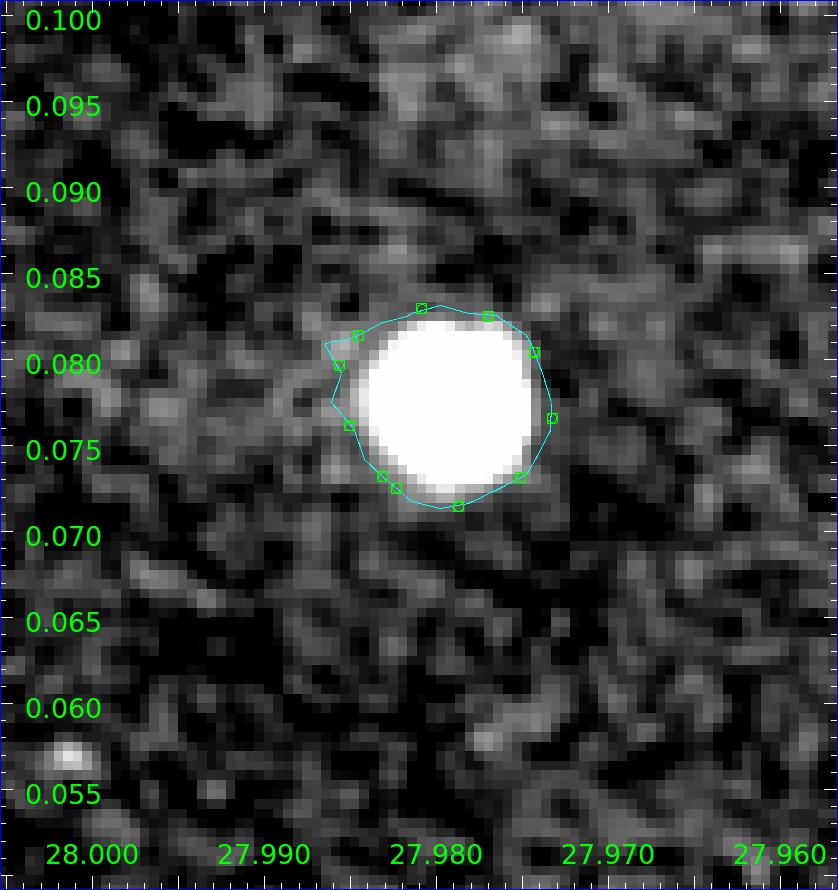}              
		%\hspace{5cm}
		\caption{Group 6} \label{fig:mag_train4}
\end{figure}

\begin{figure}
		\centering
		\includegraphics[width=0.2\textwidth]{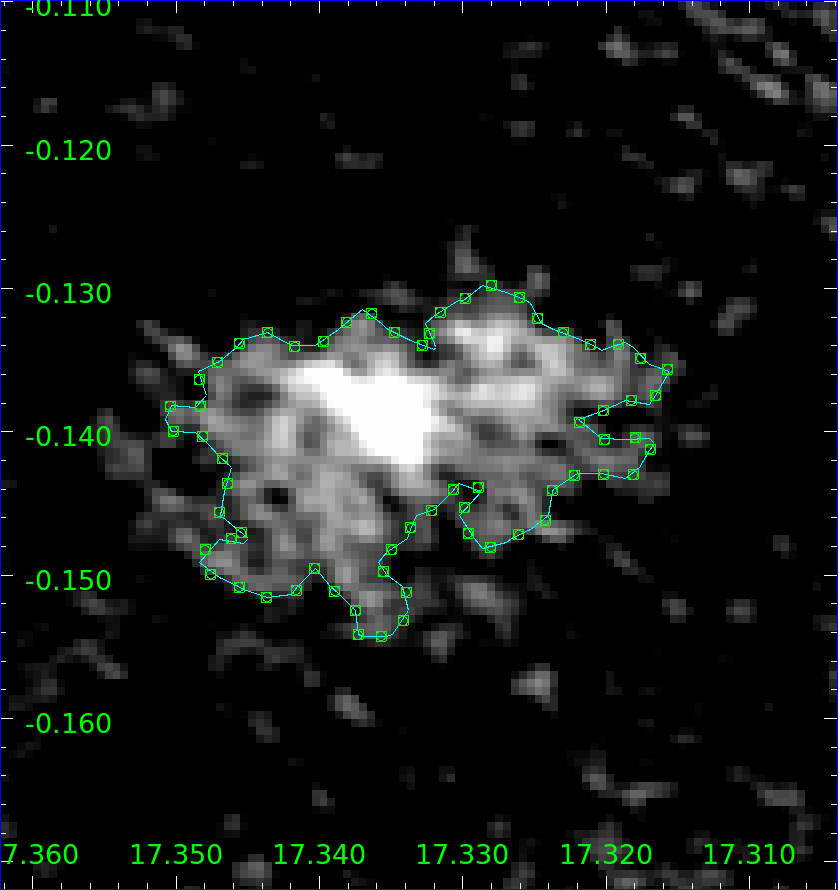} 
		\includegraphics[width=0.2\textwidth]{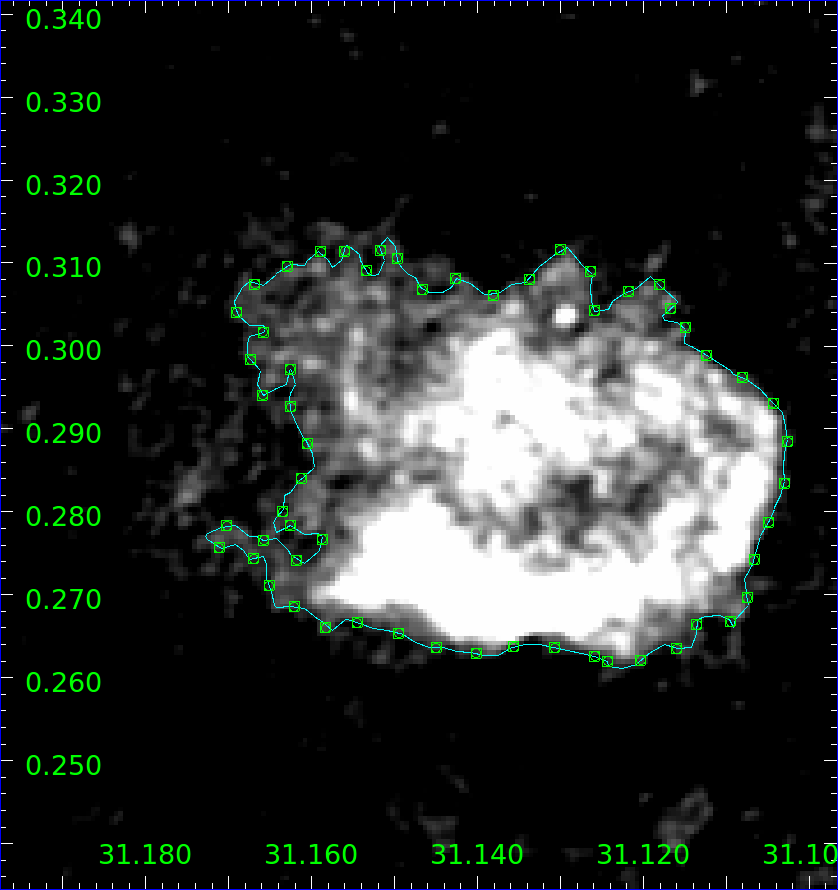}
		\includegraphics[width=0.2\textwidth]{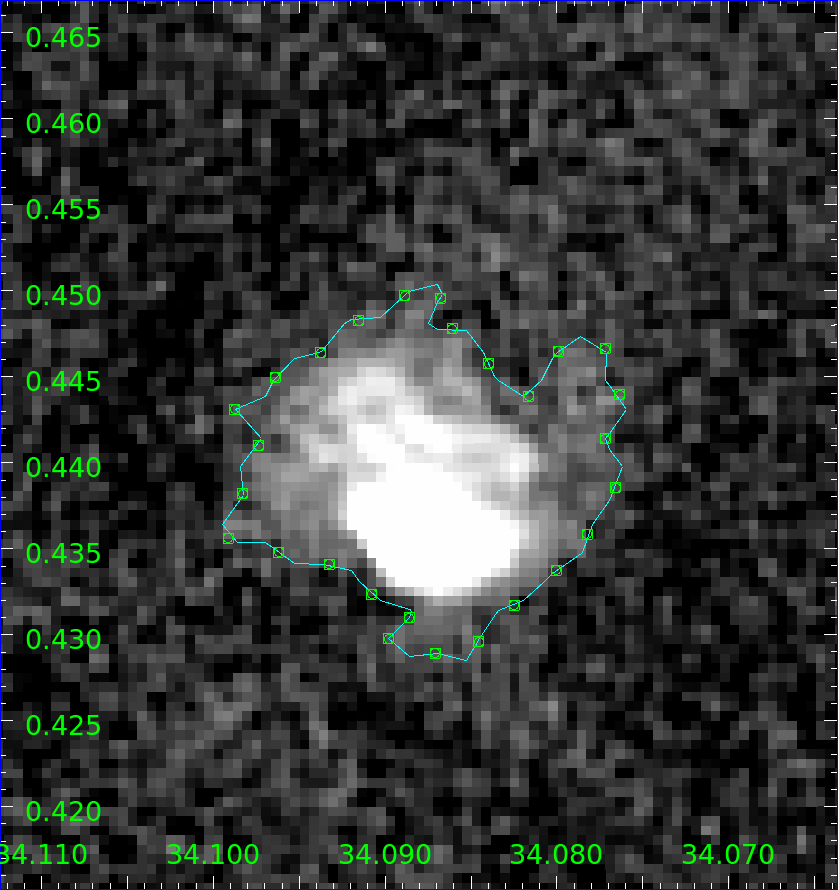}          
		\includegraphics[width=0.2\textwidth]{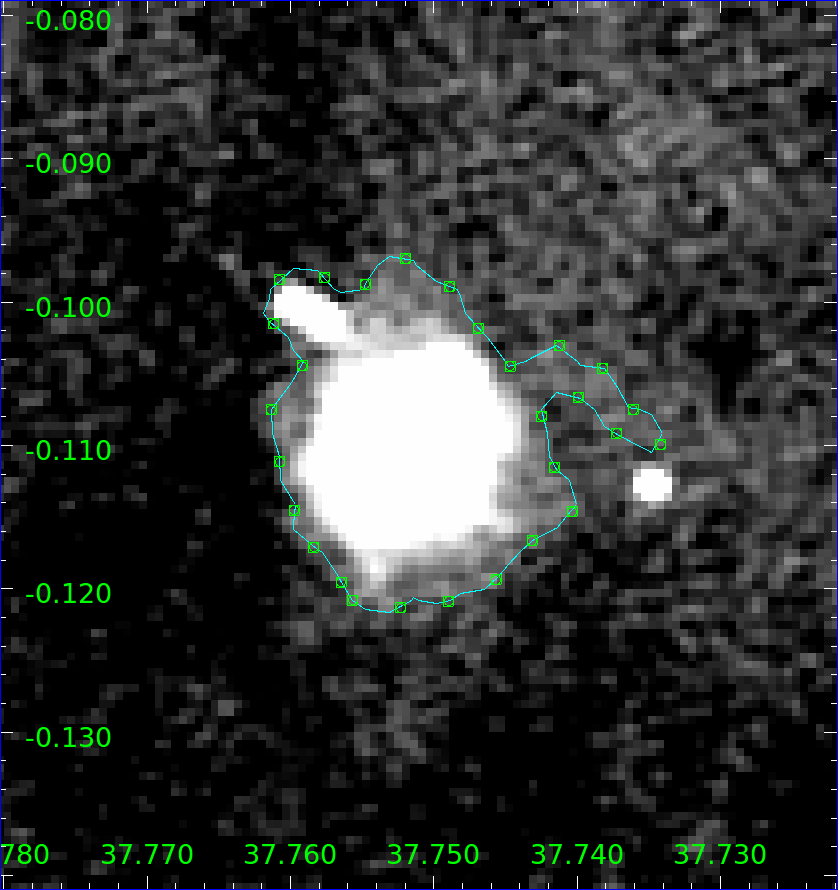}     
		%\hspace{5cm}
		\caption{Group 4} \label{fig:mag_train1}
\end{figure}

\begin{figure}
		\centering
		\includegraphics[width=0.2\textwidth]{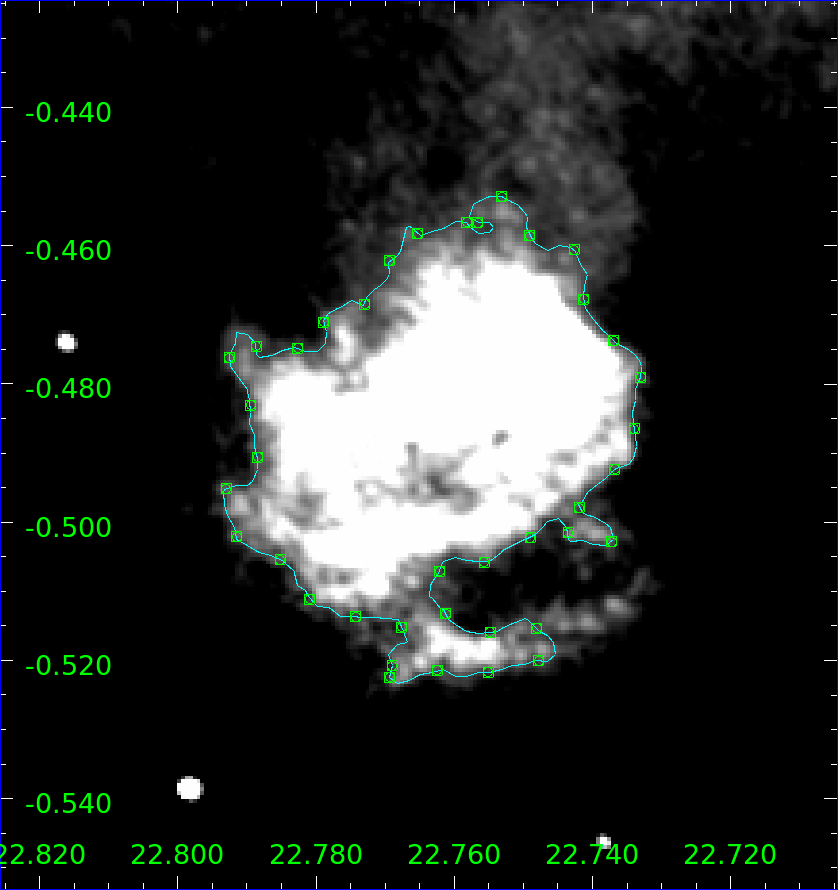} 
		\includegraphics[width=0.2\textwidth]{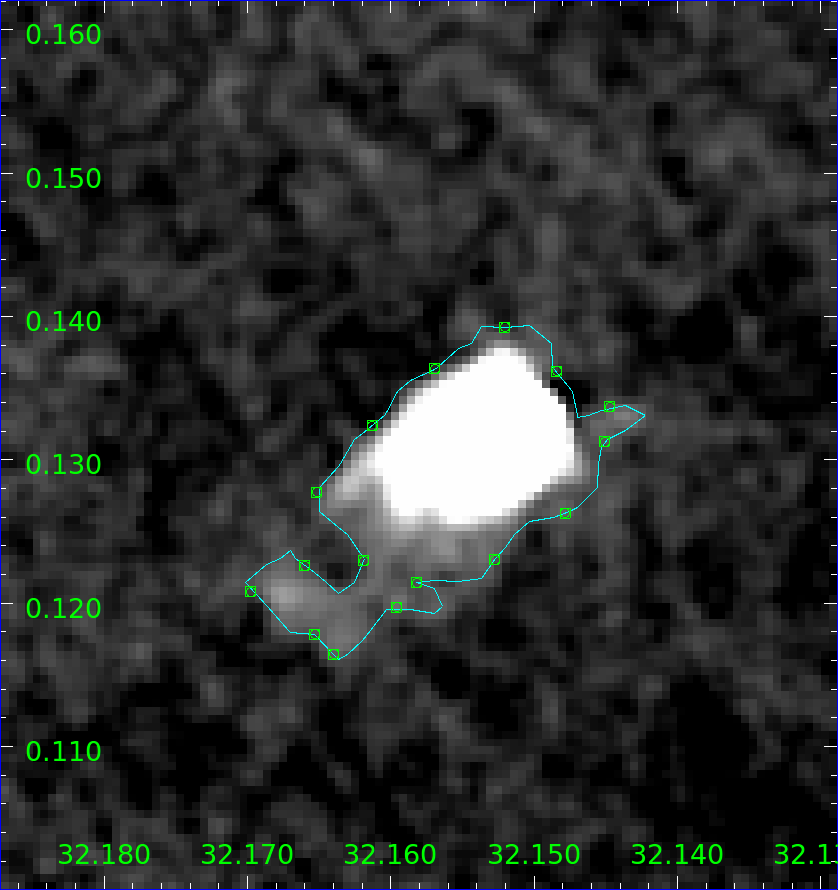}
		\includegraphics[width=0.2\textwidth]{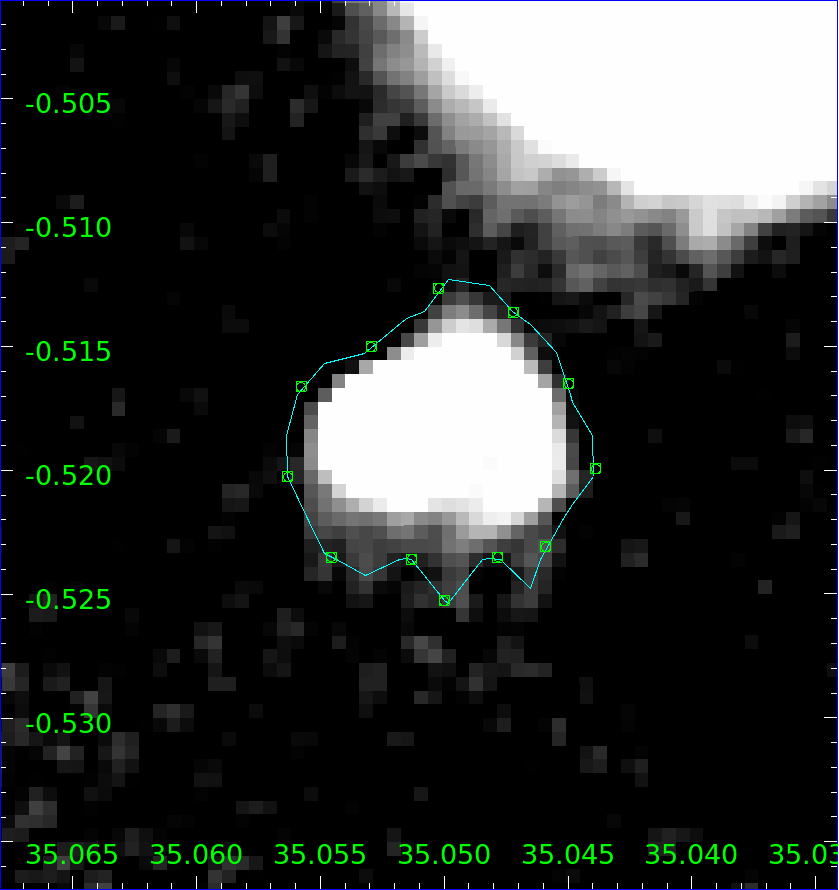}          
		\includegraphics[width=0.2\textwidth]{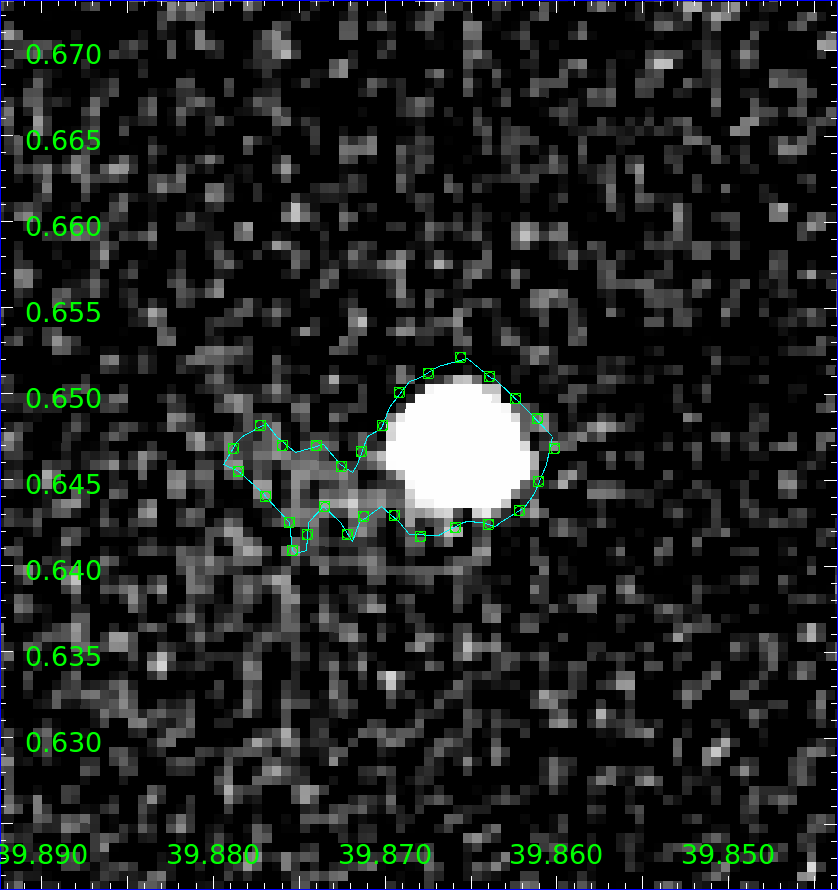}  
		%\hspace{5cm}
		\caption{Group 5} \label{fig:mag_train3}
\end{figure}

\clearpage
\newpage
\onecolumn
\section{Extracted Shapes from Synthetic Observations with Artificial Guassian Noise}
\label{app:so_images}

Synthetic Observations (SOs) of the numerical simulation in \citep{2018MNRAS.477.5422A}. The 77 images are shown with coordinates in parsecs. Headings identify the snapshot age and projection viewing angle (t = $\theta$, p = $\phi$). In each image, random Gaussian noise has been added to each pixel value, following the distribution from an example MAGPIS 1.4\,GHz image tile. The boundary shown is that of the 1$\sigma$ above the mean noise level contour. 

%\clearpage

%\begin{landscape}

\begin{center}

\begin{longtable}{cccc} 

\parbox[h]{0.2\textwidth}{\includegraphics[width=0.2\textwidth]{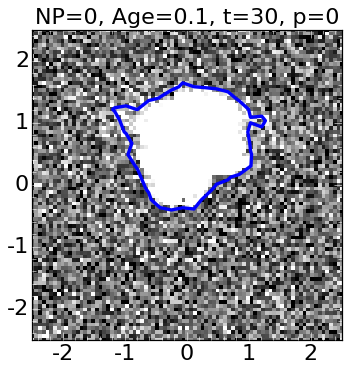}} &
\parbox[h]{0.2\textwidth}{\includegraphics[width=0.2\textwidth]{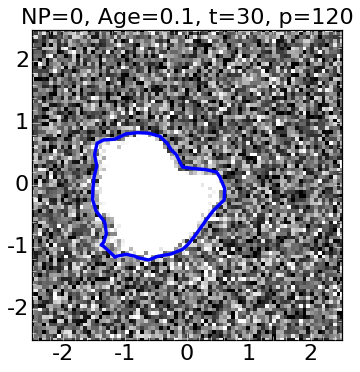}} &
\parbox[h]{0.2\textwidth}{\includegraphics[width=0.2\textwidth]{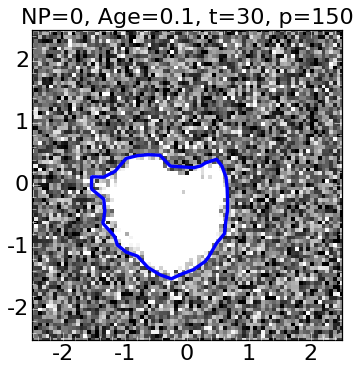}} &
\parbox[h]{0.2\textwidth}{\includegraphics[width=0.2\textwidth]{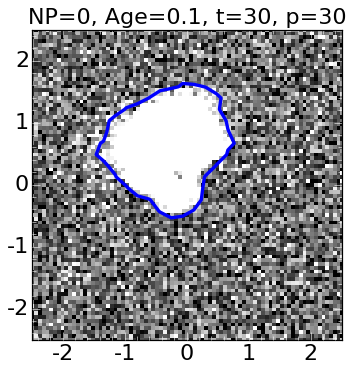}} \\
\parbox[h]{0.2\textwidth}{\includegraphics[width=0.2\textwidth]{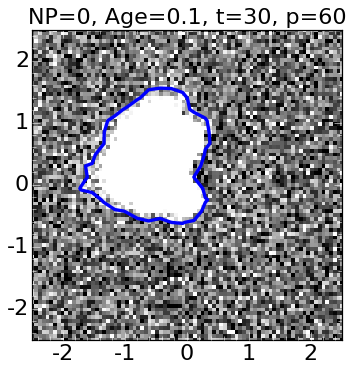}} &
\parbox[h]{0.2\textwidth}{\includegraphics[width=0.2\textwidth]{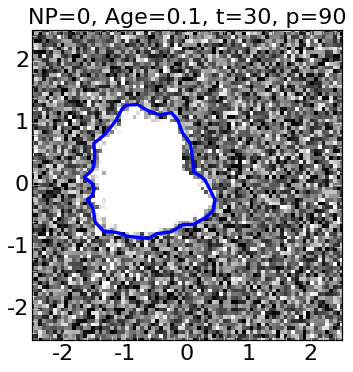}} &
\parbox[h]{0.2\textwidth}{\includegraphics[width=0.2\textwidth]{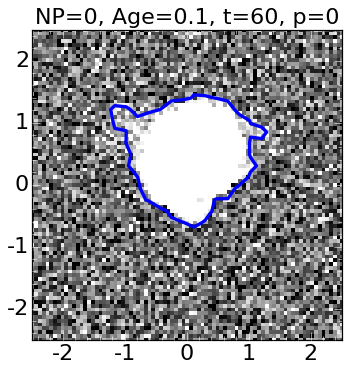}} &
\parbox[h]{0.2\textwidth}{\includegraphics[width=0.2\textwidth]{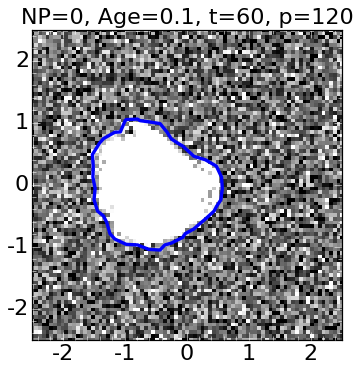}} \\
\parbox[h]{0.2\textwidth}{\includegraphics[width=0.2\textwidth]{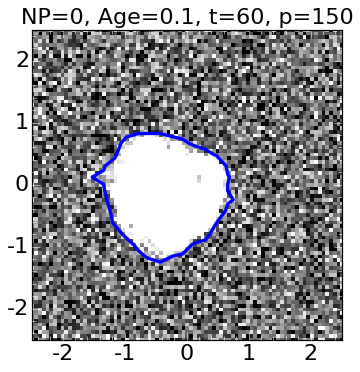}} &
\parbox[h]{0.2\textwidth}{\includegraphics[width=0.2\textwidth]{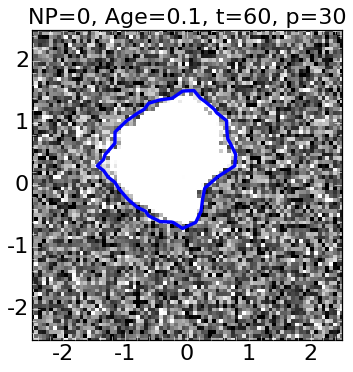}} &
\parbox[h]{0.2\textwidth}{\includegraphics[width=0.2\textwidth]{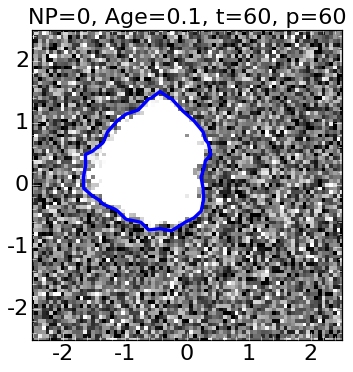}} &
\parbox[h]{0.2\textwidth}{\includegraphics[width=0.2\textwidth]{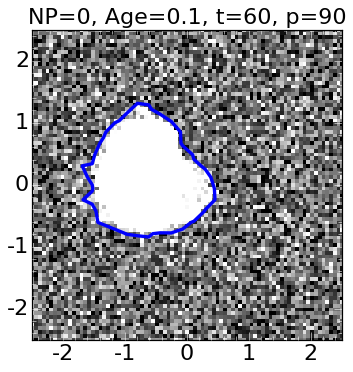}} \\
\parbox[h]{0.2\textwidth}{\includegraphics[width=0.2\textwidth]{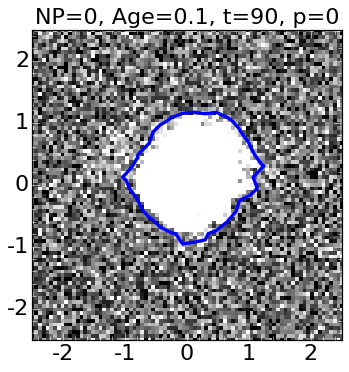}} &
\parbox[h]{0.2\textwidth}{\includegraphics[width=0.2\textwidth]{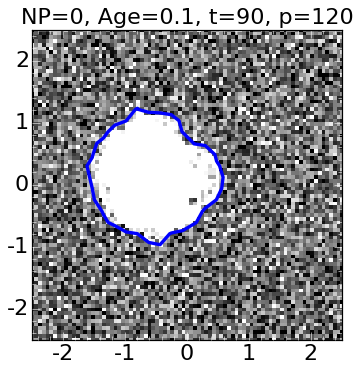}} &
\parbox[h]{0.2\textwidth}{\includegraphics[width=0.2\textwidth]{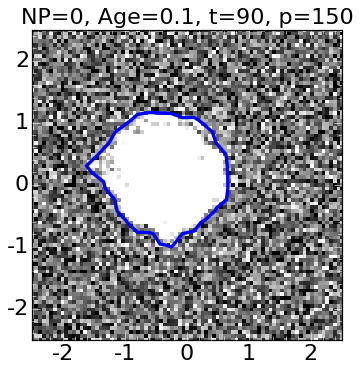}} &
\parbox[h]{0.2\textwidth}{\includegraphics[width=0.2\textwidth]{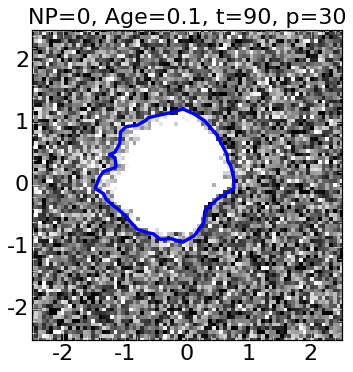}} \\
\parbox[h]{0.2\textwidth}{\includegraphics[width=0.2\textwidth]{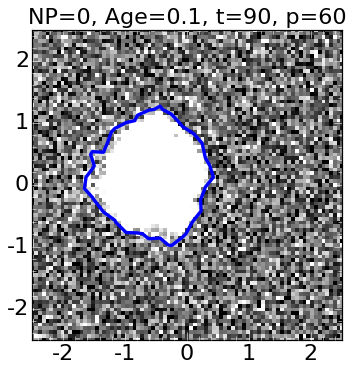}} &
\parbox[h]{0.2\textwidth}{\includegraphics[width=0.2\textwidth]{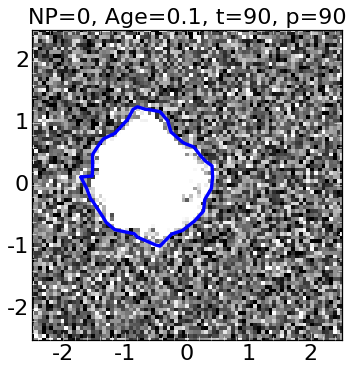}} &
\parbox[h]{0.2\textwidth}{\includegraphics[width=0.2\textwidth]{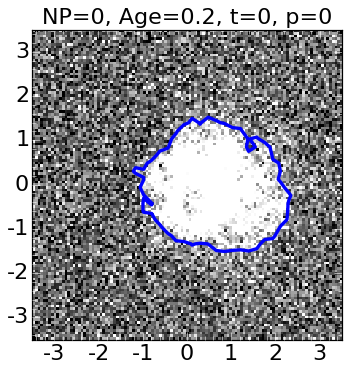}} &
\parbox[h]{0.2\textwidth}{\includegraphics[width=0.2\textwidth]{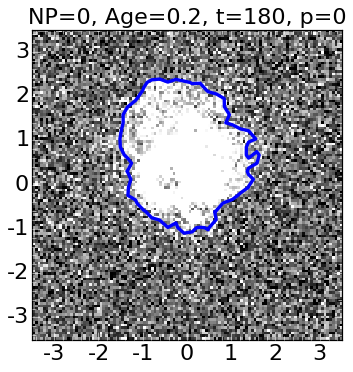}} \\
\parbox[h]{0.2\textwidth}{\includegraphics[width=0.2\textwidth]{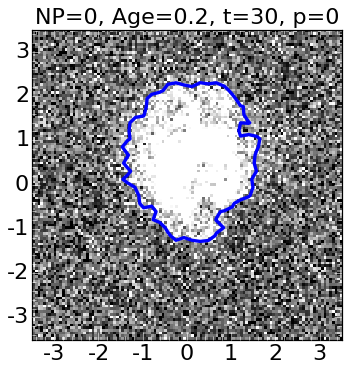}} &
\parbox[h]{0.2\textwidth}{\includegraphics[width=0.2\textwidth]{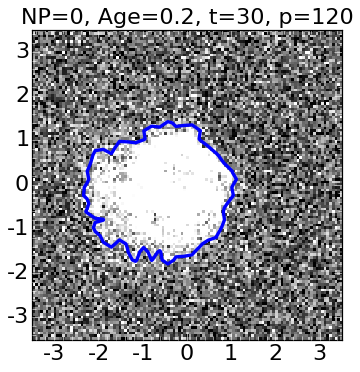}} &
\parbox[h]{0.2\textwidth}{\includegraphics[width=0.2\textwidth]{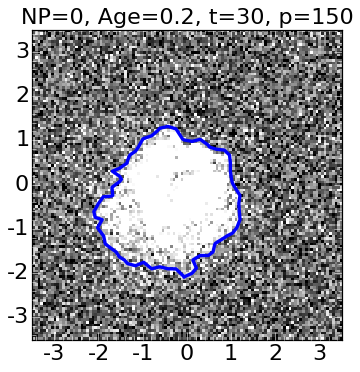}} &
\parbox[h]{0.2\textwidth}{\includegraphics[width=0.2\textwidth]{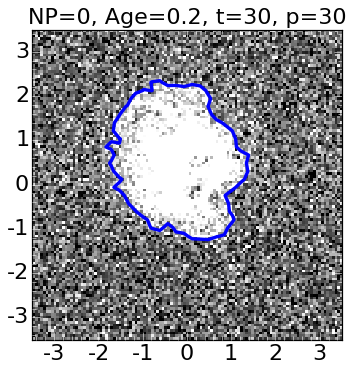}} \\
\parbox[h]{0.2\textwidth}{\includegraphics[width=0.2\textwidth]{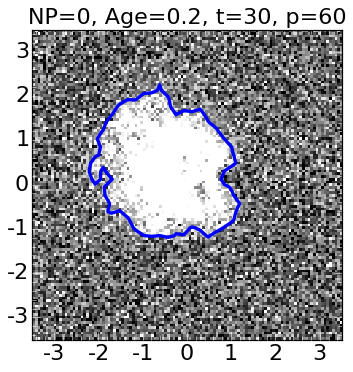}} &
\parbox[h]{0.2\textwidth}{\includegraphics[width=0.2\textwidth]{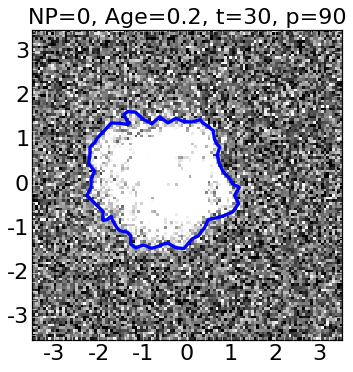}} &
\parbox[h]{0.2\textwidth}{\includegraphics[width=0.2\textwidth]{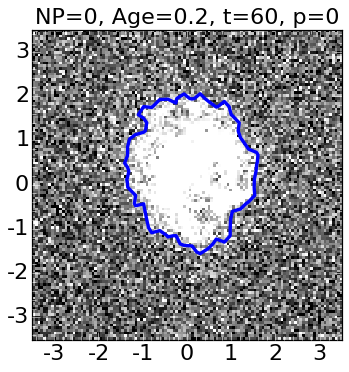}} &
\parbox[h]{0.2\textwidth}{\includegraphics[width=0.2\textwidth]{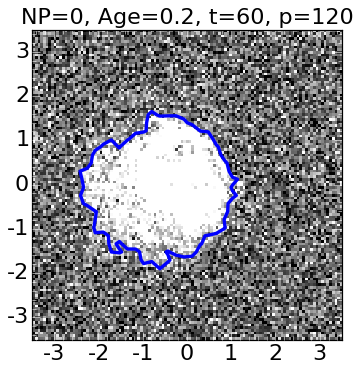}} \\
\parbox[h]{0.2\textwidth}{\includegraphics[width=0.2\textwidth]{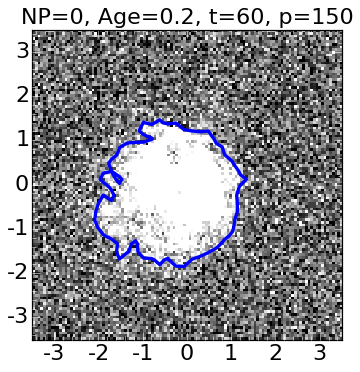}} &
\parbox[h]{0.2\textwidth}{\includegraphics[width=0.2\textwidth]{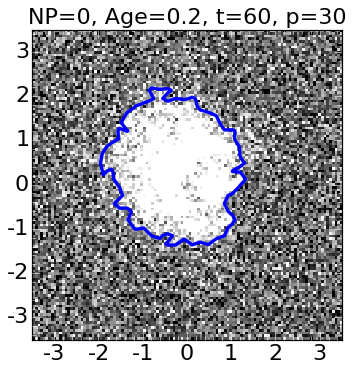}} &
\parbox[h]{0.2\textwidth}{\includegraphics[width=0.2\textwidth]{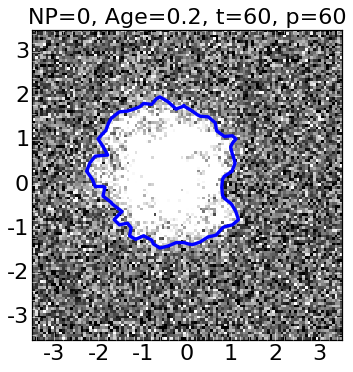}} &
\parbox[h]{0.2\textwidth}{\includegraphics[width=0.2\textwidth]{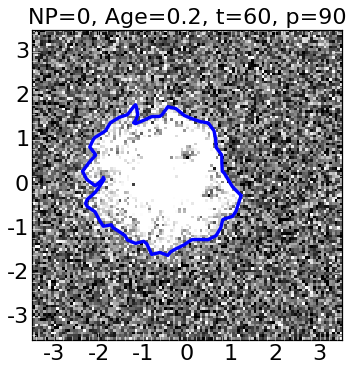}} \\
\parbox[h]{0.2\textwidth}{\includegraphics[width=0.2\textwidth]{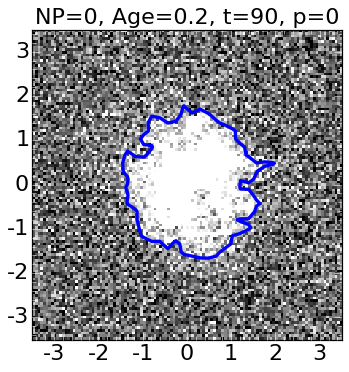}} &
\parbox[h]{0.2\textwidth}{\includegraphics[width=0.2\textwidth]{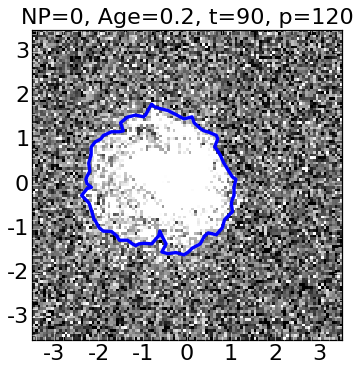}} &
\parbox[h]{0.2\textwidth}{\includegraphics[width=0.2\textwidth]{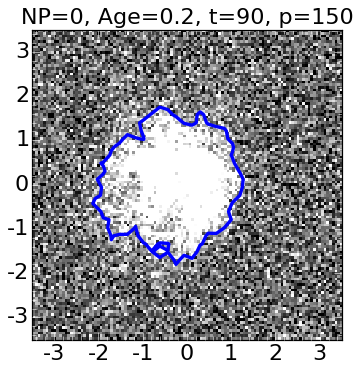}} &
\parbox[h]{0.2\textwidth}{\includegraphics[width=0.2\textwidth]{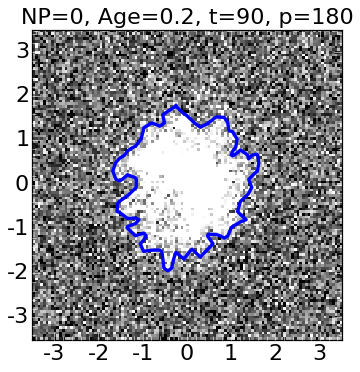}} \\
\parbox[h]{0.2\textwidth}{\includegraphics[width=0.2\textwidth]{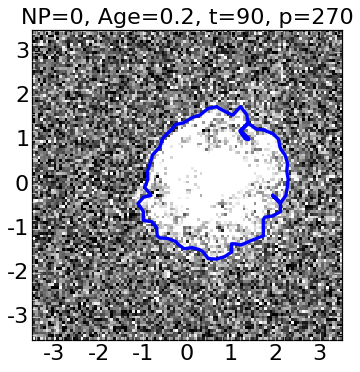}} &
\parbox[h]{0.2\textwidth}{\includegraphics[width=0.2\textwidth]{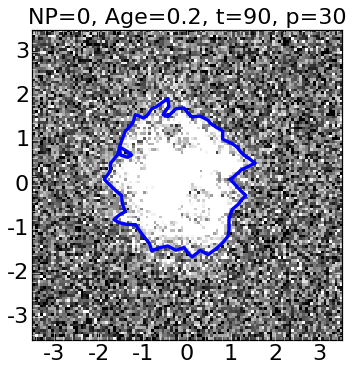}} &
\parbox[h]{0.2\textwidth}{\includegraphics[width=0.2\textwidth]{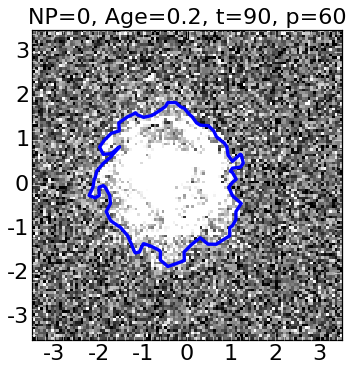}} &
\parbox[h]{0.2\textwidth}{\includegraphics[width=0.2\textwidth]{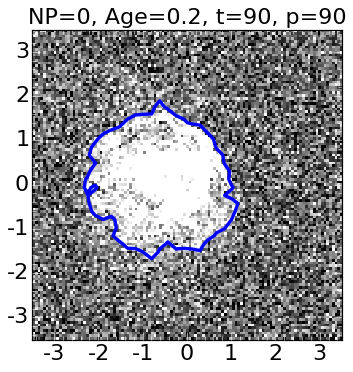}} \\
\parbox[h]{0.2\textwidth}{\includegraphics[width=0.2\textwidth]{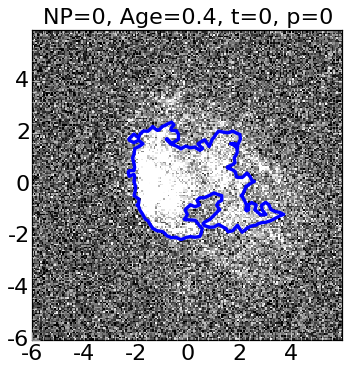}} &
\parbox[h]{0.2\textwidth}{\includegraphics[width=0.2\textwidth]{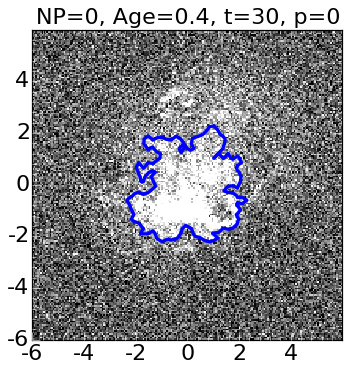}} &
\parbox[h]{0.2\textwidth}{\includegraphics[width=0.2\textwidth]{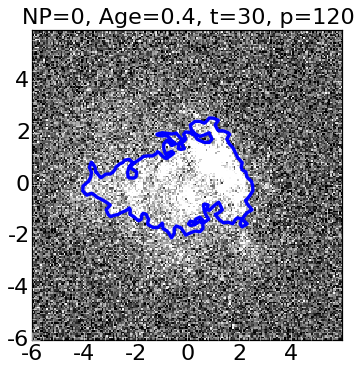}} &
\parbox[h]{0.2\textwidth}{\includegraphics[width=0.2\textwidth]{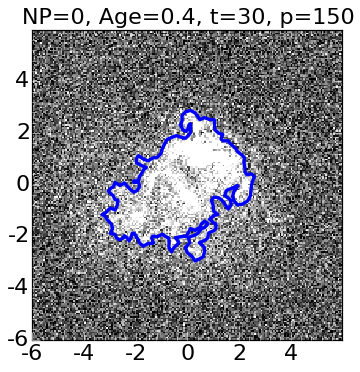}} \\
\parbox[h]{0.2\textwidth}{\includegraphics[width=0.2\textwidth]{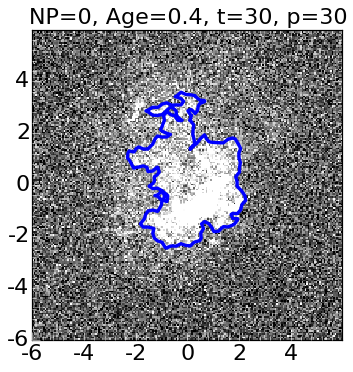}} &
\parbox[h]{0.2\textwidth}{\includegraphics[width=0.2\textwidth]{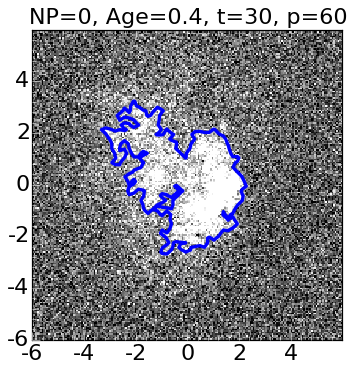}} &
\parbox[h]{0.2\textwidth}{\includegraphics[width=0.2\textwidth]{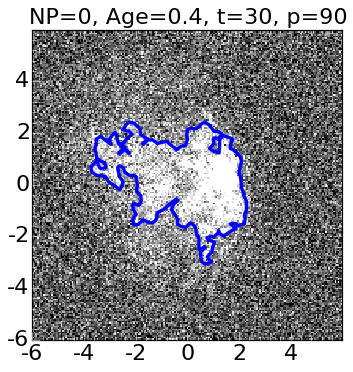}} &
\parbox[h]{0.2\textwidth}{\includegraphics[width=0.2\textwidth]{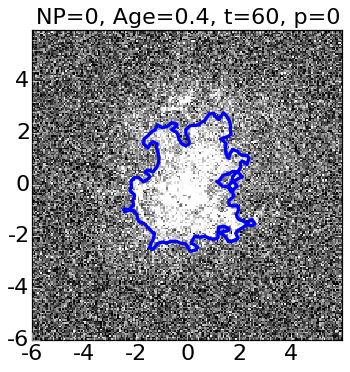}} \\
\parbox[h]{0.2\textwidth}{\includegraphics[width=0.2\textwidth]{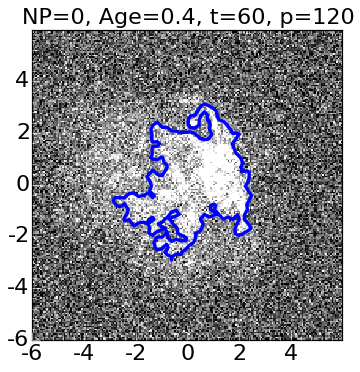}} &
\parbox[h]{0.2\textwidth}{\includegraphics[width=0.2\textwidth]{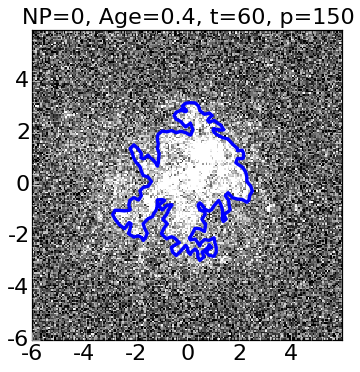}} &
\parbox[h]{0.2\textwidth}{\includegraphics[width=0.2\textwidth]{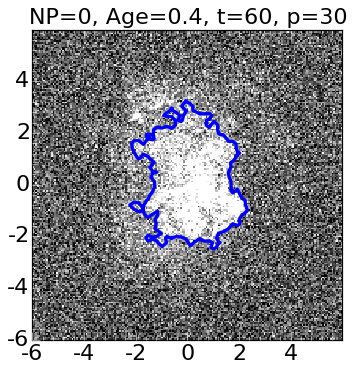}} &
\parbox[h]{0.2\textwidth}{\includegraphics[width=0.2\textwidth]{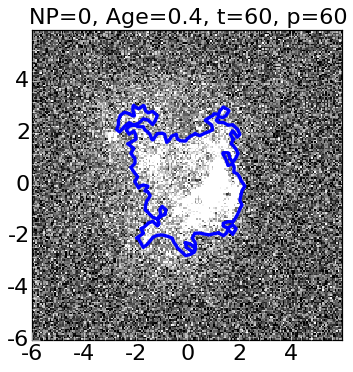}} \\
\parbox[h]{0.2\textwidth}{\includegraphics[width=0.2\textwidth]{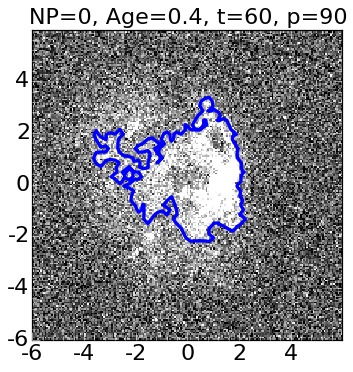}} &
\parbox[h]{0.2\textwidth}{\includegraphics[width=0.2\textwidth]{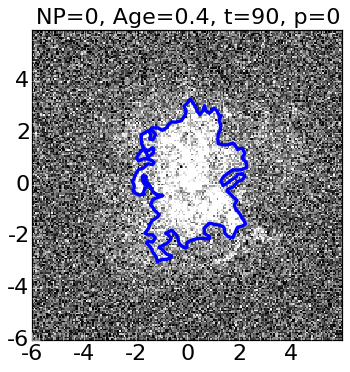}} &
\parbox[h]{0.2\textwidth}{\includegraphics[width=0.2\textwidth]{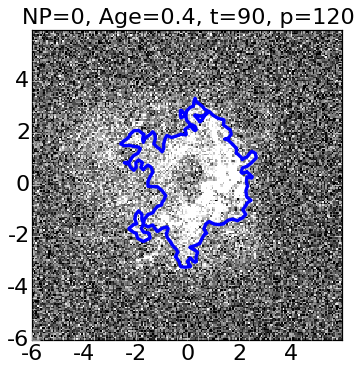}} &
\parbox[h]{0.2\textwidth}{\includegraphics[width=0.2\textwidth]{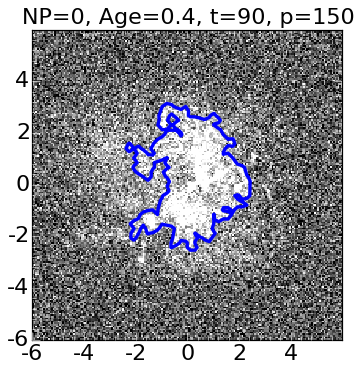}} \\
\parbox[h]{0.2\textwidth}{\includegraphics[width=0.2\textwidth]{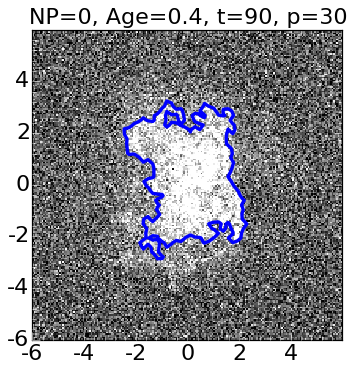}} &
\parbox[h]{0.2\textwidth}{\includegraphics[width=0.2\textwidth]{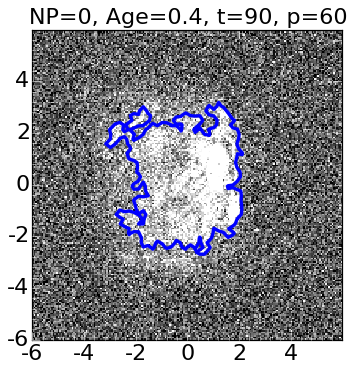}} &
\parbox[h]{0.2\textwidth}{\includegraphics[width=0.2\textwidth]{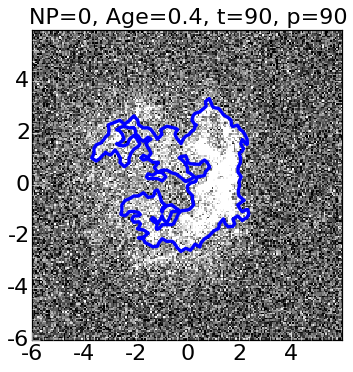}} &
\parbox[h]{0.2\textwidth}{\includegraphics[width=0.2\textwidth]{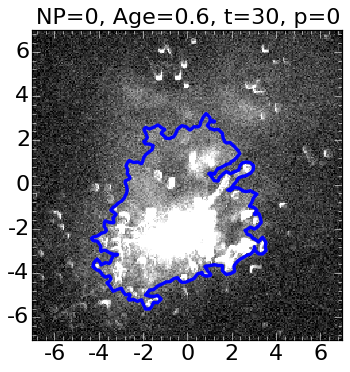}} \\
\parbox[h]{0.2\textwidth}{\includegraphics[width=0.2\textwidth]{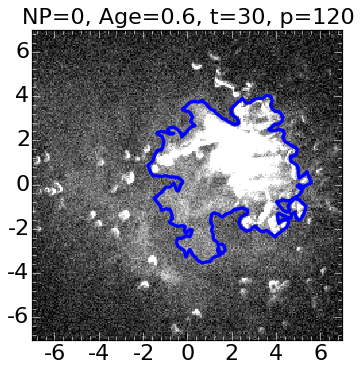}} &
\parbox[h]{0.2\textwidth}{\includegraphics[width=0.2\textwidth]{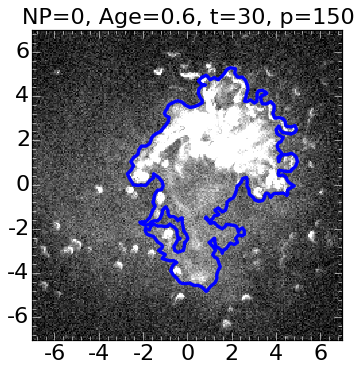}} &
\parbox[h]{0.2\textwidth}{\includegraphics[width=0.2\textwidth]{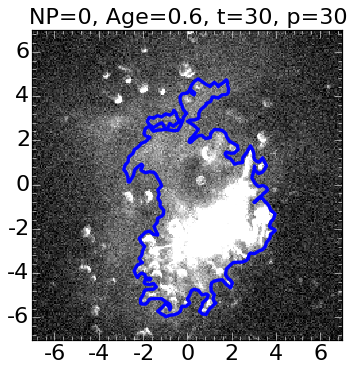}} &
\parbox[h]{0.2\textwidth}{\includegraphics[width=0.2\textwidth]{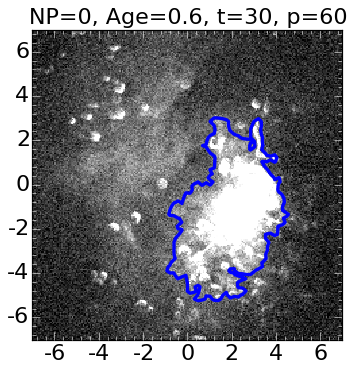}} \\
\parbox[h]{0.2\textwidth}{\includegraphics[width=0.2\textwidth]{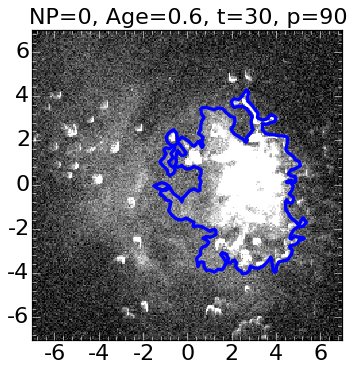}} &
\parbox[h]{0.2\textwidth}{\includegraphics[width=0.2\textwidth]{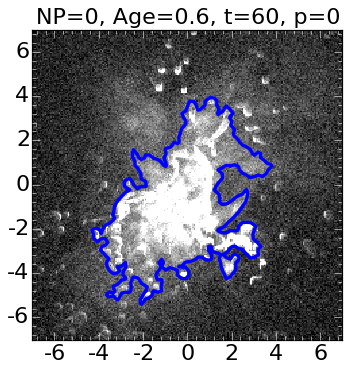}} &
\parbox[h]{0.2\textwidth}{\includegraphics[width=0.2\textwidth]{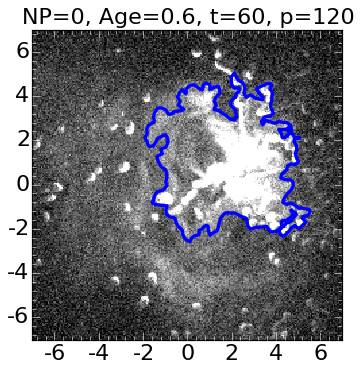}} &
\parbox[h]{0.2\textwidth}{\includegraphics[width=0.2\textwidth]{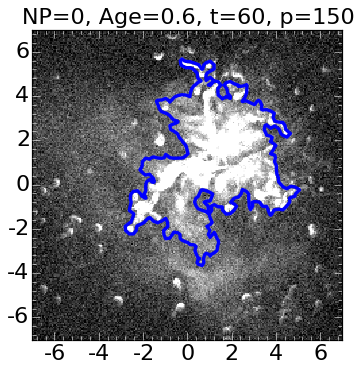}} \\
\parbox[h]{0.2\textwidth}{\includegraphics[width=0.2\textwidth]{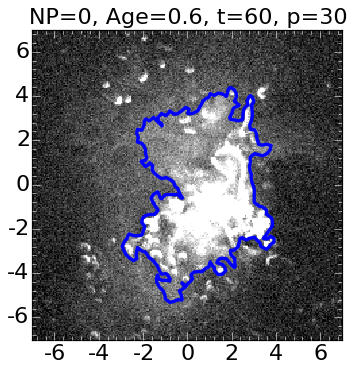}} &
\parbox[h]{0.2\textwidth}{\includegraphics[width=0.2\textwidth]{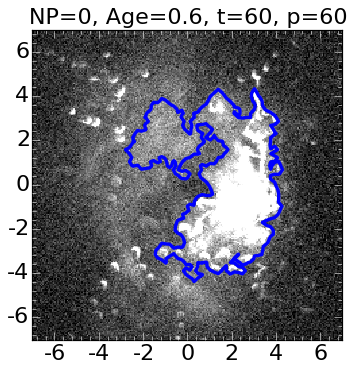}} &
\parbox[h]{0.2\textwidth}{\includegraphics[width=0.2\textwidth]{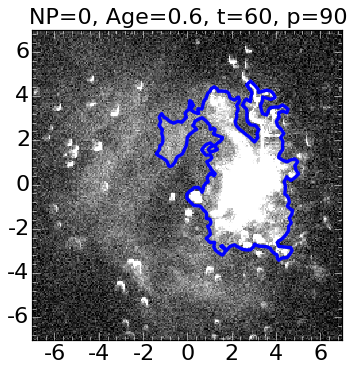}} &
\parbox[h]{0.2\textwidth}{\includegraphics[width=0.2\textwidth]{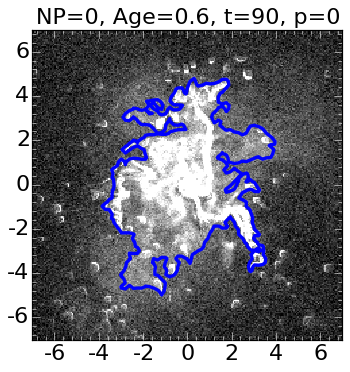}} \\
\parbox[h]{0.2\textwidth}{\includegraphics[width=0.2\textwidth]{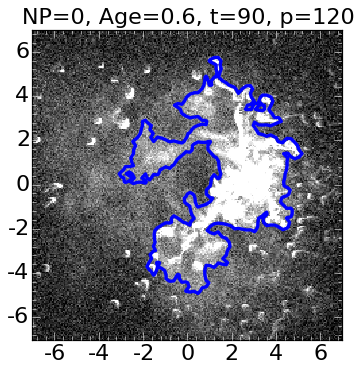}} &
\parbox[h]{0.2\textwidth}{\includegraphics[width=0.2\textwidth]{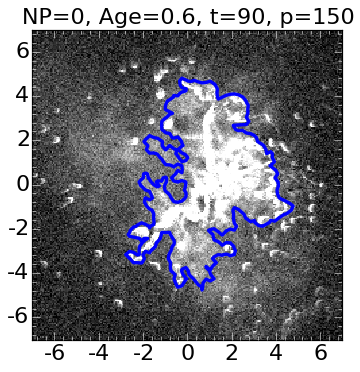}} &
\parbox[h]{0.2\textwidth}{\includegraphics[width=0.2\textwidth]{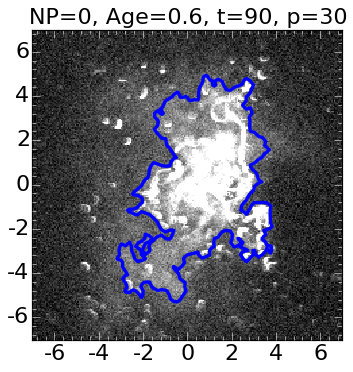}} &
\parbox[h]{0.2\textwidth}{\includegraphics[width=0.2\textwidth]{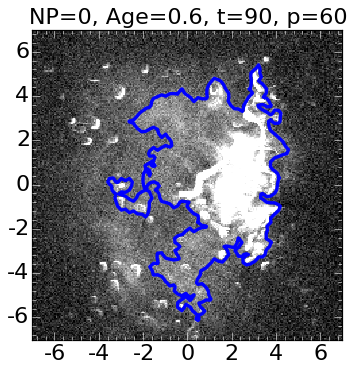}} \\
\parbox[h]{0.2\textwidth}{\includegraphics[width=0.2\textwidth]{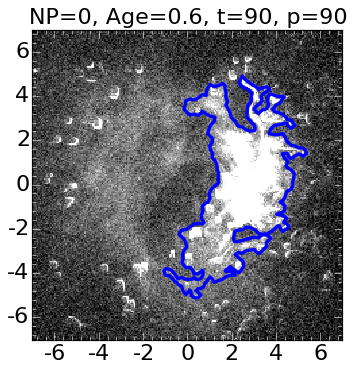}} &

\end{longtable}

\end{center}

%\end{landscape}

%\clearpage
%\newpage
%\input{ap_so_gauss_images}

%%%%%%%%%%%%%%%%%%%%%%%%%%%%%%%%%%%%%%%%%%%%%%%%%%

% Don't change these lines
\bsp	% typesetting comment
\label{lastpage}
\end{document}